\documentclass[11pt]{article}
\usepackage{amsmath}
\usepackage{latexsym}
\usepackage{epsf}
\def\IR{\relax{\rm I\kern-.18em R}}
\def\I1{\relax{\rm 1\kern-.40em 1}}
\def\IZ{\relax{\rm Z\kern-.40em Z}} 
\begin{document}
\thispagestyle{empty}
\begin{flushright} LPTENS-08/45 \end{flushright}
\vskip 1cm
\begin{center}
{\bf \Large PHYSICS BEYOND THE STANDARD MODEL}\\
\vskip 1cm

{\Large JOHN ILIOPOULOS}
\vskip 1cm

Laboratoire de Physique Th\'eorique \\ de L'Ecole Normale Sup\'erieure \\
75231 Paris Cedex 05, France

\vskip 5cm

We review our expectations in the last year before the LHC commissioning.

\vskip 3cm

Lectures presented at the 2007 European School of High Energy Physics
\vskip 1cm
Trest, August 2007
\end{center}

\newpage

\tableofcontents
\newpage

\section{The standard model}
\label{secSM}
One of the most remarkable achievements of modern theoretical physics has been the construction of the standard model for weak,
 electromagnetic and strong interactions. It is a gauge theory based on the group $ U(1) \otimes SU(2) \otimes  SU(3) $ which 
is spontaneously broken to $ U(1) \otimes SU(3) $. This relatively simple model epitomises our present knowledge of elementary
 particle interactions. It is analysed in detail in this School, so  I present only a short summary.

The model contains three types of fields:
 
\emph{(i)	The gauge fields.} There are twelve spin-one boson fields which belong to the adjoint, {\it i.e.} the
 $(1, 8) \oplus (3, 1) \oplus  (1, 1) $ representation of $SU(2) \otimes SU(3)$. The first eight are the gluons which mediate 
strong interactions between quarks and the last four are ($W^+, W^-, Z^0$ and $\gamma$), the vector bosons of the electroweak 
theory. The remarkable point is that the gauge fields are purely geometrical objects. Their number and their properties are uniquely determined by the gauge structure of the theory. In particular, they always belong to the adjoint representation. Once the group is given, everything is fixed.

\emph{(ii) Matter fields.} The basic unit is the ``family'' consisting exclusively of  spinor 
fields. Until a few years ago we believed that the neutrinos were massless and
this allowed us to use only fifteen two-component complex fields, which,
under $SU(2) \otimes SU(3)$, form the representation $(2, 1) \oplus (1, 1)
\oplus (2, 3) \oplus 
(1, 3) \oplus (1, 3)$. In other words, we had no right-handed neutrino. It is still probable that this is correct, but since I shall not discuss the neutrino masses in any detail here, I will include a $\nu_{e_R}$  with the basic unit. If it is not needed, it will end up being a free field. The prototype is the electron family:

\begin{equation}
\label{eqfam}
\left.  \begin{array}{cccccc}
\left( \begin{array}{c}
\nu_e\\e^- \end{array} 
\right)_L ; & \nu_{e_R} ; & e_R ; &
\left( \begin{array}{c}
u_i \\ d_i \end{array}
\right)_L ; & u_{i_R} ; & d_{i_R} \\
  &  &  &  &  &  \\
(2,1) & (1,1) & (1,1) & (2,3) & (1,3) & (1,3)
\end{array} 
 \right\} i=1,2,3  
\end{equation} 

\vskip 0.3cm
\noindent For the muon family $\nu_e \rightarrow \nu_{\mu}; e \rightarrow \mu; u_i \rightarrow c_{i}$ and $d_i \rightarrow s_i $ and, 
similarly, for the tau family, $\nu_e \rightarrow \nu_{\tau}; e \rightarrow \tau; u_i \rightarrow t_{i}$ and $d_i \rightarrow 
b_i $. At first sight, the introduction of massive neutrinos puts leptons and quarks on equal footing and, indeed, neutrino masses are also described by a $3 \times 3$ mass-matrix. As a consequence, the number of parameters one should measure in order to determine the Standard Model Lagrangian is considerably increased and this opens a new chapter in experimental research. Although the chapter is far from being closed, the smallness of the resulting values suggests that their origin may belong to physics beyond the Standard Model. 

A remarkable property is that the sum of the electric charges in each family vanishes. This turns out to be necessary 
for the cancellation of triangle anomalies in the Ward identities of axial currents and hence, for the construction of a 
renormalisable theory. On the other hand we shall see later that the same property plays a crucial role in grand unified 
theories. This family structure is an as yet unexplained feature of the theory. It gives strong predictions for the existence 
of new species of particles. For example, the experimental discovery of the tau lepton signaled the opening of the third 
family and  gave a prediction for the existence of two new quarks,  $t$ and  $b$.   
 The so-called ``family problem'' is that the total number of families is not restricted by the theory. In fact, we know of no 
good reason why any, beyond the first one, should exist. This, of course, is
related to the fact that the matter fields, contrary to the gauge bosons, are
not geometrical objects. Given the group, we can consider arbitrary
representations with arbitrary spins. It so happens that Nature seems, up to
now, to use only fundamental representations, doublets and triplets, with spin
one-half fermions, but simplicity is the only reason we can think of. 

\emph{(iii) Higgs scalar fields.} We would be very happy if we could live with only the first two kinds of fields but, in fact,
 we need a third one, the scalar Higgs fields. Their nonzero vacuum expectation values break the gauge symmetry spontaneously,
 thus providing masses to the gauge bosons $W^+, W^- $ and $Z^0$ as well as the fermions. In the standard model this is 
accomplished with a complex doublet of scalar fields. At the end one neutral spin-zero boson survives as a physical particle. 
There are no severe restrictions on its mass. The data favour a rather low value, close to the experimental limit which is currently 114 GeV. If $m_{\phi} \geq 1 $TeV a sector of the theory becomes strongly interacting and 
perturbation theory breaks down. In the absence of any concrete experimental evidence, one is left to speculate on the number 
of Higgs particles as well as on their elementary or composite nature. Whichever the ultimate answer to these questions may be,
 we can say that the Higgs sector is at present the least understood and probably the most interesting sector of gauge 
theories. An important aim for LHC is precisely the experimental probe of this sector. I shall come back to this question presently.

Our confidence in this model is amply justified on the basis of its ability to accurately describe the bulk of our present day 
data and, especially, of its enormous success in predicting new phenomena. A short list of these successes includes:

(i) The weak neutral currents: Not only their existence, but also their main properties were predicted. In general we would 
expect, for every flavour, a parameter that determines the strength of the neutral current relatively to the charged one and 
another to fix the ratio of the vector and axial parts. In the simplest model, in which the breaking comes through isodoublet 
scalars, they are all expressible in terms of a single one, the angle $\theta_W $. This is brilliantly confirmed by the fact 
that the values of $\theta_W $ measured in various experiments coincide.

(ii) The charmed particles were predicted and they were found to decay predominantly to strange particles, thus confirming 
the theoretical prediction.

(iii) As we mentioned already, each family must be complete. Therefore the discovery of a new lepton ($\tau $) was interpreted 
as the opening of a third family. Indeed the $b$- and $t$-quarks were discovered. 

(iv) The experimental discovery of the intermediate vector bosons $W$ and $Z$, with the accurately predicted masses and decay 
properties, has been one of the most remarkable achievements of  accelerator technology and experimental high energy 
physics.

(v) The large amount of data which have been accumulated in recent years concerning high energy and large $p_T$ physics are 
correctly described by the Standard Model, including quantum chromodynamics, the gauge theory of strong interactions. In the above kinematic region 
asymptotic freedom has set in and perturbation theory is meaningful.

\section{Waiting for the L.H.C.}
\subsection{An impressive global fit}

\begin{figure}
\centering
\epsfxsize=10cm
\epsffile{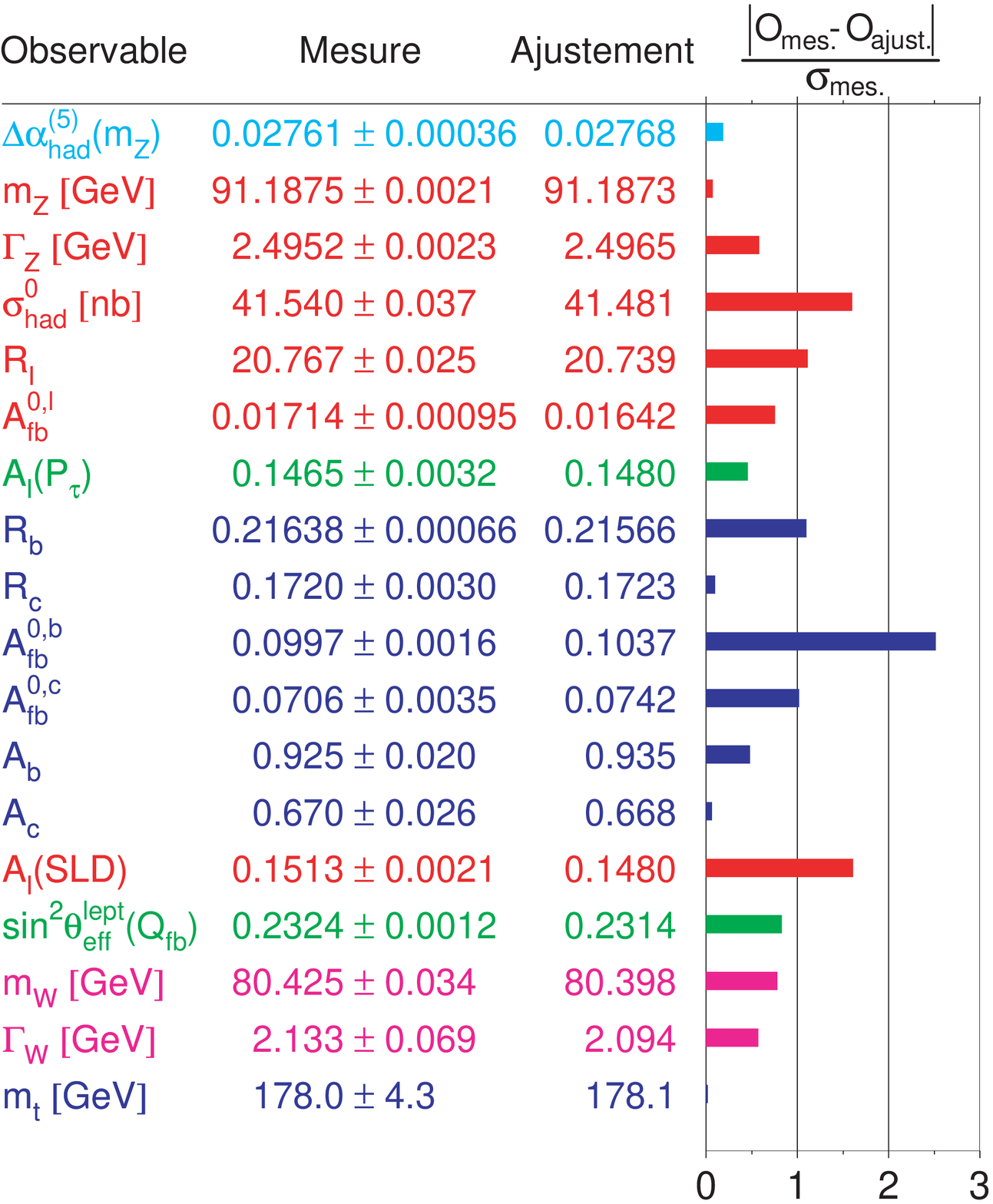}
\caption {Various physical quantities measured and computed.} \label{globfit}
\end{figure}

All these spectacular successes of the standard model are in fact successes of renormalised perturbation theory. 
Indeed what we have learnt was how to apply the methods which had been proven so powerful in quantum electrodynamics, 
to other elementary particle interactions. The remarkable quality of modern High Energy Physics experiments, mostly at LEP, 
but also elsewhere, has provided us with a large amount of data of unprecedental accuracy. All can be fit using the 
Standard Model with the Higgs mass as the only free parameter. Let me show some examples: Figure \ref{globfit} indicates the overall quality of such 
a fit. There are a couple of measurements which lay between 2 and 3 standard deviations away from the theoretical predictions, but
it is too early to say whether this is accidental, a manifestation of new physics, or the result of incorrectly combining incompatible
experiments.

\begin{figure}
\centering
\epsfxsize=6cm
\epsffile{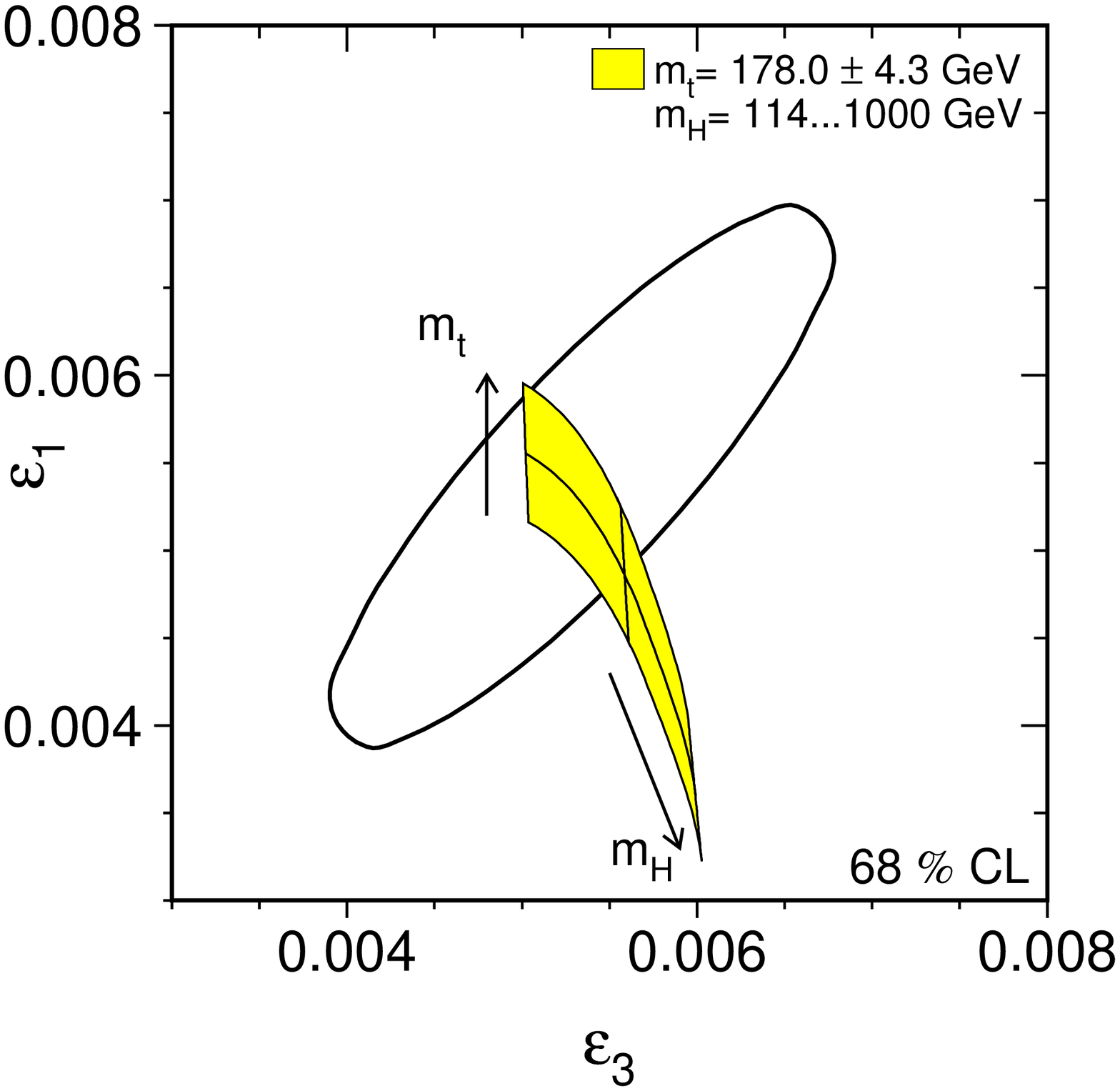}
\caption {The importance of the Standard Model radiative corrections. The
  arrows show how the prediction moves when we vary $m_t$ and $m_H$, in
  particular if we use the most recent lower value for  $m_t$.} \label{epsilon}
\end{figure}

Another impressive fit concerns the strong interaction effective coupling constant as a function of the momentum scale (Figure \ref{alphas})\footnote{The precision has been considerably increased recently. See M. Davier et al arXiv 0803.0979.}. This fit already shows the importance of taking into account the radiative corrections, since, in the tree approximation, $\alpha_s$ is, obviously, a constant. Similarly, Figure \ref{epsilon} shows the importance of the weak radiative corrections in the framework of the Standard Model. Because of the special Yukawa couplings, the dependence of these corrections on the fermion masses is quadratic, while it is only logarithmic in the Higgs mass. The $\epsilon$ parameters are designed to disentangle the two. The ones we use in Figure \ref{epsilon} are defined by:

\begin{figure}
\centering
\epsfxsize=6cm
\epsffile{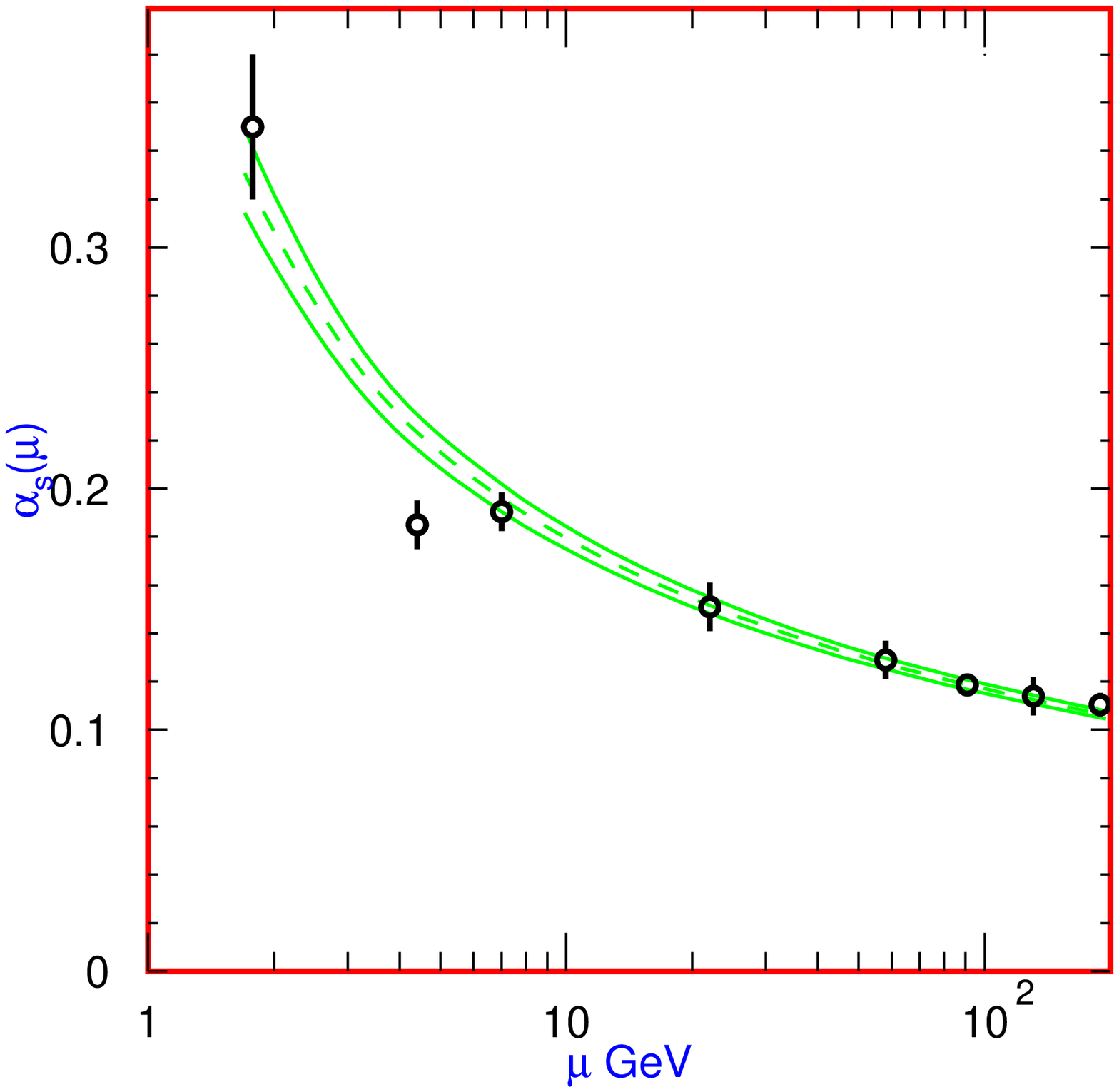}
\caption {The variation of $\alpha_s$ with the momentum scale. The renormalisation group prediction and the experimental points.} \label{alphas}
\end{figure}

\begin{equation}
\label{epspar1}
\epsilon_1=\frac{3G_Fm_t^2}{8\sqrt{2}\pi^2}-\frac{3G_Fm_W^2}{4\sqrt{2}\pi^2}\tan^2\theta_W \ln \frac{m_H}{m_Z}+...
\end{equation}

\begin{equation}
\label{epspar3}
\epsilon_3=\frac{G_Fm_W^2}{12\sqrt{2}\pi^2}\ln \frac{m_H}{m_Z}-\frac{G_Fm_W^2}{6\sqrt{2}\pi^2} \ln \frac{m_t}{m_Z}+...
\end{equation}
where the dots stand for subleading corrections. As you can see, the $\epsilon$'s vanish in the absence of weak interaction radiative corrections, in other words, $\epsilon_1=\epsilon_3=0$ are the values we get in the tree approximation of the Standard Model but including the purely QED and QCD radiative corrections. We see clearly in Figure \ref{epsilon} that this point is excluded by the data. The latest values for these parameters are $\epsilon_1=5.4 \pm 1.0$ and $\epsilon_3=5.34 \pm 0.94$. 

\begin{figure}
\centering
\epsfxsize=6cm
\epsffile{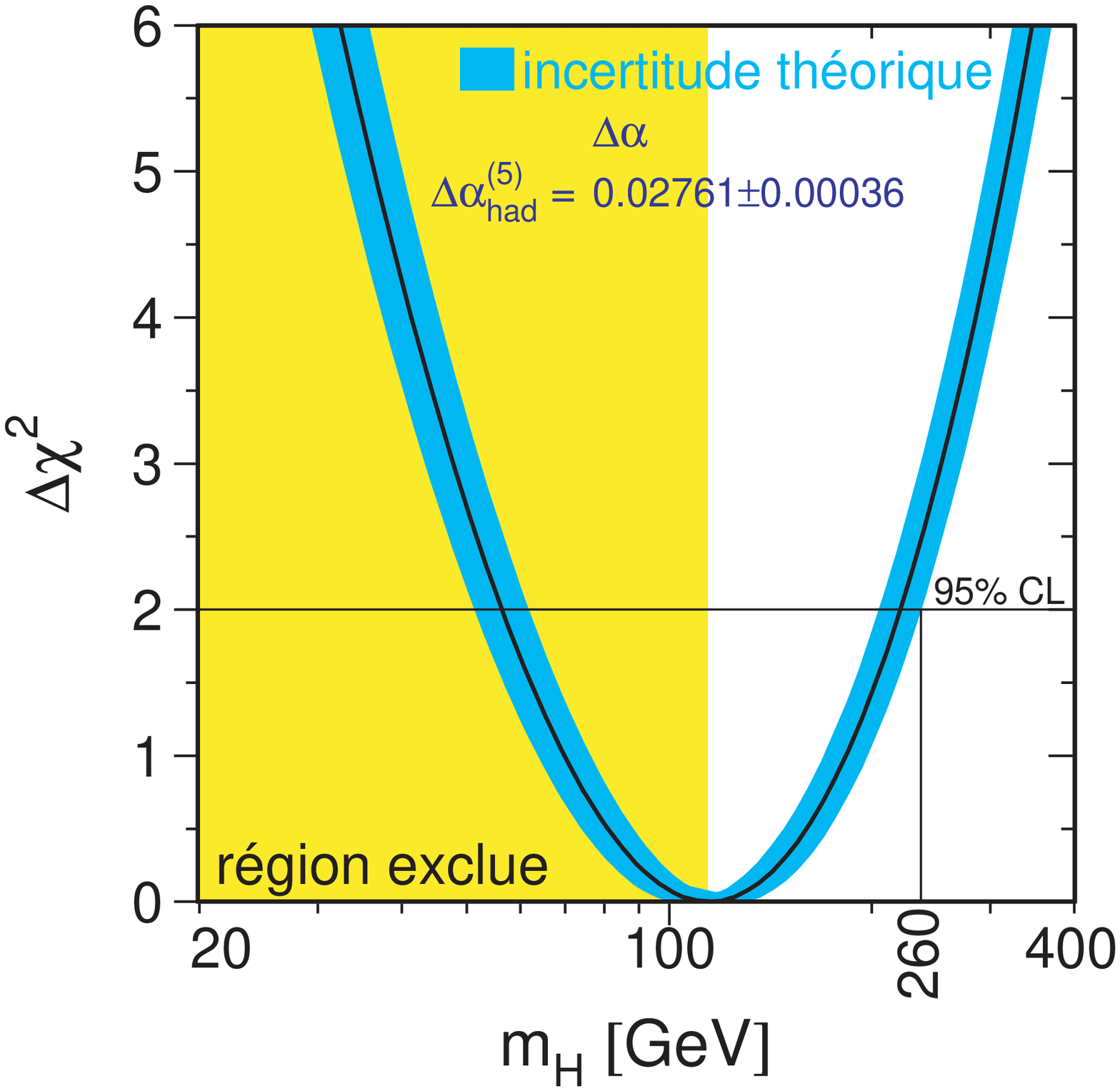}
\caption {The predicted values for the Standard Model Higgs mass using all available data. The shaded region is excluded by direct searches.} \label{limhiggs}
\end{figure}

Using all combined data we can extract the predicted values for the Standard
Model Higgs mass which are given in Figure \ref{limhiggs}. The data clearly
favour a low mass ($\leq 200$ GeV) Higgs, although, this prediction may be less
solid than what Figure \ref{limhiggs} seems to indicate. 

\subsection{Bounds on the Higgs mass}

A more detailed discussion of the quality of the fit has been presented in
this School, so I go directly to what can be expected at the LHC. First thing
LHC expects to find is, of course, the Higgs boson. Let me then ask a very general question: In the framework of the Standard
Model, is it possible to predict the value of the Higgs mass? The question can
be rephrased as follows: The model contains a large number of arbitrary
parameters which must be determined by experiment. Is it possible to find some
relation among some of these parameters such that it remains stable against
higher order corrections and it is not  spoiled by arbitrary counterterms? The only parameter which has not yet been directly measured is the
Higgs boson mass, or, alternatively, the coupling constant $\lambda$ of the
Higgs self-interaction. An example of such relation which has been extensively
studied is of the form:

\begin{equation}
\label{ncrel1}
m_Z/m_H=C
\end{equation}
with $C$ a constant. At the classical level such relation is indeed obtained
if one formulates the model in a suitably chosen space with non-commutative
geometry which allows for a unified picture of both Higgs and gauge fields. In
the tree approximation we have: 

\begin{equation}
\label{relation}
C=\frac{m_Z}{m_H}=\frac{\sqrt{g_1^2+g_2^2}}{\sqrt{8\lambda}}
\end{equation}

So, the question is: is there any renormalisation scheme, no matter how
complicated in practice, in which the relation (\ref{ncrel1}), or (\ref{relation}) does not
receive an infinite counterterm? The general theory of renormalisation tells
us that, if such a relation is stable, it corresponds to a zero of the $\beta$-function for the combination of the
coupling constants which appears at the r.h.s. of (\ref{relation}). The
important point is that the first coefficient of the $\beta$-function is
universal, 
independent of the renormalisation scheme. For the
purely bosonic sector of the theory a simple
one-loop calculation gives:

\begin{equation}
\label{smbeta1}
\begin{split}
16\pi^2\beta_{g_1} & =g_1^3\frac{1}{10}\\
16\pi^2\beta_{g_2} & =-g_2^3\frac{43}{6}\\
16\pi^2\beta_{\lambda} &
=12\lambda^2-\frac{9}{5}g_1^2\lambda-9g_2^2\lambda+\frac{27}{100}g_1^4+\frac{9}{10}g_1^2g_2^2+\frac{9}{4}g_2^4
\end{split}
\end{equation}

 Notice that $\lambda$ must be positive, otherwise the classical Higgs
potential is unbounded from below. For the combination (\ref{relation}) we obtain:

\begin{equation}
\label{smz1}
\begin{split}
\beta_z & =\beta_{\eta_1}+\beta_{\eta_2}=\\
 & =\frac{-\lambda w}{16\pi^2 \rho z}\left [ \left
 (\frac{27}{100}\rho^2+\frac{9}{10}\rho+\frac{9}{4}\right )z^2-\left
 (2\rho^2+\frac{54}{5}\rho -\frac{16}{3}\right )z+12(\rho +1)^2\right ]
\end{split}
\end{equation}
where we defined

\begin{equation}
\label{smeta1}
\eta_1=\frac{g_1^2}{\lambda} ~~;~~ \eta_2=\frac{g_2^2}{\lambda}~~;~~ z=\eta_1+\eta_2~~;~~\rho=\frac{\eta_1}{\eta_2}~~;~~w=\eta_1 \eta_2
\end{equation}

It is easy to check that the quadratic form in the r.h.s. of (\ref{smz1})
never vanishes for real and positive $z$ and $\rho$. This implies that the
relation (\ref{relation}) will be violated in one loop, no matter which
renormalisation scheme one is using. Including the fermions does not avoid
this result and a similar conclusion can be drawn for any
relation of this type. The conclusion is that the set of parameters of the
Standard Model appears to be irreducible. This does not mean that it is
impossible to predict the mass of the Higgs boson. It only means that the
origin of such a relation should come from physics beyond the Standard Model
in which the latter is embedded in a larger scheme with tighter structure and
richer particle content.

Given this result, let us see what, if any, are the theoretical constraints. The Standard Model Higgs mass is given, at the classical level, by $m_H^2=2\lambda v^2$, with $v$ the vacuum expectation value of the Higgs field. $v$ is fixed by the value of the Fermi coupling constant $G_F/\sqrt 2 =1/(2v^2)$ which implies $v\approx$246 GeV. Therefore, any constraints will come from the allowed values of $\lambda$. A first set of such constraints is given by the classical requirement:

\begin{equation}
\label{clconstr}
1>\lambda >0 ~~~~\Rightarrow ~~~~ m_H<400-500 GeV
\end{equation}
 
The lower limit for $\lambda$ comes from the classical stability of the
theory. If $\lambda$ is negative the Higgs potential is unbounded from below
and there is no ground state. The upper limit comes from the requirement of
keeping the theory in the weak coupling regime. If $\lambda \geq 1$ the Higgs
sector of the theory becomes strongly interacting and we expect to see
plenty of resonances and bound states rather than a single elementary
particle. 

Going to higher orders is straightforward, using the renormalisation group
equations. The running of the effective mass is determined by that of
$\lambda$. Keeping only the dominant terms and assuming $t=\log (v^2/\mu^2)$ is
small ($\mu \sim v$), we find 

\begin{figure}
\centering
\epsfxsize=12cm
\epsffile{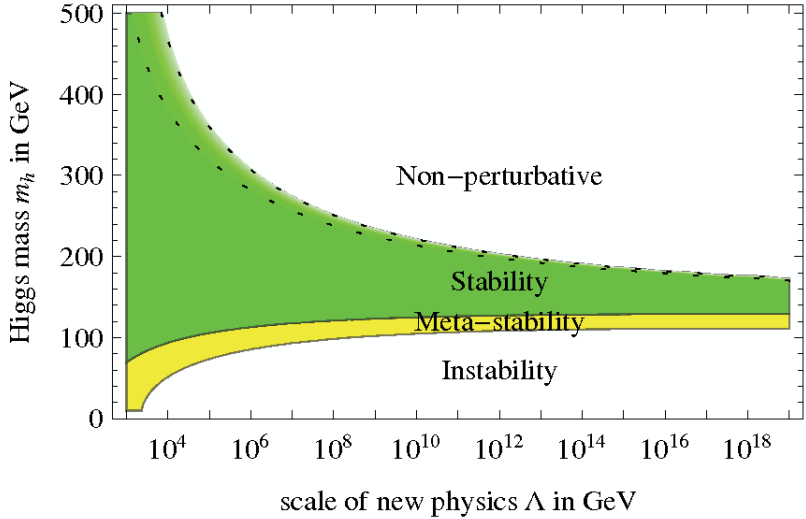}
\caption {Bounds on the Higgs mass.} \label{bounds}
\end{figure}

\begin{equation}
\label{quconstr1}
\frac{d\lambda}{dt}=\frac{3}{4\pi^2}[\lambda^2+3\lambda h_t^2-9h_t^4+...]
\end{equation}
where $h_t$ is the coupling of the Higgs boson to the top quark. The dots
stand for less important terms, such as the other Yukawa couplings to the
fermions and the couplings with the gauge bosons. This equation is correct as
long as all couplings remain smaller than one, so that perturbation theory is
valid, and no new physics beyond the standard model becomes important. Now we
can repeat the argument
on the upper
and lower bounds for $\lambda$ but this time taking into account the full
scale dependence $\lambda (\mu)$. We thus obtain for the Higgs mass an upper bound given by
the requirement of weak coupling regime ($\lambda (\mu)<1$) all the way up to
the scale $\mu$, and a lower bound
by the requirement of vacuum stability ($\lambda (\mu)>0$), again up to $\mu$. Obviously, the
bounds will be more stringent the larger the assumed value of $\mu$. Figure
\ref{bounds} gives the allowed region for the Higgs mass as a function of the
scale for scales up to the Planck mass. We see that for small $\mu \sim 1$TeV,
the limits are, essentially, those of the tree approximation equation (\ref{clconstr}),
while for $\mu \sim M_P$ we obtain only a narrow window of allowed masses
130GeV$<m_H<$200GeV, remarkably similar to the experimental results.

\subsection{New Physics}

{\it Looking at all the
data, from low energies to the Tevatron, we have learnt that perturbation
theory is remarkably successful, outside the specific regions where strong
interactions are important.} 

\begin{figure}
\centering
\epsfxsize=12cm
\epsffile{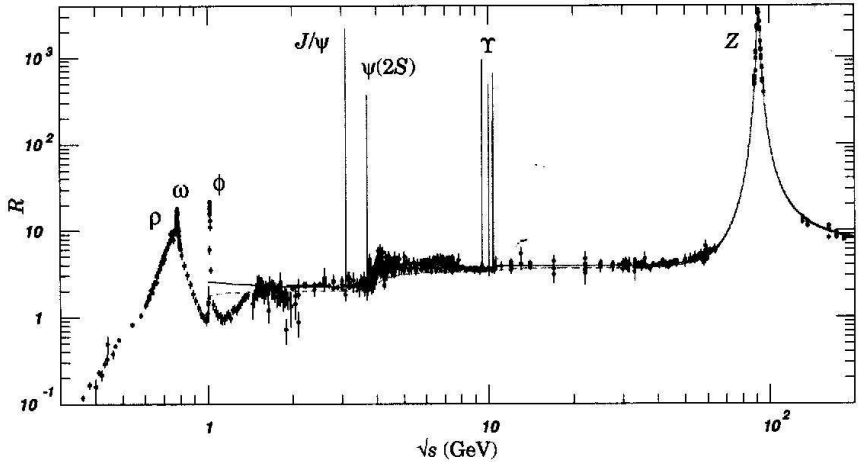}
\caption {The ratio $R$ of $e^+ +e^-$ total cross section to
hadrons normalised to that of $e^+ +e^- \rightarrow \mu^+ +\mu^-$ as a
function of the centre-of-mass energy.} \label{R}
\end{figure}

Let me explain this point better: At any given model with a coupling constant
$g$ we expect to have a weak coupling region $g\ll 1$, in which weak coupling
expansions, such as perturbation theory, are reliable, a strong coupling
region with $g\gg 1$, in which strong coupling expansions may be relevant, and
a more or less large gray region $g\sim 1$, in which no expansion is
applicable. The remarkable conclusion is that this gray
area appears to be extremely narrow. And this is achieved by an enlargement of
the area in which weak coupling expansion applies. The perturbation expansion is reliable,
not only for very small couplings, such as $\alpha_{em}\sim 1/137$, but also
for moderate QCD couplings $\alpha_s \sim 1/3$, as shown in Figure
\ref{alphas}. This is extremely important because without this property no
calculation would have been possible. If we had to wait until $\alpha_s$ drops
to values as low as $\alpha_{em}$ we could not use any available
accelerator. Uncalculable QCD backgrounds would have washed out any
signal. And this applies, not only to the Tevatron and LHC, but also to LEP. A
global view of the weak and strong coupling regions is given in Figure \ref{R}
which shows the $R$-ratio, {\it i.e.} the $e^+ +e^-$ total cross section to
hadrons normalised to that of $e^+ +e^- \rightarrow \mu^+ +\mu^-$ as a
function of the centre-of-mass energy. The lowest order perturbation value for
this ratio is a constant, equal to $\Sigma Q_i^2$, the sum of the squares of
the quark charges accessible at this energy. We see clearly in this Figure the
areas of applicability of perturbation theory: At very low energies, below 1
GeV, we are in the strong coupling regime characterised by resonance
production. The strong interaction effective coupling constant becomes of
order one (we can extrapolate from Figure \ref{alphas}), and perturbation
breaks down. However, as soon as we go slightly above one GeV, $R$ settles to
a constant value and it remains such except for very narrow regions when new
thresholds open. In these regions the cross section is again dominated by
resonances and perturbation breaks down. But these areas are extremely well
localised and  threshold effects do not spread outside these small regions. 

I want to exploit this experimental fact and argue that the available
precision tests of the Standard Model allow us to claim with confidence that
new physics will be unravelled at the LHC. The argument is based on the fact
that, whichever new physics may appear at an as yet unaccessible energy scale, 
it influences physics at present energy through the higher order radiative
corrections. Therefore, precision measurements at the LEP and Tevatron scales
allow us to guess new physics at the LHC scale. The argument assumes the
validity of perturbation theory and it will fail if the latter fails. But, as
we just saw, perturbation theory breaks down only when strong interactions
become important. But new strong interactions imply also new physics. 

Let us illustrate the argument with two examples, one with a
non-renormalisable theory and one with a renormalisable one. A quantum field
theory, whether renormalisable or not, should be viewed as an effective theory
valid up to a given scale $\Lambda$. It makes no sense to assume a theory
for all energies, because we know already that at very high energies entirely
new physical phenomena appear (example: quantum gravity at the Planck
scale). The first example is the Fermi four-fermion theory with a coupling
constant $G_F \sim 10^{-5}GeV^{-2}$. It is a non-renormalisable theory and, at
the $n$th order of perturbation, the $\Lambda$ dependence of a given quantity
$A$ is given by:

\begin{equation}
A^{(n)}=C_0^{(n)} (G_F \Lambda^2)^n+ C_1^{(n)}G_F (G_F \Lambda^2)^{n-1}+
C_2^{(n)}G_F^2 (G_F \Lambda^2)^{n-2}+....
\label{Fermiexp}
\end{equation}
where the $C_i$'s are functions of the masses and external momenta, but their
dependence on $\Lambda$ is, at most, logarithmic. Perturbation theory breaks
down obviously when $A^{(n)} \sim A^{(n+1)}$ and this happens when $G_F
\Lambda^2 \sim 1$. This gives a scale of $\Lambda \sim 300$GeV as an upper
bound  for the validity of the Fermi theory. Indeed, we know to-day that at
100GeV the $W$ and $Z$ bosons change the structure of the theory. But, in
fact, we can do much better than that. Weak interactions violate some of the
conservation laws of strong interactions, such as parity and strangeness. The
absence of such violations in precision measurements will tell us that $G_F
\Lambda^2 \sim \epsilon$ with $\epsilon$ being the experimental precision. The
resulting limit depends on the value of the $C$ coefficient for the quantity
under consideration. In this particular case it turned out that, under the
assumption that the chiral symmetry of strong interactions is broken only by
quark mass terms, the coefficient $C_0^{(n)}$ for parity and/or strangeness
violating amplitudes vanishes and no new limit is obtained. However, the
second order coefficient $C_1^{(n)}$ contributes to flavour changing neutral
current transitions and the smallness of the $K_1-K_2$ mass difference, or the
$K^0_L \rightarrow \mu^+ +\mu^-$ decay amplitude, give a limit of $\Lambda
\sim 3$GeV before new physics should appear. The new physics in this case turned out to be the charmed
particles. We see in this example that the scale $\Lambda$ turned out to be rather low and this is due to the non-renormalisable nature of the effective theory which implies a power-law behaviour of the radiative corrections on $\Lambda$. 

The second example in which new physics has been discovered through its
effects in radiative corrections is the well-known ``discovery'' of the $t$
quark at LEP, before its actual production at Fermilab. The effective theory is now the Standard Model, which is renormalisable. In this case the dependence of the radiative corrections on the scale $\Lambda$ is, generically, logarithmic and the sensitivity of the low energy effective theory on the high scale is weak (there is an important exception to this rule for the Standard Model which we shall see presently). In spite of that,  the discovery
was made possible because of the special property of the Yukawa coupling constants in the Standard Model to be proportional to the fermion mass. Therefore, the effects of the top quark in the radiative corrections are quadratic in $m_t$. The LEP precision measurements were able to extract a very accurate prediction for the top mass.  

I claim that we are in a similar situation with the precision measurements of
the Standard Model. We know that new physics will appear at the LHC scale, although
we have no unique answer on the nature of this new physics. We can only look
at various possibilities. 

The key is again the Higgs boson. As we explained above, the data favour a low mass Higgs. However,
the opposite cannot be excluded, first because it depends on the subset of the
data one is looking at\footnote{This prediction is, in fact, an average
  between a much lower value, around 50 GeV, given by the data from leptonic
  asymmetries, and a much higher one, of 400 GeV, obtained from the hadronic
  asymmetries. Although the difference sounds dramatic, the two are still
  mutually consistent at the level of 2-3 standard deviations.}, and, second,
because the analysis is done taking the minimal standard model. 

So, let me take as a first possibility the one I consider as less likely, namely
the absence of a light Higgs (by ``light'' I mean less than a few hundred
GeV). This does not necessarily mean ``no-Higgs'', because a very heavy
Higgs, above 1 TeV, is not expected to appear as an elementary particle. As we
explained above, this will be accompanied by new strong interactions. A
particular version of this possibility is the ``Technicolor'' model,  which
assumes the existence of a new type of fermions with strong interactions at
the multi-hundred-GeV scale. 
The role of the Higgs is played by a fermion-antifermion bound state. ``New
Physics'' here is precisely the discovery of a completely new sector of
elementary particles. Other strongly interacting models have been constructed
and we shall mention some of them in these lectures. The general conclusion
here is that a heavy Higgs always implies new forces whose effects are
expected to be visible at the LHC.\footnote{We can build specific models in which the
  effects are well hidden and pushed above the LHC discovery
  potential. In this case one would need very high precision measurements,
  probably with a multi-TeV $e^+-e^-$ collider.}

The possibility which seems to be favoured by the data is the presence of a ``light'' Higgs particle. In this case new strong interactions are not needed and, therefore, we can assume that perturbation theory remains valid. But then we are faced with a new problem. The Standard Model is a renormalisable theory and the dependence on the high energy scale is expected to be only logarithmic. This is almost true, but with one notable exception: The radiative corrections to the Higgs mass are quadratic on whichever scale $\Lambda$ we are using. The technical reason is that $m_H$ is the only parameter of the Standard Model which requires, by power counting, a quadratically divergent counterterm. The gauge bosons require no mass counterterm at all because they are protected by gauge invariance and the fermions need only a logarithmic one. The physical reason is that, if we put a fermion mass to zero we increase the symmetry of the model because now we can perform chiral transformations on this fermion field. Therefore the massless theory will require no counterterm, so the one needed for the massive theory will be proportional to the fermion mass and not the cut-off. In contrast, putting $m_H=0$ does not increase the symmetry of the model.\footnote{At the classical level, the Standard Model with a massless Higgs does acquire a new symmetry, namely scale invariance, but this symmetry is always broken for the quantum theory and offers no protection against the appearance of quadratic counterterms.} As a result the effective mass of the Higgs boson will be given by 

\begin{equation}
\label{mhiggseff}
(m_H^2)_{eff}=m_H^2+C\alpha_{eff}\Lambda^2
\end{equation}   
where $C$ is a calculable numerical coefficient of order one and $\alpha_{eff}$ some effective coupling constant. In practice it is dominated by the large coupling to the top quark. The moral of the story is that the Higgs particle cannot remain light unless there is a precise mechanism to cancel this quadratic dependence on the high scale. This is a particular aspect of a general problem called ``scale hierarchy''. We shall see later how such cancellation mechanisms can be implemented and what kind of New Physics they imply.

Let us summarise: In the Standard Model the absence of a light Higgs comes necessarily with new interactions. On the other hand a light Higgs is not stable without new interactions. Both are good news for LHC. Never before a new experimental facility had such a rich discovery potential and never before it was loaded with so high expectations.

\section{Grand Unification}
\label{secGUT}
In the remaining of this course I shall show some of the new physics that theorists have envisaged. Obviously, the selection of the topics reflects my own preferences.

 The hypothesis of grand 
unification states that $U(1) \otimes SU(2) \otimes SU(3)$ is the remnant of a larger, simple or semi-simple group $G$, which is 
spontaneously broken at very high energies. The scheme looks like: 

\begin{equation}
\label{eqGUT}
G \stackrel{M}{\longrightarrow}   U(1) \otimes SU(2) \otimes SU(3) \stackrel{m_W}{\longrightarrow}  U(1)_{e.m.} \otimes SU(3) 
\end{equation}

\noindent where the breaking of $G$ may be a multistage one and M is one (or several) characteristic mass scale(s). Two questions immediately 
arise concerning this idea:
(i) Is it possible? In other words are there groups which contain $U(1) \otimes SU(2) \otimes SU(3)$ as a subgroup and which can 
accommodate the observed particles?
(ii) Does it work? $i.e.$ is the observed dynamics compatible with this grand unification idea?

We shall try to answer each of these questions separately.

\subsection{The simplest G.U.T.: $SU(5)$}

In this subsection we shall answer the first question by giving a specific example of a group $G$ which satisfies our requirements.
 We first observe that $G$ must contain electromagnetism, $i.e.$ the photon must be one of the gauge bosons of $G$. This is part of the requirement that $G$ contains the group of the standard model. Another way to say the same
 thing, is to say that the electric charge operator $Q$ must be one of the generators of the algebra of $G$. Since $G$ is semi-simple, all its generators are represented by traceless matrices. It 
follows that, in any representation of $G$, we must have:

\begin{equation}
\label{eqTrQ}
Tr(Q)=0	
\end{equation}

\noindent in other words, the sum of the electric charges of all particles in a given representation vanishes.

For simplicity, let us make a first assumption: The fifteen (or sixteen)  spinors of a family 
fill a representation, not necessarily irreducible, of $G$, i.e. we assume that there are no other, as yet unobserved, particles 
which sit in the same 
representation. Property (\ref{eqTrQ}), together with the above assumption, have a very important consequence: As we have 
remarked, the  members of a family satisfy (\ref{eqTrQ}) because the sum of their charges vanishes. This, however, is 
not true if we consider leptons or quarks separately. Therefore each irreducible representation of $G$ will contain 
both leptons and quarks. This means that there exist gauge bosons of $G$ which can change a lepton into a quark, or 
vice versa. We conclude that a grand unified theory, which satisfies our assumption, cannot conserve baryon and 
lepton numbers separately. This sounds disastrous, since it raises the spectrum of proton decay. The amplitude for 
such a decay is given by the exchange of the corresponding gauge boson and therefore, it is of order $M^{-2}$, where $M$ 
is the gauge boson's mass. The resulting proton life-time will be of order:

\begin{equation}
\label{eqPlt}
\tau_p \sim \frac{M^4}{{m_p}^5}	
\end{equation}

Using the experimental limit, (for particular decay modes), of afew times $10^{-32}$  years, we can put a lower limit on $M$:

\begin{equation}
\label{eqMGUT}
M \geq 10^{15} GeV	
\end{equation}

Grand unification is not a low-energy phenomenon!

After these general remarks, let us try to find some examples: $U(1) \otimes SU(2) \otimes SU(3)$ is of rank 4 (i.e. there 
are four generators which commute: one of $U(1)$, one of $SU(2)$ and two of $SU(3)$). Therefore, let us first look for 
a grand unification group of rank 4. I list all possible candidates:
\vskip 0.3cm

$[SU(2)]^4, [SO(5)]^2, 
[G_{2}]^2, SO(8), SO(9), S_{p}(8), F4,  
[SU(3)]^2, SU(5) $
\vskip 0.3cm

The first two are excluded because they have no $SU(3)$ subgroup. The next five admit no complex 
representations, therefore they cannot accommodate the observed families where, as we already saw, the right- and 
left-handed particles do not transform the same way. (I again assume that no unobserved fermions will complete a 
given representation). Finally, in $SU(3)\otimes SU(3)$ quarks and leptons must live in separate representations because 
the leptons have no colour. But $\Sigma Q_{quarks} \neq 0$ and the same is true for leptons. This leaves
us with $SU(5)$ as the only candidate of a Grand Unified Theory (G.U.T.) group of rank 4. It is the simplest and, although, as we shall see, it has many shortcomings, it can be considered as
the ``standard model'' of grand unification.

The gauge bosons belong to the 24-dimensional adjoint representation. It is useful to decompose it 
into its $SU(2) \otimes SU(3)$ content. We find:

\begin{equation}
\label{eq24}
\begin{array}{cccccc}
24 = & [(2,3) \oplus (2, \bar{3})] &  \oplus & (1, 8) & \oplus & [(3, 1) \oplus (1, 1)]\\
 & \downarrow & & \downarrow & & \downarrow \\
 & \left( \begin{array}{c}X \\ Y \end{array} \right)  \begin{array}{c} Q=4/3 \\Q=1/3 \end{array} & & gluons  & &  W^{\pm},Z^0,\gamma
\end{array}
\end{equation}

\noindent where the first number denotes the $SU(2)$ and the second the $SU(3)$ representation. The known vector 
bosons can be identified as the eight gluons of Q.C.D. in the $(1,8)$ piece, (a singlet of $SU(2)$ and an octet 
of $SU(3)$), as well as the electroweak gauge bosons $W$, $Z$ and $\gamma $ in the $(3, 1) \oplus (1, 1)$ piece. We are left with 
twelve new ones, called $X$ and $Y$, with electric charges 4/3 and 1/3 respectively, which transform as a 
doublet of $SU(2)$ and a triplet and anti-triplet of $SU(3)$. They must be heavy, according to the limit (\ref{eqMGUT}).

Let us now come to the matter-field assignment. We shall try to put all the two-component 
spinors of a family in a representation (not necessarily irreducible) of $SU(5)$. But before doing so, we 
observe that all gauge couplings, being vectorial, conserve helicity. Therefore, we cannot put right- and 
left-handed spinors in the same representation. We go around this problem by replacing all right-handed spinors by the corresponding left-handed charge 
conjugate ones. A quick glance at the representation table of $SU(5)$ suggests to use each family in order 
to fill two (or three, if a righthanded neutrino exists) distinct, irreducible, representations: the $\bar{5}$ and the $10$. Their $SU(2) \otimes  SU(3)$ content is:

\begin{equation}
\label{eq5}
\bar 5 = (1,\bar3) \oplus (2,1)
\end{equation}

\begin{equation}
\label{eq10}
10=(2,3) \oplus (1,\bar3)\oplus (1,1) 
\end{equation}

The identification is now obvious. We often write these representations as a five-vector and a five by five antisymmetric matrix. Comparing (\ref{eqfam}) with (\ref{eq5}) and (\ref{eq10}) and using (\ref{eqTrQ}), we find:

\begin{equation}
\label{eq510} 
\bar5 = {\left( \begin{array}{c}
{d^c}_1 \\ {d^c}_2 \\ {d^c}_3 \\ e^- \\ -{\nu}_e
\end{array}
\right)}_L = {\psi}_{L_a} 
\hskip 1cm 10= {\left( \begin{array}{ccccc}
0 & {u^c}_3 & -{u^c}_2 & -u_1 & -d_1 \\
 & 0 & {u^c}_1 & -u_2 & -d_2 \\
 & & 0 & -u_3 & -d_3 \\
 & & & 0 & -e^c \\
 & & & & 0
\end{array}
\right)}_L = {\psi^{ab}}_L
\end{equation}

If we have a right-handed neutrino, it must be assigned to the singlet representation of $SU(5)$.   This is an unpleasant feature. In the absence of a ${\nu}_R$ we could say that the choice of $SU(5)$ was ``natural'', in the sense that it is the only group with an acceptable 15-dimensional representation (although not an irreducible one). As we shall see, with 16 dimensions, other choices are aesthetically more appealing. A technical remark: It is important to notice that the sum 
of these representations is anomaly-free.

Let us finally study the Higgs system. The first symmetry breaking goes through a $24$-plet of scalars $\Phi(x)$. 
It is convenient to represent the $24$ as a $5 \times 5$ traceless matrix. The vacuum expectation value which 
breaks $SU(5)$ down to $U(1) \otimes SU(2) \otimes SU(3)$ is proportional to the diagonal matrix:

\begin{equation}
\label{eql24} 
\lambda_{24} = \frac{1}{\sqrt{15}}\left( \begin{array}{ccccc}
1 & & & & \\ & 1 & & & \\ & & 1 & & \\ & & & -3/2 & \\ & & & & -3/2
\end{array} \right)
\end{equation}

$SU(3)$ is defined to act on the upper three components of the five-dimensional space and $SU(2)$ on the lower two. The potential for the $\Phi (x)$  field can be written as:

\begin{equation}
\label{eqpot24}
V(\Phi) = -\frac{1}{2}m^2Tr({\Phi}^2)+\frac{h_1}{4}[Tr({\Phi}^2)]^2+\frac{h_2}{2}Tr({\Phi}^4)
\end{equation}

The vacuum expectation value of $\Phi $ is determined by the minimum of $V(\Phi )$. It is easy to show that, for $h_1 $
and $h_2 $ positive, this minimum is precisely $V\lambda_{24}$ with

\begin{equation}
\label{eqVev}
V^2=m^2\left[h_1+\frac{7}{15}h_2\right]^{-1}
\end{equation}

Notice that if $h_2 <0$ with $h_1 > 0$ the direction of breaking is instead $SU(5) \rightarrow U(1) \otimes SU(4)$.

Can we use the same $24$-plet of Higgs in order to obtain the second breaking of the standard model? 
The answer is no for two reasons: First, the $24$ does not contain any $(2 , 1)$ piece (see eq. (\ref{eq24})) which is the 
one needed for the $U(1) \otimes SU(2) \rightarrow U(1)_{e.m.}$ breaking. Second, the $24$ does not have the required Yukawa 
couplings to the fermions. Indeed with the $\bar5$ and 10 assignment the fermions can acquire masses through 
Yukawa couplings with scalars belonging to one of the representations in the products:

\begin{equation}
\label{eqprod1}
\bar5 \otimes 10 = 5 \oplus 45 
\end{equation}

\begin{equation}
\label{eqprod2}
10 \otimes 10 = \bar5 \oplus \bar{45} \oplus \bar{50}
\end{equation}

We see that the $24$ is absent while the $5$ looks promising. If $H(x)$ is a five-plet of scalars, the 
complete potential of the Higgs fields is:

\begin{equation}
\label{eqcompot}
V_{Higgs} = V(\Phi)+V(H)+V(\Phi,H)
\end{equation}

\noindent with $V(\Phi)$ given by (\ref{eqpot24}) and

\begin{equation}
\label{eqpot5}
V(H)=-\frac{1}{2}\mu^2H^{\dagger}H+\frac{1}{4}\lambda (H^{\dagger}H)^2
\end{equation}

\begin{equation}
\label{eqpot524}
V(\Phi,H) = \alpha H^{\dagger}H Tr({\Phi}^2)+\beta H^{\dagger}{\Phi}^2 H
\end{equation}

We can show that, for an appropriate range of the parameters $m^{2}$, ${\mu}^{2}$, $h_{1}$, $h_{2}$, $\lambda$, $\alpha$ and 
$\beta$, we obtain the desired breaking.

\begin{equation}
\label{eqVevcom}
<\Phi >_0 \sim V \left(
\begin{array}{ccccc}
1 & & & & \\ & 1 & & & \\ & & 1 & & \\ & & & -3/2-\epsilon /2 & \\ & & & & -3/2+\epsilon /2 
\end{array}
\right)
\end{equation}

\begin{equation}
\label{eqvevcom}
<H>_0 \sim v \left(
\begin{array}{c}
0 \\ 0 \\ 0 \\ 0 \\ 1
\end{array}
\right)
\end{equation}

The small number $\epsilon $ in (19) is due to the mixed terms $V(\Phi, H)$ in the potential and it causes a 
breaking of $SU(2) \otimes U(1)$ already from the vacuum expectation value of $\Phi$. We must have $\epsilon \ll 1$,
otherwise the breaking of the standard model would have been of the same order as that of $SU(5)$. Using the potential (\ref{eqcompot}) we find:

\begin{equation}
\label{eqeps}
\epsilon = \frac{3\beta v^2}{20h_2 V^2}+O\left(\frac{v^4}{V^4}\right)
\end{equation}

\noindent which means that $\epsilon $ must be of the order $10^{-28}$. It is hard to see how such a number may come out for generic values of the coupling constants. This is part of the famous hierarchy problem which we encountered already in the previous lecture and, as we shall explain in the next section, plagues all grand unified theories. In the case of $SU(5)$ this problem has two aspects: The first is the general problem of the two widely separated symmetry breaking scales. We expect to have $V^2 \sim m_{\Phi}^2$ and $v^2 \sim m_{H}^2$. But the presence of the mixed terms in $V(\Phi, H)$ induces a $(m^2_H)_{eff}$ of the order of $V^2$, unless the parameters of the potential are very finely tuned. The second is related to the five-plet of Higgs $H$ which, under $U(1) \otimes SU(2) \otimes SU(3)$, is split as shown in equation (\ref{eq5}). The $SU(2)$ doublet is used for the electroweak breaking and must have a mass of the order of $v^2$. The $SU(3)$ triplet components, however, can mediate baryon number violating transitions and should be superheavy of the order of $V^2$. Again there is no natural way to obtain such a doublet-triplet splitting without a fine tuning of the parameters in the potential.  

The fermion masses are due to the vacuum expectation value of $H$. Looking back at the 
assignment (\ref{eq510}) we see that the up-quarks take their masses through (\ref{eqprod2}) while the down quarks 
and the charged leptons through (\ref{eqprod1}). 

This discussion answers the first question, namely it shows that there exist groups which 
have the required representations to be used for grand unification. Let us now turn to the second 
question, namely the study of the dynamical consequences of G.U.T.s.

\subsection{Dynamics of G.U.T.s}
\label{secdynGUT}

\subsubsection{Tree-level $SU(5)$ predictions}

Let us first examine the dynamical predictions of $SU(5)$ at the Lagrangian level without taking into 
account higher order effects. There are several such predictions:

(i) The first concerns the coupling constants. $SU(5)$ is a simple group and hence it has only one coupling 
constant $g$. On the other hand in nature we observe three distinct ones, $g_{1}$,
$g_{2}$ and $g_{3}$, corresponding to each one of the factors of the standard model $U(1) \otimes SU(2) \otimes SU(3)$. The naive prediction   would be $g_1 = g_2 = g_3$. However, we must be more careful with the relative normalisations. For non-abelian groups, like $SU(2)$, $SU(3)$, or $SU(5)$, the normalisation of the generators is fixed by the algebra, which is a non-linear relation. So the question arises only for $U(1)$. In the Standard Model the $U(1)$ generator $Y$ is related to the electric charge and the third component of weak isospin $t_3$ 
by $Q=t_3-Y$. For the embedding of $U(1) \otimes SU(2) \otimes SU(3)$ into $SU(5)$, all generators must be normalised the same way. 
Let us choose the normalisation by requiring $Tr(J_i J_j)=R{\delta}_{ij}$ where $R$ is a constant which may depend on the representation we use to compute the trace, but it is 
independent of $i$ and $j$. Let us now compute $Tr ({t_3}^2)$ using, for example, the electron family. We find $Tr({t_3}^2) = 2$. Similarly we find $Tr(Y^2) = 10/3$. Therefore we see
that for the embedding, the $U(1)$ generator must be rescaled by $Y \rightarrow cY$  with, $c^2 = 5/3$. Therefore, the tree-level prediction of any grand unified theory based on a simple group is  

\begin{equation}
\label{eqtree1}
g_2=g_3=g \hskip2cm {sin}^2 \theta_W = \frac{{g_1}^2}{{g_1}^2 +{g_2}^2}=\frac{g^2 /c^2}{g^2 /c^2 + g^2}= \frac{3}{8}
\end{equation}

(ii) Fermion masses: Fermion masses are generated through the same mechanism as in the standard 
model, i.e. through Yukawa couplings with Higgs scalars. Therefore they depend on the particular Higgs system one assumes. In the minimal  $SU(5)$ model with only a 5-plet of 
Higgs we see in equations (\ref{eqprod1}) and (\ref{eqprod2}) that we have two independent coupling constants for each 
family. The up-quarks take their masses through (\ref{eqprod2}), while the down ones and the charged leptons through (\ref{eqprod1}). This last property implies the relations:

\begin{equation}
\label{eqmasrel}
m_d=m_e \hskip2cm m_s=m_{\mu}\hskip2cm m_b=m_{\tau} 
\end{equation}

It is obvious that these predictions are lost if we assume a more complicated Higgs system, for example by including higher dimensional representations. 

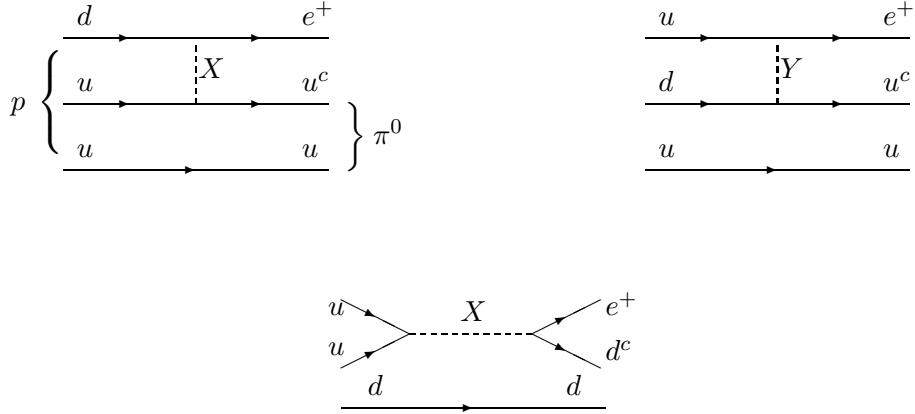
\begin{figure}
\begin{picture}(350,200)
\multiput (20,150)(220,0){2}{\vector(3,0){50}}
\multiput (70,150)(220,0){2}{\line(3,0){50}}
\multiput (20,175)(220,0){2}{\vector(3,0){25}}
\multiput (45,175)(220,0){2}{\vector(3,0){50}}
\multiput (95,175)(220,0){2}{\line(3,0){25}}
\multiput (20,200)(220,0){2}{\vector(3,0){25}}
\multiput (45,200)(220,0){2}{\vector(3,0){50}}
\multiput (95,200)(220,0){2}{\line(3,0){25}}
\multiput(70,175)(220,0){2}{\multiput(0,0)(0,4){6}{\line(0,1){2}}}
\put (10,173){$\left \{  \begin{array}{l}
  \\  \\ \\ \end{array} \right. $}
\put (115,160){$\left. \begin{array}{l}  \\  \\  \end{array} \right \} $}
\put (0,173){$p$}
\put (25,155){$u$}
\put (25,180){$u$}
\put (25,205){$d$}
\put (110,205){$e^+$}
\put (110,180){$u^c$}
\put (110,155){$u$}
\put (71,185){$X$}
\put (137,160){$\pi ^0$}
\put (245,155){$u$} 
\put (330,155){$u$}
\put (245,180){$d$}
\put (330,180){$u^c$}
\put (245,205){$u$}
\put (330,205){$e^+$}
\put (291,185){$Y$}
\put (125,60){\vector(3,0){50}}
\put (175,60){\line(3,0){50}}
\put (125,75){\line(2,1){26}}
\put (125,101){\line(2,-1){26}}
\multiput (151,88)(4,0){12}{\line(1,0){2}}
\put (197,88){\line(2,1){26}}
\put (197,88){\line(2,-1){26}}
\put (138,81.5){\vector(2,1){0}}
\put (138,94.5){\vector(2,-1){0}}
\put (210,81.5){\vector(2,-1){0}}
\put (210,94.5){\vector(2,1){0}}
\put (135,65){$d$}
\put (210,65){$d$}
\put (120,80){$u$}
\put (120,95){$u$}
\put (225,78){$d^c$}
\put (225,95){$e^+$}
\put (170,93){$X$}
\end{picture}
\caption {Some diagrams contributing to proton decay} \label{figprdec}
\end{figure}

(iii) Baryon and lepton number violation: $X$, $Y$, or heavy Higgs boson exchanges lead to baryon and lepton number 
violation. In Figure \ref{figprdec} we depict some diagrams contributing to proton decay. In $SU(5)$ the main decay 
mode is expected to be $p \rightarrow \pi^0 e^+$ with a branching ratio of the order of $30-40\%$  followed by $p \rightarrow \omega e^+$ or $p \rightarrow \rho^0 e^+$. The neutrino modes, such as
$p \rightarrow \pi^+\bar{\nu}$, are expected to be rare ($\sim 10\% $ or less). Bound neutrons are also expected to decay with $n \rightarrow \pi^-e^+$ being the 
dominant mode. All these decay modes are easily detectable. The overall life-time depends on the masses of 
the superheavy gauge bosons $X$ and $Y$ (see eq. (\ref{eqPlt})).

Finally we remark that the $SU(5)$ Lagrangian, including 
the Yukawa couplings, is invariant under a $U(1)$ group of global phase transformations:

\begin{equation}
\label{eqB-L}
\psi^{ab}\rightarrow e^{i\theta}\psi^{ab} \hskip1.5cm \psi_a\rightarrow e^{-3i\theta}\psi_a \hskip1.5cm H_a\rightarrow e^{-2i\theta}H_a
\end{equation}

\noindent with all other fields left invariant. One can verify that this global symmetry is also anomaly free. The non-
zero vacuum expectation value of $H$ seems to break this symmetry spontaneously. This sounds disastrous since it 
normally leads to the appearance of a truly massless Goldstone boson. However we are saved because the 
symmetry is not really broken, it is simply changed. We can check immediately that, even after the 
translation of the Higgses, the linear combination $J + 4Y$ remains as a global symmetry, where $J$ is the 
generator of (\ref{eqB-L}) and $Y$ the $U(1)$ part of $SU(5)$ given by (\ref{eql24}). The conserved charge of this symmetry is the 
difference $B-L$ of baryon and lepton numbers. This conservation has some very important consequences. 
In particular, it gives some precise predictions for the nucleon decay properties. For example $p \rightarrow e^+\pi^0$ or 
$n \rightarrow e^+\pi^-$ 
are allowed but $n \rightarrow e^-\pi^+$ is not. The same is true
for $n - \bar{n}$ oscillations which violate $B-L$. As we shall see, this property remains true (or nearly true) in many 
grand unified models.

\subsubsection{Higher order effects}

The tree-level predictions (\ref{eqtree1}) or (\ref{eqmasrel}) are in violent disagreement with experiment. It is therefore 
important to study the effects of higher order. For this purpose it is necessary to go with some detail into the 
renormalisation program of the theory. Since this analysis has appeared already in many places, I shall not present it in these lectures. I shall give only the main results.

The idea is that predictions, such as (\ref{eqtree1}), are consequences of the full symmetry and can only be true at energies well above $M$ where the $SU(5)$ breaking can be neglected. Similarly for the mass relations (\ref{eqmasrel}) which follow from the equality of the Yukawa coupling constants, again a property valid only in exact $SU(5)$. In order to compare these predictions with the real world we must extrapolate to present day energies. These extrapolations can be performed using the equations of the renormalisation group. Several assumptions enter in this procedure. The most important one is connected with the very idea of grand unification. We must assume that we know all fundamental physics, in particular the entire spectrum of elementary particles, from our accelerators, to energies of $10^{16}$GeV or higher. As we shall see, the results are sensitive to the possible existence of new thresholds at high energies. 

\begin{figure}
\centering
\epsfxsize=10cm
\epsffile{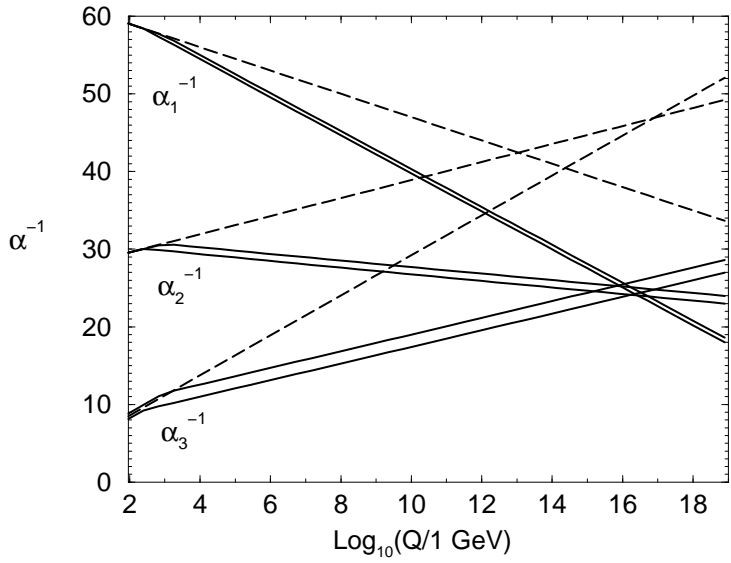}
\caption {The renormalisation group evolution of the inverse coupling
constants $\alpha_i^{-1}, i=1,2,3$ of the three Standard Model groups
$U(1)$, $SU(2)$ and $SU(3)$ without supersymmetry, (dashed lines) and
with supersymmetry, (solid lines). For the supersymmetric case the
uncertainties are also shown. The calculations include two loop effects.} \label{figunif}
\end{figure}

In their simplest form, the renormalisation group equations for the evolution of the gauge coupling constants take the form:

\begin{equation}
\label{eqRG1}
{g_i}^{-2} = {g_j}^{-2} + 2\int_{0}^{\lambda}\frac{dx}{x}[{b_i}^0 (x,\alpha)-{b_j}^0 (x,\alpha)] 
\end{equation}

\noindent where $i,j=1,2,3$ denote the three physical coupling constants of the Standard Model, ${b_i}^0$ are the one-loop coefficients of the 
corresponding $\beta$-functions, $\lambda=-M^2/p^2$ is the mass-scale measured in units of $M^2$ and $\alpha$ represents all other mass variables, always in units of $M^2$, such as the masses of the remaining physical Higgs particles. Fortunately, it turns out that the dependence on $\alpha$ is very weak. If we ignore it, we obtain two equations with one unknown, the value of $M$. We can use, as input, the experimentally measured effective strengths of strong, electromagnetic 
and weak interactions at moderate (say $p^2 \sim 10-100{GeV}^2$) energies. The consistency of the scheme is verified by comparing the two values of $\lambda$ obtained by the two independent equations (\ref{eqRG1}). It is usually presented in terms of a plot to see whether the three running coupling constants ``meet''. Alternatively, we can use one of the equations to determine $\lambda$ and the other to predict the physical value of the third coupling constant and compare with experiment. Traditionally, this is expressed as a prediction for ${sin}^2\theta_W$. A precise calculation must take into 
account the two-loop effects, as well as the breaking of $U(1) \otimes SU(2)$. In fact it turns out that the value of ${sin}^2\theta_W$ is quite sensitive to 
this last breaking. The result is presented in Figure \ref{figunif}. As we
can see, the agreement with experiment is only qualitative. The three
curves do not really come together. As we shall see in section
\ref{secsupsymexp}, the agreement becomes spectacular if we include
supersymmetry.

\begin{figure}
\centering
\epsfxsize=10cm
\epsffile{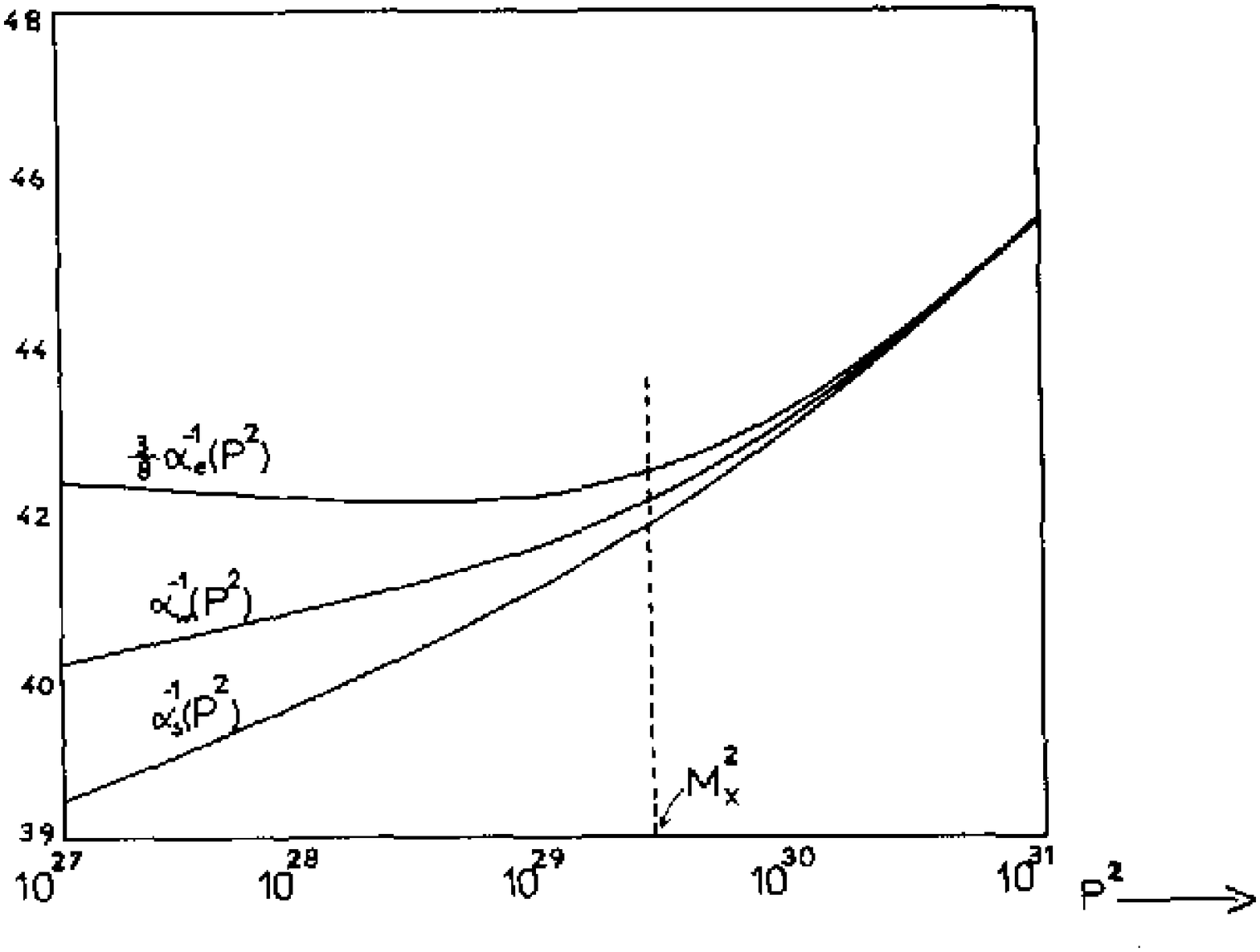}
\caption {Detail of the region around the grand unification scale using exact two-loop $\beta$-functions.} \label{gut}
\end{figure}

A final remark: Some students have asked the question of the precise meaning of the statement ``the three curves meet''. In fact, they are not supposed to meet. At any finite energy $SU(5)$ is broken down to $U(1) \otimes SU(2)\otimes SU(3)$. It is only asymptotically at infinite energy that the breaking can be neglected. At any finite energy we have three distinct coupling constants. The plots showing the coupling constants meeting at the grand unification scale are obtained in the so-called ``step function approximation'' in which we treat the mass $M$ of any particle, such as the superheavy bosons, as infinite at scales smaller than $M$ and zero at scales larger than $M$. The exact treatment is not very difficult to perform and gives curves such as the one shown in Figure \ref{gut}. 

A similar treatment can be applied to the Yukawa couplings and, therefore, to the fermion masses. The couplings to leptons and quarks, which are equal at $M^2$, evolve differently under the renormalisation group, because the leptons have no strong interactions. The study is reliable only for the relatively heavy particles ($\tau$-lepton and $b$-quark) because at low energies the strong interaction coupling constant becomes large and perturbation theory breaks down. In a simplified form, the result can be written as 

\begin{equation}
\label{eqRG2}
\frac{m_b}{m_{\tau}}=\left[ \frac{\alpha_s (Q=2m_b)}{\alpha_s (Q=M)} \right]^{\frac{12}{33-F}}
\end{equation}

\noindent where ${\alpha}_s$ is the strong interaction coupling constant evaluated at $Q = 2m_b \sim m_Y$  (the upsilon mass) and at $M\sim 10^{16}$GeV 
 respectively, and F is the number of families. For F = 3 we find a very good agreement with 
experiment, while the agreement gets much worse with increasing F. This result was obtained before LEP established the existence of precisely three families and it is considered as the second successful prediction of grand unified theories.

\subsection{Other Grand Unified Theories}

In the previous section we examined in some detail the grand unified model based on the group $SU(5)$. The 
main reason for this choice was its simplicity. In fact, as we mentioned already, this simplest model does 
not quite fit the experimental data. However, the general properties remain the same in practically all 
models and the methods we developed can be applied in a straightforward way to every other model, 
although the detailed numerical results may differ. In this section we shall briefly present some other 
``classical'' grand unified theories and we shall try to explain their respective merits.

\subsubsection{A rank 5 G.U.T.: $SO(10)$}
\label{secso10}

The $SU(5)$ model, in its simple and most attractive version, has no natural place for a right-handed neutrino. We must add it as an extra singlet. The only simple group  which can be used for grand unification without need for a singlet representation and without introducing exotic 
fermions is $SO(10)$. It is a group of rank 5, which means that the corresponding algebra has five commuting generators. For $SU(5)$ we had proven that it was the only acceptable group of rank 4. Similarly, we can prove that $SO(10)$ is the only one of rank 5. The proof goes along the same lines: we list all possible candidates and we eliminate the unacceptable ones. Some examples:

-$[SU(2)]^5$: no $SU(3)$ subgroup.

-$SO(11)$, $Sp(10)$: no complex representations, no 15- or 16-dimensional representations. 

-$SU(6)$: It has a 15-dimensional representation but its decomposition in $SU(2) \otimes SU(3)$ 
 shows that it cannot accommodate the members of a family. One finds:

$15=(2,3) \oplus (1,\bar3) \oplus (1,3) \oplus (2,1) \oplus (1,1)$

 The troublesome piece is the (1,3) which is a singlet of $SU(2)$ and triplet of colour rather than an 
anti-triplet.

The final candidate is $SO(10)$ which has a 16-dimensional irreducible representation. 
$SO(10)$ contains $SU(5)$ as a subgroup and the 16-plet decomposes under $SU(5)$ into:
	
\begin{equation}
\label{eq16}
16=10 \oplus \bar5 \oplus 1
\end{equation}

\noindent i.e. we find our old $\bar5$ and 10 as well as a singlet. The obvious interpretation of this last one is a 
right-handed neutrino (or ${\nu^c}_L$).

The salient features of this G.U.T. are the following: The gauge bosons belong to a 45-dimensional (adjoint) representation which under $SU(5)$ decomposes as:

\begin{equation}
\label{eq45}
45=24 \oplus 10 \oplus \bar{10} \oplus 1
\end{equation}

An interesting property of the model is that all members of a family, enlarged with a right-
handed neutrino, belong to a single irreducible representation, the 16-dimensional spinorial 
representation we mentioned above. In this respect the family structure seems more natural in $SO(10)$ as compared to 
$SU(5)$. On the other hand, again no explanation is offered for the observed number of families. It is also interesting to point out that $SO(10)$ has no 
anomalies.
Another interesting feature is that $B-L$ is now a gauge generator. It must be spontaneously 
broken otherwise there would remain a massless photon coupled to it. However this violation  does not lead to any observable effects in nucleon decay because the branching ratio of 
forbidden to allowed decays is predicted to be very small.

In the long journey from $SO(10)$ down to $U(1) \otimes SU(2) \otimes SU(3)$, nature may choose various 
paths. She can take the direct road (just one big break) or she may decide to go through one, or 
more, of the intermediate subgroups:

\begin{equation}
\label{eqbreak}
SO(10) \rightarrow \left\{ 
\begin{array}{c}
\rightarrow \\ \rightarrow SU(5) \rightarrow \\ \rightarrow SU(5) \otimes U(1) \rightarrow \\ \rightarrow SU(4) \otimes SU(2) \otimes SU(2) \rightarrow \\ \rightarrow SU(4) \otimes SU(2) \otimes U(1) \rightarrow \\ ...
\end{array} \right\} \rightarrow U(1) \otimes SU(2) \otimes SU(3)
\end{equation}

The Higgs system depends on the breaking pattern we choose, but, in any case, it is more 
complex than that of $SU(5)$. Several representations are necessary.

The main experimental prediction of $SO(10)$, which differs substantially from that of $SU(5)$, 
concerns the neutrino mass. The presence of $\nu_R$ allows for a Dirac mass $\bar{\nu}_R \nu_L$ and the violation of $B-L$
 allows for a Majorana term. The Dirac mass term comes presumably from a Yukawa coupling to a Higgs scalar and, therefore, it is an adjustable parameter, like any other fermion 
mass in the theory. A priori one expects a term of the same order of magnitude as the up quark 
masses. The problem then is how to make sure that the physical neutrino masses are sufficiently 
small. The main remark here is that the  Majorana mass, which comes from the superheavy breaking, is expected to be large, of the order of the $SO(10)$ 
symmetry breaking scale. The resulting neutrino mass spectrum will depend on the details of the 
Higgs system. For example if $SO(10)$ is broken through the vacuum expectation value of a 126-plet of Higgs scalars, the neutrino mass matrix for one family takes the form:

\begin{equation}
\label{eqneumass}
(\bar{\nu}_L, \bar{\nu}^c_L ) \left( 
\begin{array}{cc}
0 & m_D/2 \\ m_D/2 & M
\end{array} \right) \left(
\begin{array}{c}
\nu^c_R \\ \nu_R 
\end{array} \right)
\end{equation}

\noindent where $m_D$ and $M$ are the Dirac and Majorana masses. As explained above one expects $m_D \ll M$. 
Then the SU(2) doublets and singlets will be approximate mass eigenstates with masses $m^2/4M$ 
and $M$ respectively. For $m_D \sim $ 1 GeV and $M \sim 10^{16}$GeV we find a negligibly small mass for the 
doublet neutrino of order $10^{-7}$eV. Of course one also expects mixings among the three families 
but they are very model dependent. Let me also mention that, even if the Majorana mass term is 
forbidden in the tree approximation, (for example, if $B-L$ is not broken through the 126 but 
through a 16 representation), it may be generated in higher orders through a particular two-loop 
diagram. The value of $M$ is suppressed in this case by coupling constants and the resulting neutrino mass may be of the order of 1 eV or higher.

The moral of the story is two fold: First, the theory offers, through the spontaneous breaking 
of $B-L$, a natural mechanism to obtain very light neutrinos and second almost any desired 
pattern of masses and mixings can be reproduced by adjusting the parameters of an already 
rather complicated Higgs system.

\subsubsection{Other models}

I assume that it is by now obvious that a large variety of grand unified theories can be obtained 
by playing around with elementary group theory and Higgs representations. This partly explains 
the popularity that G.U.T.s have enjoyed for more than thirty  years. There is no point in giving a 
complete list of all models which claim agreement with data, which is usually the case for 
published models. Let me only mention some examples of ``special purpose'' models, i.e. models 
constructed specifically in order to reproduce a particular feature of the data.

{\it 1) Exceptional groups:} An aesthetically unattractive feature of all models based on unitary, 
orthogonal or symplectic algebras is that these form infinite series, so it may be hard to 
understand why any one in particular would provide the basis for a fundamental theory. 
Exceptional groups on the other hand are unique, they are just $G_2, F_4, E_6, E_7$ and $E_8$. The first one 
is excluded because it is too small to contain $SU(3)$ as a subgroup. The others could, in principle, 
be used as candidates for G.U.T.s. I shall briefly describe two of them, $E_6$ and $E_8$.

$E_6$:  It is the most attractive exceptional group for grand unification. It is the only one which admits 
complex representations and, from the group theory point of view, it can be considered as the 
natural extension of $SU(5)$ and $SO(10)$. Indeed, based on the Dynkin diagram pattern, one could 
define exceptional algebras $E_4$ and $E_5$ as isomorphic to $SU(5)$ and $SO(10)$ respectively. Of course, 
$E_6$ contains $SO(10)$ and {\it a fortiori} $SU(5)$, as subgroups. Furthermore it is ``safe'', i.e. it is 
automatically anomaly-free. Its fundamental representation has 27 dimensions and, under $SO(10)$ 
and $SU(5)$, decomposes as:

\begin{equation}
\label{eqE61}
27 \stackrel{SO(10)}{\longrightarrow} 16 \oplus 10 \oplus 1 \stackrel{SU(5)}{\longrightarrow} 10 \oplus \bar5 \oplus 1 \oplus 5 \oplus \bar5 \oplus 1
\end{equation}

The 78-dimensional adjoint representation decomposes as:
	
 \begin{equation}
\label{eqE62}
\begin{array}{ll}
78 & \stackrel{SO(10)}{\longrightarrow} 45 \oplus 16 \oplus \bar{16} \oplus 1 \\ & \stackrel{SU(5)}{\longrightarrow} 24 \oplus 10 \oplus \bar{10} \oplus 1 \oplus 10 \oplus \bar5 \oplus 1 \oplus \bar{10} \oplus 5 \oplus 1 \oplus 1
\end{array}
\end{equation}

There are several inequivalent possibilities of constructing grand 
unified theories based on $E_6$ and, in this section, I shall mention only 
one which satisfies the following requirements:
(In a later section we shall study different uses of $E_6$ coming from 
super-string inspired models)

(i)	All fermions of one family belong to the same irreducible representation.

(ii)	All unobserved, fermions get naturally superheavy masses.

(iii) All required Higgs scalars belong to representations appearing in the product of two fermion 
representations. This last requirement means that the Higgs scalars can be viewed as fermion-anti-fermion bound states.

We shall assign the fermions of each family to the 27 fundamental representation which, 
therefore, contains new, unobserved fermions. The Higgs fields must belong, according to (iii) 
above, to one, or more, of the representations:

\begin{equation}
\label{eqE63}
27 \otimes 27 = (27 \oplus 351)_S \oplus {351}_A
\end{equation}

The important observation comes from the decomposition (\ref{eqE61}). Out of the 27 fermions of a 
family, 12 (i.e. $ 5 \oplus \bar5 \oplus 1$) can take an $SU(5)$ invariant (and {\it a fortiori} $U(1) \otimes SU(2) \otimes SU(3)$ 
invariant) mass. Therefore these fermions are expected to have masses of the order of $10^{16}$GeV 
(this is sometimes called ``the survival hypothesis''). It is easy to check that with the Higgs system 
(\ref{eqE63}) this indeed happens. This explains why only 15 light fermions are observed in each family.

The simplest symmetry breaking pattern of this model goes through $SO(10)$, although others have also been considered. The detailed predictions, including some interesting speculations concerning the fermion mass spectrum, are model dependent. 

{\it $E_8$ :} If uniqueness is an important criterion for choosing the group of grand unification, then $E_8$ is the most prominent candidate. Its unique features include: (i) It is the largest exceptional group with a finite dimensional Lie algebra. (ii) It contains $E_6$, and, thus, $SO(10)$, $SU(5)$ etc, as subgroups. (iii) It is the only simple Lie group whose lowest dimensional representation is the adjoint. This offers the possibility of putting fermions and gauge bosons in the same, lowest dimensional, representation. In fact, $E_8$ has a natural, built-in, supersymmetry, as we shall see later. It is the symmetry group which appears automatically in some superstring models. These are the good news. Now the bad news: The adjoint representation has dimension 248, so a large number of new gauge bosons and fermions is required. Similarly, the necessary Higgs representations are enormous, the simplest version using the 3875 dimensional representation. On the other hand, we can put all the 48 known fermions, together with many unknown ones, in the same 248 representation. After symmetry breaking, we can arrange to have three light $SU(5)$ families ($\bar5 \oplus 10$) and three heavier ($\sim$ 1TeV), conjugate ones ($5 \oplus \bar{10}$). All other fermions become superheavy. $E_8$ has only real representations, so the theory is ``vector-like'', i.e. there are equal numbers of right- and left-handed fermions and, before symmetry breaking, we can write the theory using only vector currents. 

{\it 2) Models with horizontal symmetries:} The motivation behind this approach is to understand the apparent repetition in the family structure. In the standard model, as well as the simplest grand unified theories, the fermions for each family form a representation and the scheme is supposed to be repeated for every new family. Obviously, such an approach does not offer any insight into the reason why nature chooses to copy itself, neither does it give any hint on the total number of copies. Following once more the standard path, physicists tried to answer this question by enlarging the symmetry with the introduction of ``horizontal'' symmetries which relate the families, thus providing, in principle, information on their number, as well as on their masses, mixing angles etc. Although discrete symmetries can and have been used, I shall only discuss some attempts to enlarge the gauge group. Continuous global symmetries must, in general, be avoided, because they offer a less rich structure and, furthermore, if spontaneously broken, they lead to the appearance of unwanted, physical, massless, Goldstone bosons. We are thus left with gauge symmetries.  Among the most attractive models, are the ones in which the horizontal symmetry is unified with the same coupling constant with the other gauge interactions. Interesting examples of $SU(n)$, $SO(n)$ or exceptional groups have been built. Concerning $SU(n)$, one can show that there is no group with an irreducible, complex, anomaly-free representation with no bizarre fermions (i.e. fermions belonging to high colour representations). So one often uses complicated patterns of irreducible representations in order to cancel anomalies. Models with high $n$ tend to have too many fermions and they are not asymptotically free beyond the unification scale, although this may not be a problem, since this scale is close to the Planck mass. Among the models with just three light families, I mention one based on $SU(11)$ and one based on $SU(9)$. The latter is asymptotically free. Similarly one can build models based on high rank orthogonal groups. For example, the 64-dimensional complex spinorial representation of $SO(14)$ yields two light and two heavy families of opposite chirality. The real 128-dimensional representation of $SO(16)$ gives four light families. The 256 representation of $SO(18)$ can give three light families plus heavy fermions which can be incorporated into a Technicolor scheme. None of these models give any characteristic 
predictions which do not depend on the details of an already quite complex Higgs system. In the absence of 
any specific experimental input, one is left with one's own prejudices for guide.

\subsection{Some general consequences of Grand Unified Theories}
As we said earlier, the idea of grand unification satisfies our prejudice for a unified description of elementary particle interactions. On the other hand, it is fair to admit that there exists, at the moment, no obvious candidate for such a unification. I shall present here two general consequences of G.U.T.s, which are independent of any particular model.

\subsubsection{The baryon asymmetry of the universe}

By placing leptons and quarks in the same irreducible representation, all grand unified models imply that baryon and lepton numbers are not separately conserved. A large experimental effort has been concentrated to detect any trace of proton instability. The result is a higher limit on its life-time. At present, for the easily detectable decay modes, such as the $\pi^0 + e^+$ one, this limit is close to $10^{33}$ years. In the absence of any direct observation of baryon number non conservation, physicists have tried to see its possible effects in cosmology. 

In traditional cosmological models baryons and antibaryons were assumed to have been created in pairs 
since the Hamiltonian conserved baryon number. The only way to obtain a non-zero baryon number was to 
put it in by hand as an initial condition. In the so-called ``symmetric'' cosmologies it was argued that, within 
some range of temperatures ($\sim $1GeV), a phase transition occurs which results in a spontaneous symmetry 
breaking and thermal radiation becomes unstable against separation of nucleons from antinucleons. The 
situation was compared to what happens in a ferromagnet where a domain structure appears. According to 
this view the observed predominance of matter over antimatter is a local effect.
The trouble with this theory is that there is no evidence for the presence of large amounts of antimatter 
anywhere in the universe. The rare traces of antinucleons detected in cosmic rays are compatible with the 
estimated production of antimatter in particle collisions and no large-scale annihilations have been observed. 
Nevertheless this was the accepted doctrine for many years. The reason is that in a symmetric cosmological 
model, where no net baryon number is put in by hand in the initial conditions, the eventual appearance of 
baryon excess requires (i) the violation of C and CP invariance, (ii) the violation of baryon number 
conservation and (iii) the departure from thermal equilibrium. The necessity for the first two conditions is 
obvious since otherwise there is no distinction between baryons and antibaryons. The significance of the 
third one is also very simple:
In a stationary universe, where all interactions are in thermal equilibrium, the particle abundances are given 
by Boltzmann's law which involves only the particle masses. But CPT invariance guarantees that baryons 
and antibaryons have equal masses and, therefore, no net baryon number can possibly be produced. Charge 
conjugation is known to be maximally violated in weak interactions. The violation of CP has been observed already in 1964 in the decays of neutral kaons. The expansion of the universe provides the necessary 
departure from thermal equilibrium. So out of the three necessary conditions only the second one, the 
violation of baryon number, has not yet been experimentally verified. Grand unified theories provide a theoretical framework for such a violation and offer the possibility for an 
estimation of the resulting number of baryons in the universe. It is worth noticing that the suggestion for a 
baryon-antibaryon asymmetry was first made by A.D. Sakharov in 1966, much before the advent of gauge 
theories.

Let me describe here a typical scenario for baryon number generation taking as an example, the $SU(5)$ 
model. Grand unified theories, together with the standard cosmological model which is based on classical 
gravity, can, in principle, describe the early universe, at times substantially later than  the Planck time of $t_P \sim 10^{-44}$sec, corresponding to a 
temperature of $T \sim 10^{19}$.GeV. At earlier times quantum gravitational fluctuations, as well as other interactions 
unified with gravity, become strong and no reliable computations are possible. As the temperature drops the universe undergoes a series of phase transitions during which the original symmetry of the model breaks 
down to the one observed at present, namely confining Q.C.D. and electromagnetism. In the simplest $SU(5)$ 
model these transitions are the following: At $T\sim 10^{16}$GeV,  the 24-plet of Higgs develops a non-
zero vacuum expectation value and $SU(5)$ breaks into $U(1) \otimes SU(2) \otimes SU(3)$. At $T\sim $100 GeV,  the 
standard model breaking occurs and $U(1) \otimes SU(2) \otimes SU(3)$ breaks to  $ U(1) \otimes SU(3)$. Still later, at $T \sim $100 MeV, we have the confining transition of Q.C.D. during which the quarks and gluons get confined inside 
the hadrons. At even later stages of the universe evolution, we have more transitions which result into 
nucleosynthesis, hydrogen atom recombination, etc. In this simple scenario we must add the phase 
transitions for inflation and reheating which I shall ignore here.

The number we wish to explain is the observed ratio of baryon number density to entropy density $kn_B/s \sim 10^{-10}$,  or, equivalently the baryon to photon ratio $n_B/n_{\gamma} \sim 10^{-9}$. We shall assume no initial baryon asymmetry at $t\sim 10^{-44}$sec, since it wouldn't have survived any subsequent inflation anyway. A net baryon number can be created 
either through two-particle collisions or, at $T \sim 10^{16}$ GeV, as the result of heavy boson decays. The latter can 
be either gauge bosons or Higgs bosons. In more complicated models than $SU(5)$ there may exist also 
heavy fermions whose decays violate baryon number. It is therefore important to compare the rates $\Gamma_c$ and 
$\Gamma_X$ for collisions and boson decays with the expansion rate of the universe $H =\mbox{\.{R}}/R$ as a function of the 
temperature in order to determine when each of these processes drops out of thermal equilibrium. This last condition is essential so that the products of a decay have no chance, by inverse reaction, to recreate the 
initial state. 

The detailed calculation requires the numerical integration of the Boltzmann equation for any specific model. It 
turns out that the collision term is relatively unimportant. In the $SU(5)$ model the contribution of the heavy gauge 
bosons is also small for essentially two reasons: (i) Their couplings to fermions are real and CP violation can be 
introduced only through the fermion mass matrix. This requires the presence of all three families, $i.e.$ we must go to a 
multi-loop diagram. (ii) A simple estimation shows that, by the time the gauge boson interactions drop out of thermal 
equilibrium, the Boltzmann factor has considerably reduced their strength. The dominant mechanism seems to be the 
decays of the superheavy Higgs scalar bosons of the 5-representation. In fact, several 5's may be necessary. In any case, 
since the Higgs sector is the less well-understood aspect of a gauge theory, no precise, quantitative model-independent 
conclusions can be drawn. It is nevertheless remarkable that even a qualitative agreement between theory and 
observation can be reached.

\subsubsection{Magnetic Monopoles}

The second general feature of grand unified theories I want to present is the appearance of magnetic monopoles. This discussion will allow us to introduce the idea of duality, which plays also a crucial role in the theory of strings and branes. 

{\it Abelian magnetic monopoles.}
The empty-space Maxwell's equations possess an obvious invariance under the following interchange of electric and 
magnetic fields:

\begin{equation}
\label{eqdual1}
\mbox{\bf E}\rightarrow \mbox{\bf B} \hskip 2cm \mbox{\bf B} \rightarrow  -\mbox{\bf E}
\end{equation}

As a matter of fact, this invariance is much larger and covers the entire $U(1)$ group of transformations $\mbox{\bf E}+i\mbox{\bf B}\rightarrow e^{i\phi}(\mbox{\bf E}+i\mbox{\bf B})$. However, as we shall see presently, only the discreet subgroup (\ref{eqdual1}) could possibly survive in the presence of sources.  

In a compact relativistic notation we can write the equations as:

\begin{equation}
\label{eqmaxw1}
\partial_{\nu}F^{\mu \nu} = 0 \hskip 2cm \partial_{\nu}\tilde{F}^{\mu \nu} = 0
\end{equation}

\noindent where $ \tilde{F}^{\mu \nu}$ is the dual tensor to $F^{\mu \nu}$ defined as $\tilde{F}^{\mu \nu}=\frac{1}{2}\epsilon^{\mu \nu \rho \sigma}F_{\rho \sigma}$ and the transformation (\ref{eqdual1}) can be written as a duality transformation:

\begin{equation}
\label{eqdual2}
F^{\mu \nu}\rightarrow  \tilde{F}^{\mu \nu} \hskip 2cm \tilde{F}^{\mu \nu}\rightarrow -F^{\mu \nu}
\end{equation}

It is an empirical fact that the presence of matter destroys this symmetry. Indeed, we have electric charges and electric 
currents $j_{\mu}=(\rho,\mbox{\bf j})$  but no corresponding magnetic ones. Maxwell's equations are:
	
\begin{equation}
\label{eqmaxw2}
\partial_{\nu}F^{\mu \nu} = -j^{\mu}  \hskip 2cm \partial_{\nu}\tilde{F}^{\mu \nu} = 0
\end{equation}

We can try to restore the symmetry, but then we must introduce a magnetic current $k_{\mu}=(\sigma,\mbox{\bf k})$ and write the equations (\ref{eqmaxw2})
 as:

\begin{equation}
\label{eqmaxw3}
\partial_{\nu}F^{\mu \nu} = -j^{\mu}  \hskip 2cm \partial_{\nu}\tilde{F}^{\mu \nu} = -k^{\mu}
\end{equation}

\noindent which are invariant under (\ref{eqdual2}), provided $j^{\mu}$ and $k^{\mu}$ transform as:

\begin{equation}
\label{eqdual3}
j^{\mu}\rightarrow k^{\mu} \hskip 2cm k^{\mu}\rightarrow -j^{\mu}
\end{equation}

For example, if the electric current results from the motion of point electric charges, the magnetic current $k^{\mu}$ will result from the motion of point magnetic charges, $i.e.$ magnetic monopoles. 

The introduction of magnetic charges looks at first sight like a trivial generalisation of quantum electrodynamics. However, it is easy to see that this is not so. The usual quantisation procedure is set up in terms of the vector potential $A^{\mu}$ rather than the electric and magnetic fields $F^{\mu \nu}$. The latter is given by $F^{\mu \nu}=\partial^{\mu}A^{\nu}-\partial^{\nu}A^{\mu}$. This last relation implies the vanishing of $\partial_{\nu}\tilde{F}^{\mu \nu}$ and thus the absence of any magnetic current. We conclude that, in a theory with magnetic monopoles, the vector potential cannot be a well-defined function of the space-time point $x$. 

In order to get a feeling of what is going on, let us view an isolated magnetic monopole of magnetic charge $g$ sitting at the origin, as one end of a solenoid in the limit when the latter is infinitely long and infinitely thin. The line occupied by the solenoid, say the negative $z$-axis, is called ``the Dirac string''. An observer will see a magnetic flux coming out of the origin as if a monopole were present. He won't realize that the flux is coming back through the solenoid, because, by assumption, this is infinitely thin. The total magnetic field is singular on the string and is given by:

\begin{equation}
\label{eqmonf1}
\mbox{\bf B}=\frac{g}{r^2} \mbox{\boldmath $\hat{r}$}+4\pi g\theta (-z)\delta (x)\delta (y) \mbox{\boldmath $\hat{z}$} 
\end{equation}

\noindent with $\mbox{\boldmath $\hat{r}$}$ and $\mbox{\boldmath $\hat{z}$}$ being the corresponding unit vectors. The first term represents  the field of the monopole which has the usual point-particle singularity, while the second is the singular contribution of the string. We can construct a vector potential {\bf A} whose curl is {\bf B}. Of course, we expect also {\bf A} to be singular on the negative $z$-axis. A simple computation, using spherical coordinates, gives:

\begin{equation}
\label{eqmonf2} 
\mbox{\bf A}=\frac{g}{r}\frac{1-\cos \theta}{\sin \theta} \mbox{\boldmath $\hat{\phi}$}
\end{equation}

\noindent where $\mbox{\boldmath $\hat{\phi}$}$ is the unit vector in the $\phi$-direction. {\bf A} can be taken to represent the field of the monopole and indeed this is true everywhere except on the negative $z$-axis. Since, by our previous argument, we know that a string-like singularity must exist, the form (\ref{eqmonf2}) is the best we can do. Obviously, the choice of the negative $z$-axis as the position of the string is arbitrary and we could have placed the solenoid along any line from the origin to infinity. 

So far the discussion was purely classical. Quantum mechanics brings a subtle difference. In classical electrodynamics the vector potential $A_{\mu}$ is not measurable, only the components {\bf E} and {\bf B} of $F_{\mu \nu}$ are. In quantum mechanics however, we can detect directly the presence of $A_{\mu}$ by the Bohm-Aharonov effect. By moving around electrically charged test particles we can discover the magnetic flux coming back through the string. The corresponding change in the phase of the wave function will be:

\begin{equation}
\label{eqphase}
\psi \rightarrow e^{4\pi ieg}\psi
\end{equation}

\noindent where $4\pi g$ is the flux and $e$ the charge of the test particle. The usual interference experiment will detect the phase 
change and hence the presence of the string, unless $\exp(4\pi ieg)$ equals one or
	
\begin{equation}
\label{eqquant}
eg=0,\pm \frac{1}{2}, \pm 1,...
\end{equation}

Condition (\ref{eqquant}) is the famous Dirac quantisation condition. If it is satisfied the string is undetectable by any 
conceivable experiment and we have obtained a real magnetic monopole. On the other hand, it shows that if there 
exists a magnetic monopole in the world, all electric charges must be quantised, $i.e.$ they must be multiplets of an 
elementary charge $e_0$ . Similarly all magnetic charges must be multiplets of an elementary magnetic charge $g_0$ such that 
$2e_0 g_0$ is an integer. A particle that has both electric and magnetic charge is called ``a dyon''. 

{\it The 't Hooft-Polyakov monopole.}
In the Abelian case we saw that magnetic monopoles give rise to singular vector potentials. We shall now turn to non-Abelian theories. We have good reasons to believe that the electromagnetic gauge group is part of a bigger group 
which is spontaneously broken through the Higgs mechanism. The simplest such theory, although not the one chosen 
by nature, is the Georgi-Glashow $SO(3)$ model. It is a theory without weak neutral currents in which the only gauge 
bosons are $W^+, W^-$ and $\gamma$. We introduce a triplet of Higgs scalars $\mbox{\boldmath $\Phi$}$ and we write the Lagrangian as:

\begin{equation}
\label{eqGG1}
{\mathcal L}=-\frac{1}{4}\mbox{\boldmath $G$}_{\mu \nu}\cdot \mbox{\boldmath $G$}^{\mu \nu}+\frac{1}{2}({\mathcal D}_{\mu}\mbox{\boldmath $\Phi$})\cdot ({\mathcal D}^{\mu}\mbox{\boldmath $\Phi$}) -V(\mbox{\boldmath $\Phi$})  
\end{equation}

\begin{eqnarray}
\label{eqGG2}
\mbox{\boldmath $G$}_{\mu \nu} & = & \partial_{\mu}\mbox{\boldmath $W$}_{\nu}-\partial_{\nu}\mbox{\boldmath $W$}_{\mu}-e\mbox{\boldmath $W$}_{\mu}\times\mbox{\boldmath $W$}_{\nu} \nonumber \\
{\mathcal D}_{\mu}\mbox{\boldmath $\Phi$} & = & \partial_{\mu}\mbox{\boldmath $\Phi$} -e\mbox{\boldmath $W$}_{\mu} \times \mbox{\boldmath $\Phi$} \\
V(\mbox{\boldmath $\Phi$}) & = & \frac{\lambda}{4}({\mbox{\boldmath $\Phi$}}^2 - v^2)^2 \nonumber
\end{eqnarray}

We have written the scalar potential V in a form which exhibits explicitly the minimum away from the origin in 
field space and we have not included any fermions for simplicity.

From (\ref{eqGG1}) we can compute the corresponding Hamiltonian density.

\begin{equation}
\label{eqGGH}
{\mathcal H}= \frac{1}{2}\sum_{i=1}^{3}[ \mbox{\boldmath ${\mathcal E}$}^i \cdot \mbox{\boldmath ${\mathcal E}$}^i + \mbox{\boldmath ${\mathcal B}$}^i \cdot \mbox{\boldmath ${\mathcal B}$}^i + ({\mathcal D}_0 \mbox{\boldmath $\Phi$})\cdot ({\mathcal D}_0 \mbox{\boldmath $\Phi$}) +({\mathcal D}^i \mbox{\boldmath $\Phi$})\cdot ({\mathcal D}^i \mbox{\boldmath $\Phi$})] + V(\mbox{\boldmath $\Phi$})
\end{equation}

\noindent where we have defined the non-Abelian ``electric'' and ``magnetic'' fields as $\mbox{\boldmath ${\mathcal E}$}^i = -\mbox{\boldmath $G$}^{0i}$ and $\mbox{\boldmath ${\mathcal B}$}^i = -\epsilon^{ijk} \mbox{\boldmath $G$}_{jk}$ and the bold-face vectors refer again to the internal symmetry space. The important point about ${\mathcal H}$ is that it is the sum of positive semidefinite terms. Therefore, the minimum energy solution will be that for which ${\mathcal H}=0$. On the other hand, ${\mathcal H}$ is invariant under local $SO(3)$ rotations in the  internal symmetry space. However, the only symmetric solution, $i.e.$ the field configuration

\begin{equation}
\label{eqGGsymsol} 
\mbox{\boldmath $W$}_{\mu}=0 \hskip 2cm \mbox{\boldmath $\Phi$}=0
\end{equation}

\noindent gives ${\mathcal H}=v^2$ and thus corresponds to infinite total energy. The zero energy solution must make each term of (\ref{eqGGH}) vanish. An example of such a solution is

\begin{equation}
\label{eqGGHigsol} 
\mbox{\boldmath $W$}_{\mu}=0 \hskip 2cm \mbox{\boldmath $\Phi$}=v \mbox{\boldmath $\hat{k}$}
\end{equation}

\noindent with $\mbox{\boldmath $\hat{k}$}$ being the unit vector in the third direction in the internal symmetry space. Obviously, since ${\mathcal H}$ is gauge invariant, any gauge transform of (\ref{eqGGHigsol}) will give another zero-energy solution. In particular, there is nothing sacred about choosing $\mbox{\boldmath $\Phi$}$ to point along the third direction.

The solution (\ref{eqGGHigsol}), or any transform of it, exhibits the well-known Higgs phenomenon. The symmetry $SO(3)$ is spontaneously broken since the invariance of the solution is reduced to the group of rotations around the third axis, $i.e.$ $U(1)$. Two of the vector bosons acquire a mass and it is natural to identify the third one, which remains massless, with the photon. 

Up to now we have found two sets of solutions of the equations of motion given by the Lagrangian density (\ref{eqGG1}). One $SO(3)$ symmetric solution given by (\ref{eqGGsymsol}) and a whole family of asymmetric ones given by (\ref{eqGGHigsol}) and all its gauge transforms. The first corresponds to infinite total energy while all the second ones to zero energy. They describe the family of stable vacuum states. A natural question is the following: Are there any finite, non-zero-energy, non-trivial, particle-like solutions? The condition of finite total energy implies that  ${\mathcal H}$ must vanish at large distances, therefore, asymptotically, any such solution will approach one belonging to the family (\ref{eqGGHigsol}). 

There is no general method for finding the solutions of coupled, non-linear, partial differential equations. What is usually done, is to guess a particular form of the solution and to simplify the equations. In doing the guesswork one often tries first to guess the symmetries of the 
solution. Since we are looking for a stable particle-like solution (the magnetic monopole) the solution must be time-independent. Furthermore it will be left invariant by a group of transformations which is a subgroup of the symmetries 
of the equations of motion. In the rest frame of the particle the latter is $G = SO(3) \otimes SO(3) \otimes P \otimes R$ where the first 
$SO(3)$  corresponds to spatial rotations, the second to the internal symmetry, $P$ denotes parity and $R$ the transformation $\mbox{\boldmath $\Phi$} \rightarrow -\mbox{\boldmath $\Phi$}$
. In guessing the form of the monopole solution we shall try to enforce as much of the symmetry $G$ as possible. 
Invariance under spatial rotations will force $\mbox{\boldmath $\Phi$}$ to be asymptotically constant and $G^{ij} \sim r^{ij}$. It is easy to verify that this solution has zero total magnetic charge like the Higgs vacuum solution of eq. (\ref{eqGGHigsol}). On the other hand, since our 
solution must approach at large distances one of the internal symmetry breaking ones, we cannot enforce the second 
$SO(3)$ either. Finally, both $P$ and $R$ change the sign of the magnetic charge and cannot be included. Let us choose, 
therefore, to impose invariance under $SO(3) \otimes PR$ where $SO(3)$ is the diagonal subgroup of $SO(3) \otimes SO(3)$ and $PR$ is 
the product of the two. We seek solutions of the general form:

\begin{eqnarray}
\label{eqmonsol}
\Phi_{\alpha} & = & H(evr)\frac{x_{\alpha}}{er^2} \nonumber \\
W^0_{\alpha} & = & 0 \\
W^i_{\alpha} & = & -\epsilon_{\alpha ij}\frac{x_j}{er^2}[1-K(evr)] \nonumber
\end{eqnarray}

\noindent with $H$ and $K$ functions of a single variable. Space and internal symmetry indices are mixed and the ansatz (\ref{eqmonsol}) is 
spherically symmetric in the sense that a spatial rotation can be compensated for by an internal symmetry rotation.  Plugging (\ref{eqmonsol}) into the equations of motion 
we obtain a system of coupled ordinary differential equations for $H$ and $K$ which can be solved, at least numerically. It is easy to verify that this solution does describe a magnetic monopole. We can compute the associated magnetic field and we find asymptotically

\begin{equation}
\label{eqmagfi}
B^i \rightarrow -\frac{1}{e}\frac{x^i}{r^3}
\end{equation}

\noindent which corresponds to a magnetic charge:

\begin{equation}
\label{eqmagfch}
g=-\frac{1}{e}
\end{equation}

\noindent thus obtaining $|eg| = 1$. Notice that here $e$ is the charge of the boson which has isospin = 1. If we had included 
isospin 1/2 fields, for example isospinor fermions, the symmetry group would have been $SU(2)$ and the 
smallest electric charge in the theory would have been $|e/2|$ . We thus recover the Dirac quantisation 
condition which states that the minimum magnetic charge is given by $|e_{min}g_{min}| = 1/2$. We can also compute the total energy of the solution which can be interpreted as the classical approximation to 
the mass of the monopole. We find:

\begin{equation}
\label{eqmonmas}
M_{monopole}=\frac{4\pi v}{e}f(\lambda /e^2)
\end{equation}

\noindent with $f$ a given function of the ratio of the coupling constants. It turns out that for $x =\lambda /e^2$ ranging from zero to ten $f$ 
stays of order one ($f(0) = 1$, $f(10)\simeq 1.44$). Since the mass of the massive vector bosons after spontaneous symmetry 
breaking is of order $ev$, it follows that

\begin{equation}
\label{eqmonmas1}
M_{monopole} \sim M_{vector\ boson}/\alpha \sim 10^2 M_{vector\ boson}
\end{equation}

B. Julia and A. Zee have generalised the solution (\ref{eqmonsol}) by choosing a non-vanishing $W_0^{\alpha}$. They obtained dyon solutions with masses satisfying $M_{dyon} \geq v(e^2 + g^2)^{1/2}$. In general, we can obtain multi-charged solutions with electric charge $ne$ and magnetic charge $mg$, with $n$ and $m$ integers. The Dirac quantisation condition is again verified.

The asymptotic form of the Higgs field is obtained from (\ref{eqmonsol}) by solving for $H$. We find 

\begin{equation}
\label{eqmonHig}
\Phi^{\alpha}_{as}= v\frac{x^{\alpha}}{r}
\end{equation}

\begin{figure}[here]
\begin{picture}(350,120)
\put (175,60){\circle{40}}
\put (175,90){\vector(0,1){20}}
\put (175,30){\vector(0,-1){20}}
\put (205,60){\vector(1,0){20}}
\put (145,60){\vector(-1,0){20}}
\put (195,80){\vector(1,1){14}}
\put (195,40){\vector(1,-1){14}}
\put (155,80){\vector(-1,1){14}}
\put (155,40){\vector(-1,-1){14}}
\end{picture}
\caption { The asymptotic form of the Higgs field for the monopole solution} \label{figmon}
\end{figure}
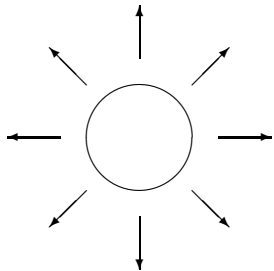

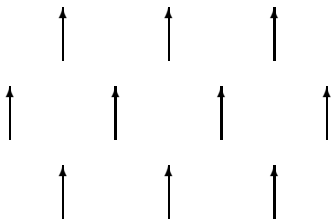
\begin{figure}[here]
\begin{picture}(350,120)
\multiput (115,50)(40,0){4}{\vector(0,1){20}}
\multiput (135,80)(40,0){3}{\vector(0,1){20}}
\multiput (135,20)(40,0){3}{\vector(0,1){20}}
\end{picture}
\caption {The Higgs field in the vacuum configuration} \label{figvac}
\end{figure}

\noindent $i.e.$ $\Phi_{as}$  points as in Figure \ref{figmon}. This corresponds to a (singular) gauge transformation of the Higgs vacuum configuration (\ref{eqGGHigsol})
 of Figure \ref{figvac}. 

Topologically we can view $\Phi_{as}$  as mapping the surface at spatial infinity $S^2$ onto the corresponding surface of 
internal space which is also $S^2$. Since it is not possible to continuously deform the map of Figure \ref{figmon} to the constant map of 
Figure \ref{figvac}, we conclude that the monopole configuration is topologically stable and cannot decay to the vacuum. We can 
understand this physically as the consequence of conservation of magnetic charge. 

We also note that the solution (\ref{eqmonsol}) is regular everywhere, including the origin. The presence of the 
Higgs field makes possible the existence of a magnetic monopole solution which not only does not have the Dirac string singularity, but also it is smooth all the way to $r$ = 0.  Therefore the monopoles are non-singular, finite energy, static solutions of the classical equations of motion. It 
is reasonable to assume, although there is no rigorous proof for this, that they survive quantisation and correspond to real, physical particles. The fact that they have a regular internal structure has a very important physical consequence: the mass of the 't Hooft-Polyakov monopole is calculable, while that of the Dirac one is not. The spectrum of the stable one-particle states will consist of: (i) a massless photon, (ii) a pair of charged vector bosons $W^{\pm}$ with mass of order $ev$ and electric charge $\pm e$, (iii) a neutral scalar boson with mass of order $\sqrt{\lambda} v$ {\it and} (iv) a monopole anti-monopole pair with mass of order $v/e$ and magnetic charge $\pm g = \mp (1/e)$. 

An interesting simplification occurs in the so-called ``BPS'' (Bogomolny-Prasad-Sommerfield) limit which consists of $\lambda \rightarrow 0^+$ keeping $v$ fixed and maintaining the form of the solution (\ref{eqmonsol}). In this limit the physical scalar boson becomes massless, like the photon and gives rise to a new long-range force. We shall come back to this point presently.

{\it Other gauge groups.}
In quantum electrodynamics magnetic monopoles are a curiosity. They may exist, in which case they have 
important physical consequences like the quantisation of electric charge, but nobody forces us to introduce them. The 
theory makes perfect sense without them. The monopole configuration is a singular one for the gauge potential $A^{\mu}(x)$. 
On the other hand, we just saw that, if electromagnetism is part of an $SO(3)$ gauge theory with  a triplet Higgs fields, 
magnetic monopoles are, probably, part of the physical spectrum. Of course,
$SO(3)$, or $SU(2)$, are not the right invariance groups for physics because they have no room for weak neutral 
currents. Could we apply the same reasoning to the gauge group of the standard model which is $SU(2) \otimes U(1)$ 
broken to $U(1)$? Does the standard model imply the existence of stable magnetic monopoles? The answer is 
no even if one enlarges the Higgs content. The reason is that the stability argument of the previous section 
does not apply to the standard model. The argument was based on the fact that there is no $SO(3)$ continuous, 
non-singular gauge transformation which can deform the monopole configuration of eq. (\ref{eqmonsol}) to the vacuum 
one of eq. (\ref{eqGGHigsol}). However, it is easy to see that this is no more true when the gauge group is $SU(2) \otimes U(1)$. The 
presence of the $U(1)$ factor makes the 't Hooft-Polyakov magnetic
monopole unstable. Therefore, the natural application of these ideas
is the framework of grand unified theories which we shall see
shortly. Let me here present briefly the results of the general case
which was investigated by P. Goddard, J. Nuyts and D. Olive. 

Let us assume that a gauge theory of a simple group $G$ is
spontaneously broken through the Higgs mechanism to a subgroup
$H$. (In the 't Hooft-Polyakov case $G$ was $SO(3)$ and $H$ was
$U(1)$). Goddard, Nuyts and Olive  showed that we can again find
magnetic monopole and dyon solutions with the following properties: If
$H$ is not just $U(1)$, the ``electric'' (and ``magnetic'') charges
are not given by a single number because the fields form multiplets of
$H$. It turns out that  the electric charges are described in terms of
the group $H$, while the magnetic ones in terms of its dual, denoted
by $H^*$. The distinction between the two sounds like a mathematical
detail because, for all practical purposes, they are locally
isomorphic, like $SO(3)$ and $SU(2)$. However, we have seen already
that such details may be relevant. For example, $SU(2)$ admits
half-integer charges while  $SO(3)$ does not and this affects the Dirac
quantisation condition. We shall encounter this problem again when
studying monopoles in the grand unified model $SU(5)$. More precisely,
the values of the electric charges can be viewed as vectors $\vec{q}$
in the weight lattice of $H$ and the magnetic charges as $\vec{g}$ in that of $H^*$. The quantisation condition now reads:

\begin{equation}
\label{eqDiracgen1}
e \vec{q} \cdot \vec{g} = 2 \pi N \hskip 0.3cm ; \hskip 0.3cm N \in \IZ
\end{equation}

{\it Monopoles and instantons in gauge theories.}
Up to now we have assumed that a gauge theory based on a simple group $G$ is described in terms of a single coupling constant $g$. This is certainly true in perturbation theory. However, with monopoles we entered already the non-perturbative regime and it is well-known that a second coupling constant appears, the $CP$-violating $\theta$ angle. It corresponds to a new effective term in the Lagrangian  given by:

\begin{equation}
\label{eqtheta1}
 L_{\theta}=\frac{\theta e^2}{32 \pi}\int d^4 x F^a _{\mu
\nu}\tilde{F}^{a\mu \nu}=\theta N  
\end{equation}
 
\noindent where $N \in \IZ$ is an integer denoting the number of
instantons in the configuration. This term is absent in perturbation
theory because it is easy to show that the integrant is equal to a
four derivative and, therefore, the integral vanishes for field
configurations which vanish at infinity. In the presence of
non-trivial boundary conditions (``instantons'') we obtain the
topological answer of (\ref{eqtheta1}). In the mathematical literature
$N$ is called ``the Pontryagin index''. 

$\tilde{F}$ is a pseudo-tensor, so this term, when added to the usual
Yang-Mills Lagrangian,  strongly
violates $CP$, so for Q.C.D. we have very severe limits on the value
of $\theta$. The fact that $N$ is an integer implies that $\theta$ is
an angle, in the sense that all physical results must be periodic in
$\theta$, $\theta \rightarrow \theta +2\pi$.

We can repeat the previous analysis and look for monopole and dyon
solutions in a gauge theory including $L_{\theta}$. For a monopole, we obtain a shift
in the value of the charge given by: 

\begin{equation}
\label{eqtheta2}
q=\frac{\theta e^2}{8 \pi^2}g
\end{equation}

\noindent and, for a dyon, taking into account the Dirac condition:

\begin{equation}
\label{eqtheta3}
q=ne+\frac{\theta e}{2 \pi}m \hskip 0.3cm ; \hskip 0.3cm g=\frac{4
\pi}{e}m \hskip 0.3cm ; \hskip 0.3cm n,m \in \IZ
\end{equation}

It is useful to introduce a complex representation and write:

\begin{equation}
\label{eqcomplt1}
\tau=\frac{\theta }{2 \pi}+i\frac{4
\pi}{e^2}\hskip 0.3cm ; \hskip 0.3cm q+ig=e(n+m\tau)
\end{equation}

The BPS bound can be written using this notation as $M \geq ve|n+m\tau|$.

{\it Monopoles in Grand Unified Theories.}
In the previous 
section we argued in favour of grand unified theories in which one starts with a simple group $G$ and, after 
several spontaneous symmetry breakings, one ends up with $SU(3) \otimes U(1)$. Since $SU(2)$ can be embedded in 
any simple $G$, we expect to find stable monopole solutions in every grand unified theory. This question was first studied by C.P. Dokos and T.N. Tomaras in 1979.

Indeed, we can give a general proof of this statement. The presence of the $U(1)_{e.m.}$ factor in the 
unbroken gauge subgroup of the simple grand unified group guarantees the existence of smooth, finite-energy, topologically stable, particle-like solutions of the equations of motion with quantised magnetic 
charge. The proof of stability is the direct generalisation of the simple topological argument given in the 
previous section. Let us exhibit the grand unified monopoles in the case of the $SU(5)$ prototype model.

Without going into details, let us try to find the quantum 
numbers of the $SU(5)$ monopole using only simple symmetry arguments. Following Dokos and Tomaras, we can first determine the 
minimum value of magnetic charge $g_{min}$. Since the minimum value of electric charge is $e_{min} = e/3$ (the down-
quark) we would expect, na\"{\i}vely, to have $g_{min} = 3/(2e)$. However this is incorrect. The reason is, again, 
our sloppy way in dealing with group theory, not paying much attention to the difference 
between a group and its Lie algebra. Being more careful, Dokos and Tomaras have established that, in the case of $SU(5)$, the 
unbroken group $H$ is only locally isomorphic to $SU(3) \otimes U(1)$. The group $H$ is really the set of $SU(5)$ 
matrices of the form:

\begin{equation}
\label{eqh}
\left( \begin{array}{ccc}
ue^{-i\alpha} & & \\
 & e^{3i\alpha} & \\
 & & 1
\end{array}
\right)
\end{equation}

\noindent with $u$ an element of $SU(3)$. The mapping from $SU(3) \otimes U(1)$ to $H$, defined by $(u, e^{i\alpha}) \rightarrow$ diag$(ue^{-i\alpha}, e^{3i\alpha}, 1)$ is 
3 to 1, since the three elements $(u, e^{i\alpha}), (ue^{2\pi i/3}, e^{i(\alpha +2\pi/3)}), (ue^{4\pi i/3}, e^{i(\alpha +4\pi /3})$ of $SU(3) \otimes U(1)$ are mapped to the 
same element of $H$. In other words, $H$ is the group $SU(3) \otimes U(1)/Z_3$, with $Z_3$ being the centre of $SU(3)$. As a 
consequence of this, using the same phase argument of the previous section, we can show that $g_{min}=[2\cdot 3\cdot e_{min}]^{-1}=[2e]^{-1}$ with $|e|$ being again the electron charge. We thus recover the Dirac quantisation condition.

Let us next identify the $SU(2)$ subgroup of $SU(5)$ which will be used to construct the symmetry group 
of the monopole solution. It is clear that it cannot be a subgroup of $SU(3)_{colour}$ because it must contain 
electromagnetism. On the other hand it cannot be the $SU(2)$ of the standard model as we have already shown. 
Therefore, we are led to a choice of the form:

\begin{equation}
\label{eqmongen}
\mbox{\boldmath $T$}=\frac{1}{2} \left( \begin{array}{cccc}
0 & & & \\ & 0 & & \\ & & \mbox{\boldmath $\tau $} & \\ & & & 0
\end{array}
\right)
\end{equation}

\noindent for the $SU(2)$ generators $\mbox{\boldmath $\tau $}$. It follows that the monopole magnetic field will have both ordinary magnetic and 
colour magnetic components.

From now on, the analysis proceeds like in the previous section. We can identify the asymptotic form of 
the solution and obtain a system of ordinary differential equations. The monopole mass in the classical 
approximation will be of the order of $10^2 M$, with $M$ being the mass of the vector bosons of $SU(5)$, $i.e.$ the 
magnetic monopoles of grand unified theories are superheavy with masses of order $10^{17}$GeV, far 
beyond the reach of any accelerator. On the other hand the lightest monopole is stable. It is amusing to 
notice that the only stable elementary particles in a grand unified theory are the photon, which is massless, 
the lightest neutrino, the electron and the monopole. Neutrino decay is 
forbidden by angular momentum conservation because the neutrino is the lightest spin-1/2 particle. The 
stability of the electron and that of the monopole are guaranteed by the conservation of electric and magnetic charge. Let me also mention that, although baryons in grand unified theories  may have long life-times,  magnetic monopoles act as catalysers: in their presence baryons decay promptly.

Since monopoles are stable, whichever might have been produced in the course of the evolution of the Universe 
have a reasonable chance of being around to day. The problem of estimating the monopole abundance is therefore 
reduced into that of estimating the rate of monopole production and that of subsequent monopole-antimonopole 
annihilation.
At the very early Universe, when temperature was sufficiently high, we expect 
all symmetries, which are spontaneously broken today, to be restored. At
$T > T_c \sim M$ we have the full $SU(5)$ symmetry. But for magnetic monopoles to exist $SU(5)$ must be broken. 
Therefore there were no monopoles when $T > T_c$ . They are produced during the phase transition but the 
production mechanism is model dependent. If the phase transition is second order the vacuum expectation 
value of the Higgs field undergoes large fluctuations and a domain structure is established. The resulting 
domain walls contain topological defects in the Higgs field orientation which give rise to magnetic 
monopoles. A rough estimation of their density gives $d_{monopole} \sim  \xi^{-3}$ where $\xi$ is the correlation length. Although 
its precise estimation is difficult, it is certainly less than the horizon length at temperature $T$. The latter is 
of order $CM_P T^{-2}$, where $M_P$ is the Planck mass and $C$ depends on the number of massless degrees of freedom 
in thermal equilibrium at temperature $T$. We thus obtain a bound for the initial monopole density of the 
order $d_{mon}T^{-3} \geq 10^{-10}$. A similar result is obtained even in the case of a first order phase transition.  Order of magnitude estimations  show that the annihilation rate in an expanding Universe does not substantially reduce this number. We 
are thus left with a monopole density $d \sim 10^{-10}T^3$, $i.e.$ comparable to the baryon density. This is obviously 
absurd if we take into account their enormous mass.

Several mechanisms have been proposed to reduce the number of monopo\-les surviving but the most 
attractive proposal today is based on the inflation scenario. During the exponential expansion any initial 
monopole density is ``inflated away'' and reduced to essentially zero. During an eventual subsequent 
reheating new monopoles could be produced but the production rate depends crucially on the reheating 
temperature. The final result is model dependent, but acceptably low densities can be obtained.

{\it The Montonen-Olive duality conjecture.}
All our quantitative understanding of four-dimensional quantum field theories comes from perturbation theory. The enormous success of the Standard Model testifies to that. On the other hand we know that such an understanding is necessarily limited. Many exciting physical phenomena are not accessible to perturbation. The obvious example is confinement in Q.C.D. The perturbative spectrum of the theory is described by the Fock space of states of quarks and gluons, but we know that the asymptotic states are those of hadrons. In a qualitative way, we attribute this property to the behaviour of the Q.C.D. effective coupling strength as a function of scale. At short distances the coupling is weak and perturbative calculations are reliable. At large distances we enter the strong coupling regime and perturbation theory breaks down. If we had an exact, analytic solution of the quantum field theory, we would have found the hadrons as complicated functionals of the quark and gluon fields, describing static, finite energy, particle-like configurations. Physicists have often tried to guess effective descriptions of Q.C.D. in terms of collective variables valid at large distances, where the description in terms of quark and gluon fields is inadequate.  

The model with non-abelian magnetic monopoles may provide a simple example of such a situation. We recall that in the BPS limit, the perturbative, one-particle states of the system, $i.e.$ those corresponding to the elementary fields in the original Lagrangian, are the massless photon, the massless scalar and the two spin-one bosons of mass $M=ev$ and electric charge $\pm e$. The strength of the interaction is given by $e$. In addition the spectrum contains non-perturbative one-particle states, the magnetic monopoles. Contrary to the gauge bosons, they are not point-particles but extended objects. Their masses are $M_{mon}=gv$ and they have a magnetic charge $g=\pm 1/e$. The electric charge is conserved as a consequence of Noether's theorem, because it corresponds to an unbroken symmetry of the Lagrangian. The magnetic charge is conserved as a consequence of the topological structure of the monopole solution. All this reminds us of the duality symmetry of Maxwell's equations  (\ref{eqdual2}) and (\ref{eqdual3}). 

The complete quantum mechanical properties of this model are not known. For small $e$ we can use perturbation theory. Notice that in this case $g$ is large and the monopoles are heavy. When $e$ grows large we enter the strong coupling regime and perturbation theory breaks down. In this case $g$ is small and the monopoles are light. C. Montonen and D. Olive made the following conjecture: This model admits two ``dual equivalent'' field theory formulations in which electric (Noether) and magnetic (topological) quantum numbers exchange roles. The Lagrangian (\ref{eqGG1}) in the BPS limit gives the first one. The electrically charged gauge bosons are the elementary fields, $e$ is the coupling constant and the monopoles are the extended objects. Obviously, this description should be useful when $e$ is small. In the dual equivalent description the monopoles are the elementary fields,  together with the photon and the massless scalar, $g$ is the new coupling constant and the electrically charged gauge bosons correspond to extended, soliton-like, solutions. This will be the useful description for large $e$. This is the conservative version of the conjecture. In fact, Montonen and Olive went a step further: They conjectured that, for this particular model, the two field theories are identical, $i.e.$ the new Lagrangian is again given by  (\ref{eqGG1}) with $g$ replacing $e$. The monopole fields and the photon form a triplet under a new $SO(3)$ gauge group. This may sound strange because the monopole solution appears to be gauge equivalent to a spherically symmetric one and, na\"{\i}vely, we expect the monopole to have zero spin. In fact, the situation is more involved and a complete calculation of the total angular momentum of the solution is not easy. We shall come back to this point later. 

A proof of this conjecture would require a complete solution of the field theory, a rather unlikely state of affairs in any foreseeable future. We can only verify its calculable consequences. So far, such checks have been successful. Let me mention only one: We can generalise  the classical solution  (\ref{eqmonsol}) to one describing any number of well separated monopoles and/or antimonopoles at rest. Studying the instantaneous acceleration of such states, J.S. Manton has found the long-range part of the classical force between two monopoles. The surprising result is:

\begin{equation}
\label{eqforce}
F=-\frac{2g^2}{r^2} \hskip 0.5cm \mbox {opposite charge} \hskip 0.5cm ; \hskip 0.5cm F=0 \hskip 0.5cm \mbox {same charge}
\end{equation}

\noindent $i.e.$ twice the naively expected attraction for opposite charges and no force at all for same charges. How does this agree with the known forces between the electrically charged particles of the model? It does! The classical force between the two charged vector bosons receives two contributions. The one photon exchange graph gives a Coulomb force of $\pm e^2/r^2$, attraction for opposite charges and repulsion for like charges. The one scalar exchange graph is always attractive, like all even spin exchanges. It is given by the tri-linear term in (\ref{eqGG1}), after translation of the scalar field. The result is $-e^2/r^2$, $i.e.$ it doubles the attractive part of the photon and cancels the repulsive one. Therefore, in a completely independent calculation, we derive the result (\ref{eqforce}) with $e$ replacing $g$.

\section{ Beyond gauge theories}
\label{secsupsym}

\subsection{The trial of scalars}

The purpose of this section is not to destroy, but to fulfil. It is our firm belief, shared by most physicists, that 
gauge theories have come to stay. ``Beyond'' here does not mean that we propose to replace gauge theories by 
something else, but rather to embed them into a larger scheme with a tighter structure and higher predictive 
power. There are several reasons for such a search.

As we said in section \ref{secSM}, gauge theories contain two and possibly three independent worlds. The world of 
radiation with the gauge bosons, the world of matter with the fermions and, finally, in our present 
understanding, the world of Higgs scalars. In the framework of gauge theories these worlds are essentially 
unrelated to each other. Given a group $G$ the world of radiation is completely determined, but we have no way 
to know a priori which and how many fermion representations should be introduced; the world of matter is, to 
a great extent, arbitrary.

This arbitrariness is even more disturbing if one considers the world of Higgs scalars. Not only their 
number and their representations are undetermined, but their mere presence introduces a large number of 
arbitrary parameters into the theory.  Notice that this is independent of 
our computational ability, since these are parameters which appear in our fundamental Lagrangian. What 
makes things worse, is that these arbitrary parameters appear with a wild range of values. For example, in the 
standard model, the ratio of Yukawa couplings for different fermions equals the ratio of fermion masses. But $m_t/m_e \sim 10^5$ and $m_t/m_\nu > 10^{11}$. It is hard to admit that such  numbers are fundamental parameters.

The situation becomes even more dramatic in grand unified theories where one may have to adjust 
parameters with as many as twenty-eight significant figures. This is the problem of gauge hierarchy which is 
connected to the enormously different mass scales at which spontaneous symmetry breaking occurs. The 
breaking of $G$ into $U(1) \otimes SU(2) \otimes SU(3)$ happens at $M \sim 10^{16}$ GeV. This means that a certain Higgs field $\Phi$ 
acquires a non-zero vacuum expectation value $V \sim 10^{16}$ GeV. The second breaking, that of $U(1) \otimes SU(2)$, 
occurs at $\mu \sim 10^{2}$ GeV, $i.e.$ we must have a second scalar field $H$ with $v \sim 10^{2}$ GeV. But the combined 
Higgs potential will contain  terms of the form $\Phi^2 H^2$ (see eq. (\ref{eqpot524})). Therefore, after the first breaking, the $H$-
mass will be given by:
  
\begin{equation}
\label{eqgh}
m^2_H = {\mu}^2 + O(\alpha V^2)
\end{equation}

\noindent where $\mu $ is the mass appearing in the symmetric Lagrangian. On the other hand I want to remind you that $v^2 \sim m^2_H$, so unless there is a very precise cancellation between ${\mu}^2$ and $O(\alpha V^2)$, a cancellation which should extend 
to twenty-eight decimal figures, $v$ will turn out to be of order $V$ and the two breakings will come together, in 
other words the theory is not able to sustain naturally a gauge hierarchy. This grand-fine tuning of parameters 
must be repeated order by order in perturbation theory because, unlike fermions, scalar field masses require 
quadratically divergent counterterms. The whole structure looks extremely unlikely. The problem is similar to 
that of the induced cosmological constant in any theory with spontaneous symmetry breaking. I believe that, 
in spite of its rather technical aspect, the problem is sufficiently important so that some new insight will be 
gained when it is eventually solved.

One possible remedy is to throw away the scalars as fundamental elementary particles. After all, their sole 
purpose was to provoke the spontaneous symmetry breaking through their non-vanishing vacuum expectation 
values. In non-relativistic physics this phenomenon is known to occur but the role of Higgs fields is played by 
fermion pairs (ex. the Cooper pairs in superconductivity). Let me also remind you that the spontaneous 
breaking of chiral symmetry, which is supposed to be a fundamental property of Q.C.D., does show the same 
feature, namely the ``vacuum'' is formed by quark-antiquark pair condensates and the resulting Goldstone 
boson (the pion) is again a $q\bar q$ bound state. This idea of dynamical symmetry breaking has been studied 
extensively, especially under the name of ``Technicolor''. In spite of its many attractive features, it suffers, 
up to now, from two main difficulties. First, the available field theory technology does not allow for any 
precise quantitative computation of bound state effects and everything has to be based on analogy with the 
chiral symmetry breaking in Q.C.D. Second, nobody has succeeded in producing an entirely satisfactory 
phenomenological model. Nevertheless there is still hope and the scheme has some  predictions which can be tested experimentally in the near future.

\subsection{Supersymmetry, or the Defence of scalars}

The best defence of scalars is the remark that they are not the only ones to reduce the predictive power of a 
gauge theory. As we have already seen, going through the chain radiation -fermion matter fields- Higgs 
scalars we encounter an increasing degree of arbitrariness. One possibility which presents itself is to connect 
the three worlds with some sort of symmetry principle. Then the knowledge of the vector bosons will 
determine the fermions and the scalars and the absence of quadratically divergent counterterms from the 
fermion masses will forbid their appearance in the scalar masses. 

Is it possible to construct such a symmetry? A general form of an infinitesimal transformation acting on a 
set of fields $\phi^i(x), i = 1,. . ., m$ can be written as:

\begin{equation}
\label{eqtrans}
\delta \phi^i(x)=\epsilon^a (T_a)_j^i \phi^j(x)
\end{equation}

\noindent where $a=1,...,n$ with $n$ denoting the number of independent transformations, in other words, $n$ is the number of generators of the Lie algebra of the group we are considering. For $U(1)$  $n=1$, for $SU(2)$ $n=3$ etc. The $\epsilon$'s are infinitesimal parameters and $T_a$ is the matrix of the representation of the fields. Usually the $\epsilon$'s are taken to be $c$-numbers in which case the transformation (\ref{eqtrans}) mixes only fields with the same spin and obeying the same statistics. It is clear that, if we want to change the spin of the fields with a transformation (\ref{eqtrans}), the corresponding $\epsilon$'s must transform non-trivially under rotations. If they have non-zero integer spin they can mix scalars with vectors, or spin-1/2 with spin-3/2 fields. This was the case with the old attempts to construct a relativistic $SU(6)$ theory, with its well-known shortcomings. If, on the other hand, the $\epsilon$'s are anti-commuting parameters, they will mix fermions with bosons. If they have zero spin, the transformations (\ref{eqtrans}) will change the statistics of the fields without changing their spin, $i.e.$ they will turn a physical field into a ghost. This is the case with the B.R.S. transformations which describe the quantisation of non-Abelian gauge theories and give rise to the appearance of the Faddeev-Popov ghosts. Here we want to connect physical bosons with physical fermions, therefore the infinitesimal parameters must be anti-commuting spinors. We shall call such transformations ``supersymmetry transformations'' and we see that a given irreducible representation will contain both fermions and bosons. It is not a priori obvious that such supersymmetries can be implemented consistently, but in fact they can. In the following I shall give a very brief description of their properties as well as their possible applications.

\subsubsection{{\it Divertimento:} Grassmann algebras}

We said that the $\epsilon$'s of (\ref{eqtrans}) must be anti-commuting spinors. We shall present here a formal way to introduce such quantities.

A finite dimensional Grassmann algebra is defined as a formal power series of $n$ generators $x_1, x_2, ..., x_n$ which anti-commute among themselves.

\begin{equation}
\label{eqGrA}   
[x_i,x_j]_+ \equiv x_ix_j+x_jx_i=0
\end{equation}

The coefficients of the power series are complex numbers. The anti-commutation rule (\ref{eqGrA}) ensures that all $x$'s satisfy $x_i^2=0$, therefore, a power series contains at most $2^n$ terms. In other words, a Grassmann algebra can be viewed as a linear vector space of $2^n$ dimensions. Similarly, given a function $f(z), z\in {\IR}^n$ which admits a power series expansion around the origin, we can define through this power series the function $f(x_1, x_2, ..., x_n)$. Obviously, any such function is in fact a polynomial.

We shall need two operations on a Grassmann algebra: differentiation from the left and integration. Since any function is a polynomial, it is sufficient to define these operations on an arbitrary monomial. 

A left derivative $d/dx$ is defined by:

\begin{equation}
\label{eqderiv}
\frac{d}{dx_i}c=0 \hskip 0.5cm ;\hskip 0.5cm \frac{d}{dx_i}x_j={\delta}_{ij}-x_j\frac{d}{dx_i} \hskip 0.5cm ;\hskip 0.5cm \left[ \frac{d}{dx_i}, \frac{d}{dx_j}\right]_+=0
\end{equation}                            

\noindent where $c$ is a complex number.

The concept of integration we shall need corresponds to that of a definite integral from $-\infty$ to $+\infty$ in real numbers. The important property of the latter is translational invariance. We therefore define our integrals of functions on a Grassmann algebra by imposing invariance under translations. Again, it is enough to give the definitions for all monomials. For example, for an algebra with only one generator $x$, the most general function is of the form $f(x)=c_1+c_2x$ and translational invariance of the integral means:
\begin{equation}
\label{eqinteg1}
\int f(x)dx=\int f(x+a)dx
\end{equation}

\noindent and suggests the definitions:

\begin{equation}
\label{eqinteg2}
\int dx=0 \hskip 0.5cm ;\hskip 0.5cm \int xdx=1
\end{equation}

The generalisation to arbitrary $n$ is obvious. 
For a function $f(x_1,..., x_n)$ we find:

\begin{equation}
\label{eqfunct}
f(x_1,..., x_n)=c_0+c_ix_i+c_{ij}x_ix_j+...+c_{1...n}x_1...x_n
\end{equation}

\begin{equation}
\label{eqinteg3}
\int f(x_1, x_2, ..., x_n)dx_1...dx_n=c_{1...n}
\end{equation}

\noindent $i.e.$ the integral of a function is the coefficient of the highest term in the power series expansion. It is in this sense that we often say that integration in a Grassmann algebra corresponds, in fact, to differentiation. The familiar result that, after integration of a Gaussian form over fermion fields, the determinant of the corresponding operator appears in the numerator rather than the denominator, follows from the definition (\ref{eqinteg3}).

A final remark concerns an operation of complex conjugation. A given Grassmann algebra may admit an 
operation $(\ )^*$ (involution) with the properties:

\begin{equation}
\label{eqinv}
(f^*)^*=f\hskip 0.3cm ;\hskip 0.3cm (f_1f_2)^*=f_2^* f_1^* \hskip 0.3cm ;\hskip 0.3cm (f_1+f_2)^*= f_1^*+f_2^* \hskip 0.3cm ;\hskip 0.3cm (cf)^*=\bar{c}f^*
\end{equation}

\noindent where $\bar{c}$ is the complex conjugate of $c$. A function $f$ which satisfies $f^* = f$ is called ``real''. For an even number 
of generators, a basis that satisfies $(x_{n+k})^*=x_k \hskip 0.3cm k=1,...,n$  is called involutory.

\subsubsection{{\it Divertimento:} Majorana and Weyl spinors}

In particle physics we use most often Dirac spinors. They have four complex components and describe the degrees of freedom of a massive, spin-1/2 charged fermion, together with the corresponding anti-fermion. From the group-theory point of view, however, they are not the simplest spin-1/2 representations of the Lorentz group. They form reducible representations, which means that we can decompose them into simpler objects. In fact, we can do so in two ways: the first gives two, four-component, real spinors (Majorana spinors) and the second two, two-component, complex ones (Weyl spinors). They are both useful and we give here some formulae which allow to change from one representation to the other.

In the Majorana representation the $\gamma$-matrices have all real 
elements and satisfy 

\begin{equation}
\label{eqmaj1}
[\gamma_{\mu},\gamma_{\nu}]_+=2\eta_{\mu \nu} \hskip 0.5cm ;\hskip 0.5cm \eta_{00}=-1
\end{equation}

Furthermore, $\gamma_5^2=-1$ and $\gamma_5^T=-\gamma_5$. A Majorana spinor is real in this representation and $\bar{\psi}=\psi^T \gamma^0$. A useful relation is the 
following: Let $\psi_1$ and $\psi_2$ be two Majorana spinors and $\Gamma$ a four by four matrix. We define $\tilde{\Gamma}$ by $\tilde{\Gamma} =-\gamma_0\Gamma^T\gamma_0$. Then we can easily verify that

\begin{equation}
\label{eqmaj2}
\bar{\psi_1}\Gamma\psi_2=\bar{\psi_2}\tilde{\Gamma}\psi_1
\end{equation}

For the basic sixteen matrices $\tilde{\gamma_A} = \gamma_A$ for 1, $\gamma_5$ and $\gamma_5\gamma_{\mu}$, while $\tilde{\gamma_A} = -\gamma_A$ for $\gamma_{\mu}$ and $\gamma_{\mu}\gamma_{\nu}$.
The sixteen matrices $\gamma_A$ have squares equal to 1. For any $\gamma_A$ define $\gamma^A$ such that $\gamma_A\gamma^A$ = 1 (no 
summation). Then, the Fierz rearrangement formula can be obtained:

\begin{equation}
\label{eqmaj3}
(\bar{\psi_1}\psi_2)\psi_3=-\frac{1}{4}\sum_A(\bar{\psi_1}\gamma_A\psi_3)\gamma^A\psi_2
\end{equation}

We can work out all the formulae in these notes using the Majorana representation. However, it is often 
convenient to use spinors with a definite chirality. Starting from a Dirac, four component complex spinor, we introduce  the usual left and right projections $(1 \mp i\gamma_5)/2$. In the representation in which $\gamma_5$ is diagonal, we obtain 
  the Weyl two-component formalism. For a two-component spinor $\psi_{\alpha} \hskip 0.3cm \alpha=1,2$ we write:

\begin{equation}
\label{eqwey1}
(\psi_{\alpha})^*=\bar{\psi}_{\dot{\alpha}} \hskip 0.5cm ;\hskip 0.5cm (\psi_{\dot{\alpha}})^*=\bar{\psi}_{\alpha} \hskip 0.5cm ;\hskip 0.5cm \alpha ,\dot{\alpha}=1,2
\end{equation}

Dotted and undotted indices transform according to complex conjugate representations of $SL(2, C)$. In this 
two-dimensional space we raise and lower indices using the completely antisymmetric tensors $\epsilon^{\alpha \beta}$ and $\epsilon^{\dot{\alpha} \dot{\beta}}$  ($\epsilon^{12}=1$): $\psi^{\alpha}=\epsilon^{\alpha \beta}\psi_{\beta}$. Notice also that $\epsilon^{\alpha \beta}=-\epsilon_{\alpha \beta}$ and similarly for dotted indices. We also define the usual Pauli matrices. They connect the two chiralities and have one undotted and one dotted index:

\begin{equation}
\label{eqwey2}
(\sigma_{\mu})_{\alpha \dot{\beta}}=(\I1,\mbox{\boldmath $\sigma$})_{\alpha \dot{\beta}} \hskip 0.5cm ;\hskip 0.5cm ({\sigma'}_{\mu})^{\dot{\alpha} \beta}=(\I1,-\mbox{\boldmath $\sigma$})^{\dot{\alpha} \beta}
\end{equation}

In the Weyl 
representation we can choose the $\gamma$ matrices as:

\begin{equation}
\label{eqwey3}
\gamma_{\mu}=\left( \begin{array}{cc}
0 & i\sigma_{\mu} \\ i{\sigma'}_{\mu} & 0
\end{array}
\right) \hskip 0.5cm ;\hskip 0.5cm \gamma_5=\left( \begin{array}{cc}
i & 0 \\ 0 & -i
\end{array}
\right)
\end{equation}

In this representation a Majorana spinor has the form:

\begin{equation}
\label{eqwey4}
\psi=\left( \begin{array}{c}
\psi_{\alpha} \\ \bar{\psi}^{\dot{\alpha}}
\end{array}
\right)
\end{equation}

Formula (\ref{eqwey4}) allows us to change from one formalism to the other. Some useful identities in the Weyl 
formalism are:

\begin{equation}
\label{eqwey5}
\sigma_{\mu}{\sigma'}^{\mu}=-4 \hskip 0.5cm ;\hskip 0.5cm \sigma_{\lambda}\sigma_{\mu}{\sigma'}^{\lambda}=2\sigma_{\mu} \hskip 0.5cm ;\hskip 0.5cm [(\sigma_{\mu})_{\alpha \dot{\beta}}]^*=({\sigma'}_{\mu})_{\dot{\alpha} \beta}
\end{equation}

\begin{equation}
\label{eqwey6}
\theta^{\alpha}\psi_{\alpha}=\epsilon^{\alpha \beta}\theta_{\beta}\psi_{\alpha}=-\epsilon^{\beta \alpha}\theta_{\beta}\psi_{\alpha}=-\theta_{\beta}\psi^{\beta}=\psi^{\beta}\theta_{\beta}
\end{equation}

\begin{equation}
\label{eqwey7}
\theta^{\alpha}(\sigma_{\lambda})_{\alpha \dot{\beta}}\bar{\theta}^{\dot{\beta}}\theta^{\gamma}(\sigma_{\mu})_{\gamma \dot{\delta}}\bar{\theta}^{\dot{\delta}}=\frac{1}{2}\theta^{\alpha}\epsilon_{\alpha \beta}\theta^{\beta}\bar{\theta}^{\dot{\gamma}}\epsilon_{\dot{\gamma}\dot{\delta}}\bar{\theta}^{\dot{\delta}}\eta_{\lambda \mu}
\end{equation}

\begin{equation}
\label{eqwey8}
\theta^{\alpha}\psi_{\alpha}\theta^{\beta}\chi_{\beta}=-\frac{1}{2}\theta^{\alpha}\theta_{\alpha}\psi_{\beta}\chi_{\beta}
\end{equation}

\begin{equation}
\label{eqwey9}
\theta^{\alpha}\psi_{\alpha}\lambda^{\beta}\chi_{\beta}+\theta^{\alpha}\lambda_{\alpha}\psi^{\beta}\chi_{\beta}+\theta^{\alpha}\chi_{\alpha}\lambda^{\beta}\psi_{\beta}=0
\end{equation}

\subsubsection{The supersymmetry algebra}

We want to find symmetry transformations which generalise (\ref{eqtrans}) with anti-commuting $\epsilon$'s. Let $A_m$ $m = 1,. . . , D$ denote the generators of a Lie algebra and $Q_{\alpha}$ $\alpha = 1,. . . , d$ be the elements of a $d$-dimensional representation:

\begin{equation}
\label{eqLie}
[A_m ,A_n]=f^l_{mn}A_l \hskip 0.5cm ;\hskip 0.5cm [A_m ,Q_{\alpha}]=s^{\beta}_{m \alpha} Q_{\beta}
\end{equation}

A graded superalgebra is the algebraic scheme which consists of the generators $A_m$ and $Q_{\alpha}$ if we can find a 
set of constants $r^m_{\alpha \beta}$ such that

\begin{equation}
\label{eqgralg}
[Q_{\alpha},Q_{\beta}]_+=Q_{\alpha}Q_{\beta}+Q_{\beta}Q_{\alpha}=r^m_{\alpha \beta}A_m
\end{equation}

The constraint on the $r$'s is that they must satisfy the corresponding Jacobi identities for the set of equations (\ref{eqLie}) and (\ref{eqgralg})
 to be self-consistent.

There exist theorems which give a classification of graded superalgebras analogous to the Cartan 
classification of Lie algebras, but we shall not need them here. The only superalgebra we shall use is the one 
in which the Lie algebra is the Poincar\'e algebra with generators $P_{\mu}$ and $M_{\mu \nu}$, and the grading representation (\ref{eqLie})
 is given by a Majorana spinor $Q_{\alpha}$:

\begin{equation}
\label{eqsupsym1}
[P_{\mu},Q_{\alpha}]=0 \hskip 0.5cm ;\hskip 0.5cm [Q_{\alpha},M_{\mu \nu}]=i\gamma^{\mu \nu}_{\alpha \beta}Q_{\beta}
\end{equation}

\begin{equation}
\label{eqsupsym2}
[Q_{\alpha},\bar{Q}_{\beta}]_+=-2\gamma^{\mu}_{\alpha \beta}P_{\mu}
\end{equation}

\noindent in which $\gamma^{\mu \nu}=\frac{1}{4}[\gamma^{\mu},\gamma^{\nu}]$. The defining relations (\ref{eqsupsym1}) and (\ref{eqsupsym2}) admit the Lorentz group $SL(2, C)$ as an 
automorphism. The components of $Q$ are the generators of the special supersymmetry transformations. The second relation of (\ref{eqsupsym1}) shows only that they have spin 1/2. The first is more important because it shows that they are translationally invariant. We shall come back to this point later. The anti-commutation relation (\ref{eqsupsym2}) is the fundamental relation of the new symmetry.

An obvious generalisation of (\ref{eqsupsym1}) and (\ref{eqsupsym2}) consists of starting from the Poincar\'e algebra $\otimes$ a compact internal 
symmetry $G$ with generators $A_i$. If the $Q$'s belong to a certain representation of the internal symmetry, we 
write them as $Q^m_{\alpha}$ where the index ${\alpha}$ =1,..,4 labels
the components of the Majorana spinor and $m$ is the index of the
internal symmetry, $m$=1,2,...$N$. The algebra now takes the form:

\begin{equation}
\label{eqsupsym3}
[A_i,A_j]=f^k_{ij}A_k \hskip 0.3cm ;\hskip 0.3cm [P_{\mu},Q^m_{\alpha}]=0 \hskip 0.3cm ;\hskip 0.3cm [Q^m_{\alpha},M_{\mu \nu}]=i\gamma^{\mu \nu}_{\alpha \beta}Q^m_{\beta} 
\end{equation}

\begin{equation}
\label{eqsupsym4}
[A_i,Q^m_{\alpha}]=s^m_{in}Q^n_{\alpha} \hskip 0.3cm ;\hskip 0.3cm [Q^m_{\alpha},\bar{Q}^n_{\beta}]_+=-2\delta^{mn}\gamma^{\mu}_{\alpha \beta}P_{\mu}
\end{equation}

The algebra (\ref{eqsupsym3}) and (\ref{eqsupsym4}) admits $SL(2, C) \otimes G$ as a group of automorphisms.

As an exercise, we ask the reader to rewrite the supersymmetry algebra using Weyl spinors. For example, (\ref{eqsupsym2}) becomes:

\begin{equation}
\label{eqsupsym5}
[Q_{\alpha},Q_{\beta}]_+=[\bar{Q}_{\dot{\alpha}},\bar{Q}_{\dot{\beta}}]_+=0  \hskip 0.3cm ;\hskip 0.3cm [Q_{\alpha},\bar{Q}_{\dot{\beta}}]_+=2({\sigma}^{\mu})_{\alpha \dot{\beta}} P_{\mu}
\end{equation}

\subsubsection{Why this particular algebra; or all possible supersymmetries of the $S$ matrix}

The superalgebra (\ref{eqsupsym3}) and (\ref{eqsupsym4}) combines in a non-trivial way Poincar\'e invariance with an internal symmetry. There exists a theorem which states that, for ordinary algebras, such a combination cannot be a symmetry of a unitary $S$ matrix. We can now state, without proof, the generalisation of this theorem to include superalgebras, $i.e.$ algebras which close using both commutators and anti-commutators. The remarkable result is that (\ref{eqsupsym1}) and (\ref{eqsupsym2}), or (\ref{eqsupsym3}) and (\ref{eqsupsym4}), is essentially the only admissible one. The only possible extension is to replace the vanishing anti-commutator in (\ref{eqsupsym5}) with 

\begin{equation}
\label{eqsupsym6}
[Q_{\alpha}^m,Q_{\beta}^n]_+=\epsilon_{\alpha \beta}Z^{mn}
\end{equation}

\noindent where $Z^{mn}$ are a set of central charges, $i.e.$ operators which commute with every operator in the algebra. Out of the infinitely many ways we can grade the Poincar\'e algebra, only the one we introduced, in which we used a spin 1/2 operator, may be relevant for physics. 

\subsubsection{Representations in terms of one-particle states}
\label{sec1prep}

In order to extract the possible physical consequences of supersymmetry, we must construct the representations of the algebra in terms of one-particle states, $i.e.$ the one-particle ``supermultiplets''. We start by observing that the spinorial charges commute with $P_{\mu}$ and, therefore, they do not change the momentum of the one-particle state. Furthermore, the operator $P^2$ commutes with all the operators of the algebra, which implies that all the members of a supermultiplet will have the same mass. As it is the case with the Poincar\'e algebra, we can distinguish two cases: $P^2 \neq 0$ or $P^2 =0$. 

{\it (i) Massive case.} We can go to the rest frame in which the r.h.s. of (\ref{eqsupsym2}) or (\ref{eqsupsym5}) becomes a number. Let us first forget about a possible internal symmetry and consider the case $N$=1. Eq. (\ref{eqsupsym5}) gives:

\begin{equation}
\label{eq1part1}
[Q_{\alpha},\bar{Q}_{\dot{\beta}}]_+=2M\delta_{\alpha \dot{\beta}}
\end{equation}

\noindent where $P^2 = M^2$. Equation (\ref{eq1part1}) implies that the operators $Q/\sqrt{2M}$ and $\bar{Q}/\sqrt{2M}$ satisfy the anti-commutation relations for creation and annihilation operators of free fermions. Since the index $\alpha$ can take two values, 1 and 2, and $Q_1^2 =Q_2^2=0$, starting from any one-particle state with spin $S$ and projection $S_z$, we can build a four-dimensional Fock space with states:

\begin{equation}
\label{eq1part2}
|S,S_z;n_1,n_2>=Q_2^{n_2}Q_1^{n_1}|S,S_z> \hskip 0.5cm n_1,n_2=0,1
\end{equation}

We can define a parity operation under which the Majorana spinor $Q_{\alpha}$, $\alpha =1,...4$ transforms as:

\begin{equation}
\label{eqparity}
(Q_{\alpha})_P=(\gamma^0 Q)_{\alpha}
\end{equation}

Then, the spin-parity content of the representation (\ref{eq1part2}) is:

\begin{equation}
\label{eq1part3}
(S-1/2)^{\eta} \hskip 0.3cm ;\hskip 0.3cm S^{i\eta} \hskip 0.3cm ;\hskip 0.3cm 
S^{-i\eta} \hskip 0.3cm ;\hskip 0.3cm (S+1/2)^{-\eta}
\end{equation}

\noindent where $\eta = \pm i, \pm 1$ for $S$ integer or half-integer. Some examples:

\noindent $S$=0 : a scalar, a pseudoscalar, a spinor

\noindent $S$=1/2 : a scalar, a vector, two spinors, (or a pseudo-scalar, a pseudo-vector, two spinors).

\noindent $S$=1 : a vector, a pseudo-vector, a spinor, a 3/2 spinor.

\noindent $S$=3/2 : a vector, a tensor, two 3/2 spinors.

The generalisation to include internal symmetries is straightforward. The difference is that now we have more creation operators and the corresponding Fock space has $2^{2N}$ independent states, where $N$ is the number of spinorial charges. 

{\it (i) Massless case.} Here we choose the frame $P_{\mu}=(E,0,0,E)$. The relation (\ref{eqsupsym5}) yields:

\begin{equation}
\label{eq1part4}
[Q_{\alpha},\bar{Q}_{\dot{\beta}}]_+=2E(1-\sigma_z)=4E\delta_{\alpha 2}\delta_{\dot{\beta} 2} 
\end{equation}

Only $Q_2$ and $\bar{Q}_{\dot{2}}$ can be considered as creation and annihilation operators. Starting from a one-particle state with helicity $\pm \lambda$, we obtain the state with helicity $\pm (\lambda +1/2)$. Some interesting examples:

\noindent $\lambda$ = 1/2 : one spin 1/2 and one spin 1 (both massless)

\noindent $\lambda$ = 3/2 : one spin 3/2 and one spin 2 (both massless)

If we have more than one spinorial charge, $i.e.$ $N > 1$, we obtain $N$ creation and annihilation operators. A well-
established theoretical prejudice is that, if one excludes gravitation, there exist no elementary particles with spin higher 
than one. This prejudice is based on the great difficulties one encounters if one wants to write consistent field theories with 
high spin particles. The consequence of such a prejudice is that $N$ = 4 is the largest supersymmetry which may be 
interesting for particle physics without gravitation. The reason is that $N$ = 4 contains four creation operators and allows us to 
go from a helicity state $\lambda =-1$ to that of $\lambda =+1$. Any increase in the number of spinorial charges will automatically yield 
representations containing higher helicities. Finally, if we include gravitation, the same prejudice tells us that we must 
allow for elementary particles with helicities $|\lambda|\leq 2$. The previous counting argument now gives $N$ = 8 as the maximum 
allowed supersymmetry.

A concluding remark: All representations contain equal number of bosonic and fermionic states. All states in an 
irreducible representation have the same mass.

\subsubsection{{\it Divertimento:} Representations in terms of field operators. Superspace}

Our aim is to obtain field theoretical realizations of supersymmetry, therefore we look for representations in terms of local 
fields. Such representations were first obtained by trial and error, but the most elegant method is to use the concept of 
superspace.

We want to find a representation of the supersymmetry algebra (\ref{eqsupsym1}) and (\ref{eqsupsym2}) in terms of differential operators. The natural 
parameter space for these operators has eight dimensions, four (the usual Minkowski space) associated with the operators $P_{\mu}$
 and four with the spinors $Q$ and $\bar{Q}$. The last four coordinates, however, are not numbers but elements of a Grassmann 
algebra. An element of this space will be denoted by $z^M = (x, \theta , \bar{\theta})$. This eight dimensional space is called ``superspace''.

We can formally integrate the algebra  (\ref{eqsupsym1}) and (\ref{eqsupsym2}) in order to obtain a group. A ``finite'' group element can be defined by:

\begin{equation}
\label{eqsupgroup1}
G(x, \theta , \bar{\theta})=e^{i[\theta Q+\bar{\theta}\bar{Q} -x\cdot P]} 
\end{equation}

We can multiply two such group elements and, by Hausdorf's formula we obtain:

\begin{equation}
\label{eqsupgroup2}
G(y, \xi , \bar{\xi})G(x, \theta , \bar{\theta})=G(y+x-i\xi \sigma \bar{\theta}+i\theta \sigma \bar{\xi} , \xi+\theta, \bar{\xi}+\bar{\theta})
\end{equation}

This means that the group induces a motion of the parameter space into itself:

\begin{equation}
\label{eqsupgroup3}
G(y, \xi , \bar{\xi}): (x, \theta , \bar{\theta}) \rightarrow (y+x-i\xi \sigma \bar{\theta}+i\theta \sigma \bar{\xi} , \xi+\theta, \bar{\xi}+\bar{\theta})
\end{equation}

Equation  (\ref{eqsupgroup3}) shows that supersymmetry transformations act on superspace as generalised translations. The required 
representation of the algebra (\ref{eqsupsym1}) and (\ref{eqsupsym2}) in terms of differential operators can be read off (\ref{eqsupgroup3}):

\begin{eqnarray}
\label{eqsupgroup4}
Q_{\alpha} & = & \frac{\partial}{\partial \theta^{\alpha}}-i\sigma^{\mu}_{\alpha \dot{\alpha}}\bar{\theta}^{\dot{\alpha}}\frac{\partial}{\partial x^{\mu}} \nonumber \\
\bar{Q}_{\dot{\alpha}} & = & -\frac{\partial}{\partial \bar{\theta}^{\dot{\alpha}}}+i\sigma^{\mu}_{\alpha \dot{\alpha}}\theta^{\alpha}\frac{\partial}{\partial x^{\mu}} \\
P_{\mu} & = & i\frac{\partial}{\partial x^{\mu}} \nonumber 
\end{eqnarray}

A superfield is a function of the superspace element $z^M:
\phi. . . (x, \theta , \bar{\theta})$, where the dots stand for possible Lorentz tensor or spinor 
indices. A transformation of the group acts on it as a generalised translation:

\begin{equation}
\label{eqsupgroup5}
G(y, \xi , \bar{\xi})\phi (x, \theta , \bar{\theta})=\phi (y+x-i\xi \sigma \bar{\theta}+i\theta \sigma \bar{\xi} , \xi+\theta, \bar{\xi}+\bar{\theta})
\end{equation}

The interest of the superfields derives from the fact that, like any function of Grassmann variables, they are polynomials in $\theta$ and $\bar{\theta}$. 

\begin{equation}
\label{eqsupfield1}
\phi (x, \theta , \bar{\theta})=A(x)+\theta \psi (x)+\bar{\theta} \bar{\chi}(x)+...+\theta \theta \bar{\theta} \bar{\theta} R(x)
\end{equation}

\noindent where the coefficient functions $A(x)$ (scalar), $\psi (x)$ (spinor) etc. are ordinary fields, $i.e.$ a superfield is a finite multiplet of 
fields. Using the transformation property (\ref{eqsupgroup5}) and expanding both sides in powers of $\theta$ and $\bar{\theta}$ we obtain the transformation 
properties of the coefficient functions which, under supersymmetry transformations, transform among themselves. In this 
way we have obtained representations of supersymmetry in terms of a finite number of fields.

It is easy to see that the representation (\ref{eqsupfield1}) is a reducible one. We must be able to impose covariant restrictions on the 
superfield (\ref{eqsupfield1}) in order to decompose it into irreducible representations. For example, we can verify that the condition on $\phi$ to be a real function is a covariant one. The systematic way to obtain such covariant constraints is to realize that the 
algebra (\ref{eqsupsym1}) and (\ref{eqsupsym2}) contains the algebra of the $Q$'s or the $\bar{Q}$'s as subalgebras:

\begin{equation}
\label{eqsupfield2}
[Q,Q]_+=0 \hskip 0.5cm ; \hskip 0.5cm [\bar{Q},\bar{Q}]_+=0
\end{equation}

We can therefore study the motion of the group on the corresponding cosets. We can parametrise the group elements as:

\begin{equation}
\label{eqsupfield3}
G_1(x, \theta , \bar{\theta})=e^{i(\theta Q-x\cdot P)}e^{i\bar{\theta}\bar{Q}}\hskip 0.3cm ; \hskip 0.3cm G_2(x, \theta , \bar{\theta})=e^{i(\bar{\theta}\bar{Q}-x\cdot P)}e^{i\theta Q}
\end{equation}

The formulae (\ref{eqsupfield1}) and (\ref{eqsupfield3}) give three different but equivalent ways to represent the group elements and, 
therefore, lead to three different types of superfields. Of course, by Hausdorf's formula we can shift from one to another, 
the three representations are equivalent.

\begin{equation}
\label{eqsupfield4}
G(x, \theta , \bar{\theta})=G_1(x+i\theta \sigma \bar{\theta},\theta , \bar{\theta})=G_2(x-i\theta \sigma \bar{\theta},\theta , \bar{\theta}) 
\end{equation}

\noindent and similarly for the corresponding superfields:

\begin{equation}
\label{eqsupfield5}
\phi (x, \theta , \bar{\theta})=\phi_1 (x+i\theta \sigma \bar{\theta},\theta , \bar{\theta})=\phi_2(x-i\theta \sigma \bar{\theta},\theta , \bar{\theta}) 
\end{equation}

The generators $Q$ and $\bar{Q}$, which on a superfield of type $\phi$, were represented by the operators (\ref{eqsupgroup4}), when acting on a 
superfield of type $\phi_1$ are given by:

\begin{equation}
\label{eqsupfield6}
Q=\frac{\partial}{\partial \theta}-2i\sigma \bar{\theta} \frac{\partial}{\partial x}\hskip 0.3cm ; \hskip 0.3cm \bar{Q}= \frac{\partial}{\partial \bar{\theta}}
\end{equation}

\noindent and on a superfield of type $\phi_2$ by:

\begin{equation}
\label{eqsupfield7}
Q=\frac{\partial}{\partial \theta}\hskip 0.3cm ; \hskip 0.3cm \bar{Q}=-\frac{\partial}{\partial \bar{\theta}}+2i\theta \sigma \frac{\partial}{\partial x}
\end{equation}

We see that, the same way that we were able to impose a reality constraint which was invariant for a superfield of type $\phi$, 
we can, for example, impose on a superfield of type $\phi_1$
 to be independent of $\bar{\theta}$, or on $\phi_2$ to be independent of $\theta$. In other words $\partial /\partial \bar{\theta}$ is a covariant derivative when it acts on 
a superfield of type $\phi_1$ and $\partial /\partial \theta $  on $\phi_2$. By a shift (\ref{eqsupfield5}) we define covariant derivatives for any type of superfield:

\begin{equation}
\label{eqsupfield8}
{\cal D}_{\alpha}\phi=\left(\frac{\partial}{\partial \theta ^{\alpha}}+i\sigma _{\mu \alpha \dot{\alpha}} \bar{\theta}^{\dot{\alpha}}\frac{\partial}{\partial x^{\mu}}\right)\phi \hskip 0.3cm ; \hskip 0.3cm {\bar{\cal D}}\phi=\left(-\frac{\partial}{\partial \bar{\theta}}-i\theta \sigma \frac{\partial}{\partial x}\right)\phi
\end{equation}

\begin{equation}
\label{eqsupfield9}
{\cal D}\phi_1=\left(\frac{\partial}{\partial \theta}+2i\sigma \bar{\theta} \frac{\partial}{\partial x}\right)\phi_1 \hskip 0.3cm ; \hskip 0.3cm {\bar{\cal D}}\phi_1=-\frac{\partial}{\partial \bar{\theta}}\phi_1
\end{equation}

\begin{equation}
\label{eqsupfield10}
{\cal D}\phi_2=\frac{\partial}{\partial \theta}\phi_2 \hskip 0.3cm ; \hskip 0.3cm {\bar{\cal D}}\phi_2=\left(-\frac{\partial}{\partial \bar{\theta}}-2i\theta \sigma \frac{\partial}{\partial x}\right)\phi_2
\end{equation}

These differential operators anticommute with the infinitesimal supersymmetry transformations. They will be very useful 
when we decide to construct Lagrangian field theory models.

We can sharpen this analysis and obtain all linear irreducible representations, but for the purpose of these notes, we 
shall need only the three that we mentioned above:

\begin{equation}
\label{eqchsupfield}
\phi_1(x,\theta)=A(x)+\theta \psi (x)+\theta \theta F(x)
\end{equation}

\noindent similarly for $\phi_2 (x,\bar{\theta})$ and

\begin{eqnarray}
\label{eqvecsupfield}
\phi (x, \theta , \bar{\theta})=&C&+i\theta \chi -i\bar{\theta}\bar{\chi}+\frac{i}{2}\theta \theta (M+iN)- \frac{i}{2}\bar{\theta} \bar{\theta} (M-iN) \nonumber \\
 &-&\theta \sigma_{\mu}\bar{\theta}v^{\mu} \nonumber \\
 &+&i\theta \theta \bar{\theta}_{\dot{\alpha}}(\bar{\lambda}^{\dot{\alpha}}-\frac{i}{2}\sigma_{\mu \alpha}^{\dot{\alpha}}\partial^{\mu}\chi^{\alpha})-i\bar{\theta} \bar{\theta}\theta(\lambda+\frac{i}{2}\sigma_{\mu}\partial^{\mu}\bar{\chi})\nonumber \\
 &+& \frac{1}{2}\theta \theta \bar{\theta} \bar{\theta} (D+\frac{1}{2}\Box C)
\end{eqnarray}

\noindent where a reality condition on $\phi$ has been imposed. The supermultiplet (\ref{eqchsupfield}) contains a
chiral spinor $\psi(x)$ and it is called ``chiral multiplet'' while (\ref{eqvecsupfield}) contains a vector field $v_{\mu}(x)$
and is called ``vector''. The peculiar notation in the coefficients of the expansion (\ref{eqvecsupfield})
is used because of historical reasons and also, because it leads to simpler transformation
properties under infinitesimal transformations. For the fields of the chiral multiplet (\ref{eqchsupfield}), we get:

\begin{equation}
\label{eqtransf1}
\delta A=\xi \psi \hskip 0.2cm ; \hskip 0.2cm \delta \psi=2i\sigma_{\mu}\bar{\xi}\partial^{\mu}A+2\xi F \hskip 0.2cm ; \hskip 0.2cm \delta F=i\partial^{\mu}\psi \sigma_{\mu}\bar{\xi}
\end{equation}

\noindent where $\xi$ is the parameter of the infinitesimal supersymmetry transformation. 

Two remarks concerning these representations: First, if we compare with the results obtained in \ref{sec1prep}, we see that we have more 
fields than the physical one-particle states which are contained in an irreducible representation. Therefore some of the fields above must turn out to be auxiliary fields. Their presence is 
however necessary in order to ensure linear transformation properties. Second, we notice that the field $F$ in (\ref{eqtransf1}) transforms, under supersymmetry, with a total derivative. In fact, this property turns out to be always true with the last component in the expansion of a superfield, $i.e.$ $F$ for a chiral superfield (\ref{eqchsupfield}), $D$ for a vector (\ref{eqvecsupfield}), etc. We shall use this property soon.

Before closing this section we must establish a tensor calculus in order to be able to combine irreducible 
representations together. This is essential for the construction of Lagrangian models. Here again the superfield formalism 
simplifies our task enormously. All the necessary tensor calculus is contained in the trivial observation that the product of 
two superfields is again a superfield. For example, let $S_1(x, \theta)$ and $S_2(x, \theta)$ be two superfields of type $\phi_1(x, \theta)$. We form 
the product: $S_{12}(x, \theta)=S_1(x, \theta)S_2(x, \theta)$. Expanding both members in powers of $\theta$ and identifying the coefficients we obtain:

\begin{eqnarray}
\label{eqtens1}
A_{12}(x)&=&A_1(x)A_2(x) \nonumber \\
\psi_{12}(x)&=&\psi_1(x)A_2(x)+\psi_2(x)A_1(x)  \\
F_{12}(x)&=&F_1(x)A_2(x)+F_2(x)A_1(x)-\frac{1}{2}\psi_1(x)\psi_2(x) \nonumber
\end{eqnarray}

Similarly, we can multiply superfields upon which we have acted with the corresponding covariant derivatives, eqs. (\ref{eqsupfield8}) to (\ref{eqsupfield10}) . Two, or more, superfields of different types cannot get multiplied together. Rather one should transform them 
first into superfields of the same type by using the relations (\ref{eqsupfield5}) and then multiply them. An interesting example consists 
of a chiral superfield $S(x, \theta)$ of type $\phi_1$ and its hermitian conjugate $\bar{S}(x, \bar{\theta})$, which is of type $\phi_2$. Through (\ref{eqsupfield5}) we 
transform both of them into superfields of type $\phi$:

\begin{equation}
\label{eqtens2}
S(x, \theta)=\phi(x-i\theta \sigma \bar{\theta}, \theta) \hskip 0.3cm ; \hskip 0.3cm \bar{S}(x, \bar{\theta})=\bar{\phi}(x+i\theta \sigma \bar{\theta},\bar{\theta})
\end{equation}

We now multiply $\phi$ and $\bar{\phi}$ and expand in powers of $\theta$ and $\bar{\theta} $. We can verify that the last term in the expansion contains:

\begin{equation}
\label{eqtens3}
....+\theta \theta \bar{\theta} \bar{\theta}[A\Box A^*-\frac{i}{2}\psi \sigma^{\mu}\partial_{\mu}\bar{\psi}+FF^*]
\end{equation}

The first two terms are recognised as the kinetic energy terms of a complex spin-0 field and a two-component Weyl spinor. 
The last term has no derivative on the $F$-field, which shows that $F$ will be, in fact, an auxiliary field.

The final step is to use this tensor calculus and build Lagrangian field theories invariant under supersymmetry 
transformations. As we said earlier, supersymmetry transformations can be viewed as translations in superspace. Therefore 
the problem is similar to that of constructing translationally invariant field theories. We all know that the only Lagrangian 
density invariant under translations is a trivial constant. However, what is important is to have an invariant action which is 
obtained by integrating the Lagrangian density over all four-dimensional space-time. The same must be true for 
supersymmetry. Now, the Lagrangian density will be a polynomial  in some superfields and their covariant derivatives, $i.e.$ it will be a function of the superspace point $(x, \theta , \bar{\theta})$. The action will be given by an eight-dimensional integral over superspace:

\begin{equation}
\label{eqsupaction}
I=\int {\cal L}(x, \theta , \bar{\theta})d^4xd^2\theta d^2\bar{\theta}
\end{equation}

By construction, this integral is invariant under supersymmetry. This invariance can be verified by noticing that only the last term in the expansion of ${\cal L}$ in powers of $\theta$ and $\bar{\theta}$, the one proportional to $\theta \theta \bar{\theta} \bar{\theta}$, will survive the integration (see eq. (\ref{eqinteg3})). But, as we noticed above, the variation of the last term in the expansion of any superfield, such as an $F$ or a $D$ term, equals a total derivative. Therefore, their integrals over all space-time vanish. In fact, one can always work in superspace in terms of superfields and never write down the component fields explicitly. Feynman rules can be derived and all the results of the next sections can be obtained in a more direct way. We shall not use this powerful formalism here for the sake of physical transparency. In this way the next sections can be understood by the reader who has not studied this one very carefully.

\subsubsection{A simple field theory model}

We shall discuss here the simplest supersymmetric invariant field-theory model in four dimensions, that of a self-interacting chiral multiplet $S$. We introduced already the kinetic term, eq. (\ref{eqtens3}), and the mass term $S^2+$hermitian conjugate, eq. (\ref{eqtens1}). For the interaction we choose the term $S^3+$h.c. In terms of component fields, the complete Lagrangian, after integration over the Grassmann variables $\theta$ and $\bar{\theta}$, reads:

\begin{eqnarray}
\label{eqchmodel1} 
{\cal L}=&-&\frac{1}{2}[(\partial A)^2+(\partial B)^2 +i\bar{\psi}\gamma_{\mu}\partial^{\mu}\psi-F^2-G^2]+m[FA+GB-\frac{i}{2}\bar{\psi}\psi] \nonumber \\
 &+&g[F(A^2-B^2)+2GAB-i\bar{\psi}(A-\gamma_5 B)\psi]
\end{eqnarray}

\noindent where we changed the notations in two ways: (i) we separated the real and imaginary parts of the scalar fields $A \rightarrow 1/2(A+iB)$ and $F \rightarrow 1/2(F+iG)$ and (ii), we switched to the Majorana representation for the spinor $\psi$. $m$ is a common mass for all fields and $g$ a dimensionless coupling constant. As we mentioned earlier, $F$ and $G$ are auxiliary fields and can be eliminated using the equations of motion.

\begin{equation}
\label{eqchmodel2} 
F+mA+g(A^2-B^2)=0 \hskip 0.3cm ; \hskip 0.3cm G+mB+2gAB=0
\end{equation}

\noindent in which case the Lagrangian takes the form:

\begin{eqnarray}
\label{eqchmodel3}
{\cal L}=&-&\frac{1}{2}[(\partial A)^2+(\partial B)^2 +i\bar{\psi}\gamma_{\mu}\partial^{\mu}\psi+m^2(A^2+B^2)+im\bar{\psi}\psi]\nonumber \\
 &-&mgA(A^2+B^2)-ig\bar{\psi}(A-\gamma_5 B)\psi-\frac{1}{2}g^2(A^2+B^2)^2
\end{eqnarray}

It describes Yukawa, trilinear and quartic couplings among a Majorana spinor, a scalar and a pseudoscalar. 
The consequence of supersymmetry is that all fields have a common mass and all interactions are described 
in terms of a single coupling constant. Supersymmetry implies the conservation of a spin 3/2 current, which is:

\begin{equation}
\label{eqchmodel4}
J^{\mu}=\gamma^{\lambda}\partial_{\lambda}(A-\gamma_5 B)\gamma_{\mu}\psi -(F+\gamma_5 G)\gamma_{\mu}\psi
\end{equation}

The Lagrangian (\ref{eqchmodel1}) is the most general renormalisable supersymmetric invariant theory of one chiral 
multiplet. Strictly speaking one could add a term linear in the field $F$:

\begin{equation}
\label{eqchmodel5}
{\cal L} \rightarrow {\cal L}+\lambda F
\end{equation}

Such a term does not break supersymmetry because, as we said earlier, the variation of $F$ is a total 
derivative. However, it has no effect in the model because it can be eliminated by a shift in the field $A$. The 
renormalisation of this theory is straightforward. There are several supersymmetric invariant regularisation 
schemes and a conventional one is to introduce higher derivatives in the kinetic energy. The conservation of 
the current (\ref{eqchmodel4}) yields Ward identities among different Green functions and one can easily show that they 
can be enforced in the renormalised theory. Two important consequences follow from these Ward 
identities: (i) All vacuum-to-vacuum diagrams vanish, $i.e.$ no normal ordering is required for the 
Lagrangian (\ref{eqchmodel1}). This result is a consequence of exact supersymmetry and it is valid for every 
supersymmetric theory. (ii) The vacuum expectation values of all fields vanish. No counterterms linear in $A$ 
or $F$ are needed.

The surprising result, which could not be expected by supersymmetry considerations alone, is that, in 
this model, mass and coupling constant renormalisations are absent. All Green functions, to every order in 
perturbation theory, become finite if one introduces a single common wave function renormalisation 
counterterm. This counterterm is logarithmically divergent and is present to all orders. The theory is not 
superrenormalisable.

The absence of mass-renormalisation is a general feature in all theories which contain chiral multiplets. 
Since massive vector multiplets (or higher) lead to non-renormalisable interactions, it follows that no mass 
counterterms are needed in supersymmetric invariant theories. This non-renormalisation theorem makes supersymmetry so central in all attempts to go beyond the Standard Model. On the other hand, the vanishing of the 
coupling constant renormalisation counterterm is special to this particular model.

\subsection{Supersymmetry and gauge invariance}
\subsubsection{Field Theory Models}
\label{secsupgaugmod}

A combination of supersymmetry with gauge invariance is clearly necessary for the application of these 
ideas to the real world. We shall first examine an Abelian gauge theory and we shall construct the 
supersymmetric extension of quantum electrodynamics.

If $v_{\mu}$ is the photon field and $\phi_1$ and $\phi_2$ the real and imaginary parts of a charged field, an infinitesimal 
gauge transformation is given by:

\begin{equation}
\label{eqsupgauge1}
\delta v_{\mu}=\partial_{\mu}\Lambda \hskip 0.3cm ; \hskip 0.3cm \delta \phi_1=e\Lambda \phi_2 \hskip 0.3cm ; \hskip 0.3cm \delta \phi_2=-e\Lambda \phi_1
\end{equation}

\noindent where $\Lambda$ is a scalar function. In order to extend (\ref{eqsupgauge1}) to supersymmetry we must replace $v_{\mu}$ by a whole 
vector multiplet and let us assume that the matter fields are given in terms of a charged chiral multiplet. We 
expect, therefore, to describe simultaneously the interaction of photons with charged scalars, pseudoscalars 
and spinors. It is obvious that if we keep $\Lambda$ as a scalar function, the transformation (\ref{eqsupgauge1}) is not preserved by 
supersymmetry. The gauge transformation (\ref{eqsupgauge1}) must be generalised so that $\Lambda$ is replaced by a whole chiral 
multiplet. 

The construction of the Lagrangian using the superspace techniques we developed previously is straightforward, but we shall not present it here. We give instead directly the result with some comments:

First, what kind of particles we expect to find. (i) The photon. By supersymmetry, it must come together with a fermionic partner, often called ``photino''. It will be described by a neutral, massless, Majorana spinor. (ii) The electron. Its partners are two massive, charged, spin zero particles, a scalar and a pseudoscalar. The resulting Lagrangian is given by:

\begin{eqnarray}
\label{eqsupqed1}
{\cal L}=&-&\frac{1}{4}(\partial_{\mu}v_{\nu}-\partial_{\nu}v_{\mu})^2-\frac{i}{2}\bar{\lambda}\gamma^{\mu}\partial_{\mu}\lambda+\frac{1}{2}D^2 \nonumber \\
 &-&\frac{1}{2}[(\partial A_1)^2+(\partial A_2)^2+(\partial B_1)^2 +(\partial B_2)^2 -F_1^2-F_2^2-G_1^2-G_2^2 \nonumber \\
 &+&i\bar{\psi}_1\gamma_{\mu}\partial^{\mu}\psi_1 +i\bar{\psi}_2\gamma_{\mu}\partial^{\mu}\psi_2]\nonumber \\
 &+&m[F_1A_1+F_2A_2+G_1B_1+G_2B_2-\frac{i}{2}\bar{\psi}_1 \psi_1 -\frac{i}{2}\bar{\psi}_2 \psi_2 ]\nonumber \\
 &+&ev_{\mu}[i\bar{\psi}_1\gamma^{\mu}\psi_2-A_1\partial^{\mu}A_2+A_2\partial^{\mu}A_1-B_1\partial^{\mu}B_2+B_2\partial^{\mu}B_1]\nonumber \\
 &-&\frac{1}{2}e^2v_{\mu}v^{\mu}[A_1^2+A_2^2+B_1^2+B_2^2]+eD(A_1B_2-A_2B_1)\nonumber \\
 &-&ie\bar{\lambda}[(A_1+\gamma_5B_1)\psi_2-(A_2+\gamma_5B_2)\psi_1]
\end{eqnarray}

\noindent where $v_{\mu}$ is the photon field and $\lambda$ the field of the photino. The real Majorana spinors $\psi_1$ and $\psi_2$ can be combined together to form a complex Dirac spinor, the field of the electron. $A_1$, $A_2$, $B_1$ and $B_2$ are the real and imaginary parts of two complex, charged, spin-zero fields, a  scalar and a pseudoscalar. They are the supersymmetric partners of the electron, sometimes called ``selectrons''. As before, the fields $F_1$, $F_2$, $G_1$, $G_2$ and $D$ are auxiliary. The Lagrangian (\ref{eqsupqed1}) is  invariant under ordinary gauge transformations. In fact, if we eliminate the 
auxiliary fields, we obtain the usual interaction of a photon with a charged scalar, pseudoscalar and spinor 
field including the seagull term and the quartic term among the scalar fields. Supersymmetry has introduced 
only two new elements: (i) The coupling constant in front of the quartic self-interaction of the spin-zero fields is not arbitrary, but it is equal to $e^2/2$ and (ii) new 
terms, the ones of the last line in (\ref{eqsupqed1}), appeared which describe a Yukawa-type interaction between the Majorana 
spinor (the photino) and the spin 1/2 and zero fields of the matter multiplet. The strength 
of this new interaction is again equal to the electric charge $e$. Strictly speaking, (\ref{eqsupqed1}) is not invariant under supersymmetry transformations. However a supersymmetry 
transformation can be compensated by a gauge transformation, so all physical results will be supersymmetric.

The generalisation to non-Abelian Yang-Mills theories follows exactly the same lines. For the group $SU(m)$ we have $m^2-1$  gauge bosons $W_{\mu}$ which can be written as an $m\times m$ traceless matrix. Their supersymmetric partners, the ``gauginos'', are $m^2-1$ Majorana spinors which we write as another $m\times m$ traceless matrix $\lambda$. The resulting gauge invariant Lagrangian is:

\begin{equation}
\label{eqsupym1} 
{\cal L}=Tr[-\frac{1}{4}W_{\mu \nu}^2-\frac{i}{2}\bar{\lambda}\gamma^{\mu}{\cal D}_{\mu}\lambda]
\end{equation}

\noindent where 

\begin{equation}
\label{eqsupym2} 
W_{\mu \nu}=\partial_{\mu}W_{\nu}-\partial_{\nu}W_{\mu}+ig[W_{\mu},W_{\nu}]\hskip 0.3cm ; \hskip 0.3cm {\cal D}_{\mu}\lambda=\partial_{\mu}\lambda+ig[W_{\mu},\lambda]
\end{equation}

This  Lagrangian describes the gauge invariant interaction of $m^2-1$  massless Majorana fermions belonging to  the adjoint 
representation of $SU(m)$ with the gauge fields. The surprising result is that it is automatically supersymmetric, in the same sense as for (\ref{eqsupqed1}), $i.e.$ a supersymmetry transformation can be compensated by a gauge transformation. The corresponding spin 3/2 conserved current is:

\begin{equation}
\label{eqsupym3}
J^{\mu}=-\frac{1}{2}Tr(W_{\nu \rho}\gamma^{\nu}\gamma^{\rho}\gamma^{\mu}\lambda)
\end{equation}

We observe here the appearance of something like a connection between ``radiation'', $i.e.$ the gauge fields and ``matter'' multiplets. 
We shall come back to this point later.

The introduction of additional matter multiplets in the form of chiral superfields belonging to any 
desired representation of the group presents no difficulties. An interesting result is obtained if one studies 
the asymptotic properties of these theories. The one-loop $\beta$-function for an $SU(m)$ Yang-Mills 
supersymmetric theory with $n$ chiral multiplets belonging to the adjoint representation is:

\begin{equation}
\label{eqsupym4}
\beta (g)=\frac{m(n-3)}{16\pi^2}g^3
\end{equation}

\noindent which means that, for $n<3$, the theory is asymptotically free, although it contains scalar and pseudoscalar particles. This is because, in supersymmetric theories, the quartic couplings of the spin-zero fields are not independent but they are determined by the gauge coupling constant, see the remark after eq. (\ref{eqsupqed1}).

Before closing this section I want to mention a surprising and probably deep result. Until now 
we have been considering supersymmetric theories with only one spinorial generator. We explained already 
that the generalisation to $N$ such generators is possible. We also showed by an elementary counting  that $ N = 4$ 
 is 
the largest theory we can consider outside gravitation. The remarkable
convergence properties of supersymmetric theories, which led to the
non-renormalisation theorems we presented earlier, have now even more
surprising consequences. The most astonishing result is that the $\beta$-function of an $N = 4$ 
supersymmetric Yang-Mills theory based on any group  $SU(m)$ vanishes
to all orders, the effective coupling constant is scale independent
and does not run. For $N=2$ we have an intermediate result: the
$\beta$-function receives only one loop contributions. We shall see
that these properties open the way for a non-perturbative
understanding of these theories.

\subsubsection{The breaking of supersymmetry}
\label{secsupbr}
 
Fermions and bosons are not degenerate in nature, so supersymmetry, if it is at all relevant, must be broken. 
We shall first introduce the mechanisms for spontaneous breaking, which sounds more elegant and involves 
less arbitrariness and later we shall study the  consequences of explicit breaking.

The usual mechanism for spontaneous symmetry breaking is the introduction of some spin-zero field 
with negative square mass. This option is not available for supersymmetry because it would imply an 
imaginary mass for the corresponding fermion. It is this difficulty which makes supersymmetry hard to 
break. In fact, it would have been impossible to break it spontaneously if it were not for the peculiar 
property, which we mentioned already, namely the possibility of adding to the Lagrangian a term linear in 
the auxiliary fields without breaking supersymmetry explicitly. If we restrict ourselves to chiral and vector multiplets, 
in the notation we used previously, the auxiliary fields are $F$-fields, $G$-fields, or $D$-fields. The first are scalars, the other two pseudoscalars. Let $\phi$ denote, collectively, all other spin-zero fields. We shall assume that Lorentz invariance is not broken, consequently all other fields with non-zero spins have vanishing vacuum expectation values. Using the transformation properties, such as (\ref{eqtransf1}), we can easily show that spontaneous symmetry breaking occurs only when one, or more, of the auxiliary fields acquires a non-vanishing vacuum expectation value. 

The potential of the scalar fields in the tree approximation has the form:

\begin{eqnarray}  
\label{eqspbreak1}
V(\phi)=&-&\frac{1}{2}\left[\sum F_i^2+\sum G_i^2+\sum D_i^2\right] \nonumber \\
 &+&\left[\sum F_i F_i(\phi)+\sum G_i G_i(\phi)+\sum D_i D_i(\phi)\right]
\end{eqnarray}

\noindent where the functions $F_i(\phi)$, $G_i(\phi)$ and $D_i(\phi)$ are polynomials in the physical fields $\phi$ of degree not higher than second. The equations which eliminate the auxiliary fields are:

\begin{equation}
\label{eqspbreak2}
F_i= F_i(\phi)\hskip 0.3cm ; \hskip 0.3cm G_i=G_i(\phi)\hskip 0.3cm ; \hskip 0.3cm D_i=D_i(\phi)
\end{equation}

\noindent so the potential, in terms of the physical fields, reads:

\begin{equation}
\label{eqspbreak3}
V(\phi)=\frac{1}{2}\left[\sum F_i^2(\phi)+\sum G_i^2(\phi)+\sum D_i^2(\phi)\right]
\end{equation}

The important observation is that $V$ is non-negative and vanishes only for:

\begin{equation}
\label{eqspbreak4}
F_i(\phi)=0\hskip 0.3cm ; \hskip 0.3cm G_i(\phi)=0\hskip 0.3cm ; \hskip 0.3cm D_i(\phi)=0
\end{equation}

This positivity property of the potential is easy to understand: It follows from the anti-commutation relations of the supersymmetry algebra (\ref{eqsupsym5}). Taking the trace of the last of these relations we obtain that the Hamiltonian is equal to $2H=|Q_1|^2+|Q_2|^2$ and, therefore, non-negative.

When (\ref{eqspbreak4}) are satisfied (\ref{eqspbreak2}) show that all auxiliary fields have zero vacuum expectation values and, as we said above, supersymmetry is unbroken. In other words, a supersymmetric state, if it exists, it is always stable. It follows that the only way to break supersymmetry spontaneously, at least in the classical approximation, is to arrange so that the system of algebraic equations (\ref{eqspbreak4}) has no real solution. In such a case, at least one of the auxiliary fields will have a non-vanishing vacuum expectation value and supersymmetry will be broken. 

The simplest example which exhibits this phenomenon is given by the supersymmetric extention of quantum electrodynamics, eq. (\ref{eqsupqed1}), with the addition of a term linear in the auxiliary field 
$D$. 

\begin{equation}
\label{eqqedbr1}
{\cal L}\rightarrow {\cal L}+\xi D
\end{equation}

We repeat that this term does not break supersymmetry explicitly but, $D$ being pseudoscalar, it breaks parity 
explicitly, but softly. The system of eqs. (\ref{eqspbreak4}) reads:

\begin{eqnarray}  
\label{eqqedbr2}
 & &mA_1=0\hskip 0.3cm ; \hskip 0.3cm mA_2=0\hskip 0.3cm ; \hskip 0.3cm mB_1=0\hskip 0.3cm ; \hskip 0.3cm mB_2=0 \nonumber \\
 & &e(A_1B_2-A_2B_1)+\xi=0
\end{eqnarray}

It is clear that this system has no solution. We therefore expect supersymmetry to be spontaneously broken. 
Indeed, eliminating the auxiliary fields we find nondiagonal mass terms among the scalars and pseudoscalars. 
By diagonalisation we obtain the fields $\tilde{A_1}$, $\tilde{A_2}$, $\tilde{B_1}$ and $\tilde{B_2}$ with mass terms:

\begin{equation}
\label{eqqedbr3}
-\frac{1}{2}(m^2+\xi e)({\tilde{A_1}}^2+{\tilde{B_1}}^2)-\frac{1}{2}(m^2-\xi e)({\tilde{A_2}}^2+{\tilde{B_2}}^2)
\end{equation}

This mass spectrum shows clearly that we have obtained a spontaneous breaking of supersymmetry since the 
masses of the scalar and spinor members of the chiral multiplets are no longer the same. One can 
easily verify that the corresponding Goldstone particle is the massless spinor $\lambda$.

From this point, what follows depends on the sign of the square mass terms $m^2 \pm \xi e$. If they are both positive, the story ends here. If one of them is negative, this depends on the magnitude and sign of the parameter $\xi$, the corresponding scalar fields become Higgs fields for the $U(1)$ gauge symmetry and the photon becomes massive. Therefore, the introduction of the linear term $\xi D$ can trigger the spontaneous breaking of both supersymmetry and gauge symmetry. In this case the Goldstone spinor is a linear combination of $\lambda$ and the $\psi$'s.

This method can be applied to any gauge theory provided the algebra is not semisimple, 
otherwise a linear $D$ term cannot be added. There exists a second method for spontaneous 
supersymmetry breaking, which uses only chiral multiplets, but it needs at least three of them. The physics, in particular as regards the mass spectrum, remains the same.

As we saw in the previous examples and as we know from general theorems, spontaneous 
breaking of supersymmetry results in the appearance of a zero mass Goldstone spin
1/2 fermion. This is easy to understand. We know that the Goldstone particle of a spontaneously broken symmetry has the quantum numbers of the divergence of the corresponding conserved current. Since the conserved current of supersymmetry has spin equal to 3/2, the resulting Goldstone particle has spin equal to 1/2. Our first reaction was to rejoice with this discovery because we hoped to be
able to associate it with one of the neutrinos. Alas, appearances were deceptive! First, the neutrinos don't seem to have zero masses and, second, they don't even seem to be approximate Goldstone spinors. Indeed, there is
a low-energy theorem, known as ``Adler's zero'' satisfied by any Goldstone particle. 

Let $\eta(x)$ be the field associated with the Goldstone fermion. It has the same quantum 
numbers as the divergence of $J_{\mu}$, the conserved supersymmetry current. For any two physical states $|a>$ and $|b>$ we have:

\begin{equation}
\label{eqspbreak5}
k^{\mu}<a|J_{\mu}|b>=0
\end{equation}

 When $k \rightarrow 0$ the only intermediate state which survives 
is the one $\eta$ state due to the zero-mass propagator. It follows that the amplitude $M(a \rightarrow b+\eta)$ of 
the emission (or absorption) of a Goldstone fermion with momentum $k$ satisfies the low-energy 
theorem (except for possible pole terms):

\begin{equation}
\label{eqspbreak6}
\lim_{k_{\mu} \rightarrow 0}M(k)=0
\end{equation}

This is a very powerful prediction and can be checked by studying the end-point spectrum in 
nuclear $\beta$-decay. Unfortunately, it is a wrong one. Experiments show no such suppression, 
which means that the electron neutrino is not a Goldstone fermion.

The  question now is: If the Goldstone fermion 
is not one of the neutrinos, then where is it? There are two possible answers to this question which correspond to the two possible ways to implement supersymmetry, (i) as a global symmetry and (ii) as a local, or gauge, symmetry. In the first case the Goldstone fermion is a physical particle and, since it does not seem to appear in our experiments, we must make it ``invisible''. In the second it is absorbed in a super-Higgs mechanism. Let us look at each one of them:

Several mechanisms have been proposed to hide the zero mass ``Goldstino'', but the simplest is to endow it with a new, conserved quantum number and arrange so that  all other 
particles which share this number are heavy. Such a quantum number appears 
naturally in the framework of supersymmetric theories and it is present even in models in which 
the above motivation is absent. This number is called ``$R$-parity'' and one possible definition is: 

\begin{equation}
\label{eqRpar1}
(-)^R=(-)^{2S}(-)^{3(B-L)}
\end{equation}

\noindent where $S$ is the spin of the particle and $B$ and $L$ the baryon and lepton numbers, respectively. It is 
easy to check that eq. (\ref{eqRpar1}) gives $R = 0$ for all known particles, fermions as well as bosons, while 
it gives $R=\pm 1$ for their supersymmetric partners. The $R$-transformations act as phase 
transformations on spin-zero fields, phase or $\gamma_5$-transformations on spin-1/2 fields and leave 
vector fields invariant. They have a simple interpretation in superspace:
A point $(x,\theta ,\bar{\theta})$ transforms as:

\begin{equation}
\label{eqRpar2}
(x,\theta ,\bar{\theta}) \rightarrow (x,\theta e^{-i\alpha} ,\bar{\theta}e^{i\alpha})  
\end{equation}

A vector superfield is ``neutral'' under $R$, while a chiral one may be
multiplied by a phase

\begin{equation}
\label{eqRpar3}
V(x,\theta ,\bar{\theta}) \rightarrow V(x,\theta e^{-i\alpha} ,\bar{\theta}e^{i\alpha})\hskip 0.3cm ; \hskip 0.3cm S(x,\theta)\rightarrow e^{i\beta}S(x,\theta e^{-i\alpha})
\end{equation}

The above transformation properties allow us to find those of the component fields. The phases 
$\beta$ are adjusted for each chiral multiplet. Since $R$ is conserved, the $R$-particles are produced in 
pairs and the lightest one is stable. In a spontaneously broken global supersymmetry this is the 
Goldstino, which is massless.

An important question -before going into the details of any 
model- is the identity of the Goldstino; in particular, can it be identified, say to the photino? As 
we said earlier, the mechanism of spontaneous symmetry breaking, which is at the origin of the 
existence of the Goldstino, allows us to find some properties of the latter, independently of the 
details of a particular model. In a spontaneously broken theory the spin-3/2 conserved current is 
given by:

\begin{equation}
\label{eqgoldst1}
J_{\mu}(x)=d\gamma_{\mu}\gamma_5 \eta (x)+\hat{J_{\mu}}(x)
\end{equation}

\noindent where $d$ is a parameter with dimensions (mass)$^2$, $\eta (x)$ is the Goldstino field and $\hat{J_{\mu}}(x)$ is the usual 
part of the current which is at least bilinear in the fields. In other words, the field of a Goldstone 
particle can be identified with the linear piece in the current. The conservation of $J_{\mu}(x)$ gives:

\begin{equation}
\label{eqgoldst2}
d\gamma_5 \gamma_{\mu}\partial^{\mu}\eta = \partial^{\mu}\hat{J_{\mu}}
\end{equation}

This is the equation of motion of the Goldstino. In the absence of spontaneous breaking $d = 0$ and $\partial^{\mu}\hat{J_{\mu}}=0$. In fact, to lowest order, the contribution of a given multiplet to $\partial^{\mu}\hat{J_{\mu}}$ is 
proportional to the square mass-splitting $\Delta m^2$. Thus, the coupling constant of the Goldstino to a 
spin-0-spin-1/2 pair is given by:

\begin{equation}
\label{eqgoldst3}
f_{\eta}=\pm \frac{\Delta m^2}{d}
\end{equation}

\noindent where the sign depends on the chirality of the fermion. It follows that if the Goldstino were the 
photino, $f_{\eta} \propto e$ and the (mass)$^2$-splittings would have been proportional to the electric charge. For example, if $s_e$ and $t_e$ were the charged spin-zero partners of the electron, we 
would have:

\begin{equation}
\label{eqgoldst4}
m^2(s_e)+m^2(t_e)=2m^2(e)
\end{equation}

This relation  is clearly unacceptable. The conclusion is that the photon cannot be the bosonic 
partner of the Goldstino. With a similar argument we prove that the same is true for the $Z^0$ boson, 
the Higgs or any linear combination of them. This is a model-independent result. Not only can 
we not identify the Goldstino with the neutrino, but also we cannot pair it with any of the known 
neutral particles. Therefore, strictly speaking, there is no acceptable supersymmetric extension of 
the standard model with spontaneously broken global supersymmetry. The one that comes closest to it assumes an enlargement of the gauge group 
to $U(1) \otimes U(1) \otimes SU(2) \otimes SU(3)$ thus involving a new neutral gauge boson. We shall not study it 
in detail here but we shall rather extract those features which are model independent and are 
likely to be present in any supersymmetric theory.

The most attractive way to solve the problem of the Goldstino identity is to use a super-Higgs mechanism. In the normal Higgs phenomenon we have

$(m = 0,spin = 1 + m = 0, spin = 0) = (m\neq  0, spin = 1)$

\noindent In a super-Higgs mechanism we must get:

$(m=0,spin=3/2+m=0,spin=1/2) = (m\neq 0,spin=3/2)$

\noindent $i.e.$ we need to start with a gauge spin-3/2 field and we shall end up with a massive spin-3/2 
particle. A massless spin-3/2 field is the supersymmetric partner of the
graviton in all models which attempt to incorporate gravitation and is called
``gravitino''. So, this mechanism can be naturally applied in the framework of
supergravity theories, although the details are not easy to implement. At low
energies, when gravitational interactions are decoupled, the theory will look,
presumably, like a model with explicitly broken global supersymmetry. It will
contain many arbitrary parameters, usually  mass splitings and mixing
angles. In principle, they are calculable in terms of the initial supergravity
theory, but the relation is not always clear. Most often, they just
parametrise our ignorance of the underlying symmetry breaking mechanism. It is
this general framework which has been used in most phenomenological studies. 

As a final remark, I want to have another look at eq. (\ref{eqqedbr3}). As a result of supersymmetry 
breaking the masses of the chiral multiplet members are split, but we see that some pattern 
remains. The masses squared of the spin-zero fields are equally spaced above and below those of 
the fermions. In other words, we obtained a mass-formula of the form:

\begin{equation}
\label{eqmasform}	
\sum_J (-)^{2J}(2J+1)m_J^2=0
\end{equation}

\noindent where $m_J$ is the mass of the particle of spin $J$. It turns out that such a formula 
is valid in every spontaneously broken supersymmetric model and even in some explicitly 
broken ones. We used it already in (\ref{eqgoldst4}) in order to prove the impossibility of pairing together the photon and the Goldstino. It plays an important role in model building.

\subsubsection{Supersymmetry and the Standard Model}
\label{secsupsm}

Let us now try to apply these ideas to the real world. We want to build a supersymmetric model 
which describes the low-energy phenomenology. There may be several answers to this question 
but, to my knowledge, there is only one class of models which come close to being realistic. They were discovered by P. Fayet in the seventies.
They assume a superalgebra with only one spinorial generator, consequently all particles of a 
given supermultiplet must belong to the same representation of the gauge group. This is dictated by the requirement of parity violation. It is easy to see that in a supersymmetric model with two spinorial charges each supermultiplet will contain fermions with both right and left components and it is not clear how to break this right-left symmetry. In the following 
we shall try to keep the discussion as general as possible, so that our conclusions will be generally valid.

All models of global supersymmetry use three types of multiplets:

\noindent (i) Chiral multiplets. As we said already, they contain one Weyl (or Majorana) fermion and two scalars. Chiral multiplets are used to represent the matter (leptons and quarks) fields as well as the Higgs fields of the Standard Model.

\noindent (ii) Massless vector multiplets. They contain one vector and one Weyl (or Majorana) fermion, both in the adjoint representation of the gauge group. They are the obvious candidates to generalise the gauge bosons. 

\noindent (iii) Massive vector multiplets. They are the result of ordinary Higgs mechanism in the presence of supersymmetry. A massive vector multiplet is formed by a vector field, a Dirac spinor and a scalar. These degrees of freedom are the combination of those of a massless vector multiplet and a chiral multiplet. 

The physical degrees of freedom of the particles in the minimal Standard Model with one Higgs  are:

\noindent Bosonic degrees of freedom = 28

\noindent Fermionic degrees of freedom = 90 (or 96, if we include $\nu_R$'s)

It follows that a supersymmetric extension of the Standard Model will necessarily introduce new particles. We can even go one step further: In $N$=1 supersymmetry all the particles of a given supermultiplet must belong to the same representation of the gauge group. For the various particles of the standard model this yields:

\noindent (i) The gauge bosons are one colour octet (gluons), one $SU(2)$ triplet and one singlet ($W^{\pm}$, $Z^0$, $\gamma$). No known fermions have these quantum numbers.

\noindent (ii) The Higgs scalars transform as $SU(2)$ doublets but they receive a non-zero vacuum expectation value, consequently they cannot be the partners of leptons or quarks, otherwise we would have induced a spontaneous violation of lepton or baryon number. Furthermore, we must enlarge the Higgs sector by introducing a second complex chiral supermultiplet. This is necessary for several technical reasons which are related to the fact that, in supersymmetry, the Higgs scalars must have their own spin-1/2 partners. This in turn creates new problems like, for example, new triangle anomalies which must be cancelled. Furthermore, now the operation of complex conjugation on the scalars induces a helicity change of the corresponding spinors. Therefore, we cannot use the same Higgs doublet to give masses to both up and down quarks. Finally, with just one Higgs supermultiplet, we cannot give masses to the charged partners of the $W$'s. The net result is a richer spectrum of physical Higgs particles. Since we start with eight scalars (rather than four), we end up having five physical ones (rather than just one). They are the scalar partners of the massive vector bosons and three neutral ones.

The conclusion is that, in the standard model, supersymmetry associates known bosons with unknown fermions and known fermions with unknown bosons. We are far from obtaining a connection between the three independent worlds. For this reason this extension cannot be considered as a fundamental theory. Nevertheless, the phenomenological conclusions we shall derive are sufficiently general to be valid, unless otherwise stated, in every theory based on supersymmetry. 

We close this section with a table of the particle content in the supersymmetric standard 
model. Although the spectrum of these particles, as we shall see shortly, is model dependent, 
their very existence is a crucial test of the whole supersymmetry idea. We shall argue in the 
following sections that its experimental verification is within the reach of LHC.

\vskip 1cm

\begin{tabular}{|c|c|c|c c|}  \hline \hline
SPIN-1 & SPIN-1/2 &  
\multicolumn{3}{|c|}{SPIN-0} \\ \hline \hline
Gluons & Gluinos &
\multicolumn{3}{|c|}{no partner}  \\ \hline
Photon & Photino &
\multicolumn{3}{|c|}{no partner} \\ \hline
$W^{\pm}$ & 2 Dirac Winos & $w^{\pm}$ & H &  \\
\cline{1-3}
$Z^0$ & 2 Majorana Zinos & $z$ & i & b \\ \cline{1-3}
 & & standard & g & o \\
 & & $\phi^0$ & g & s \\
 & 1 Majorana Higgsino & & s & o \\
 & & pseudoscalar  & & n \\
 & & ${\phi^{0}}'$ & & s \\ \hline
 & Leptons &
\multicolumn{3}{|c|}{Spin-0 leptons} \\ \hline
 & Quarks &
\multicolumn{3}{|c|}{Spin-0 quarks} \\ \hline
\end{tabular}
\vskip 1cm
\noindent {\bf Table 1} The particle content of the supersymmetric Standard Model

\noindent The assignment is conventional. In any particular model the physical particles may be linear combinations of those appearing in the Table.
\vskip 1cm

\subsubsection{Supersymmetry and Grand Unified Theories}
\label{secsupGUT}

Supersymmetry is the only known scheme which allows, even in principle, a connexion between the Poincar\'e symmetry of space-time and internal symmetries. It provides a framework for the unification of the various worlds of gauge theories. As we shall see in the next section, it is a necessary ingredient in all attempts to construct a consistent theory of quantum gravity. I firmly believe that supersymmetry will turn out to be part of our world. To my mind, the only question is that of scale. How badly is supersymmetry broken? As we said earlier, supersymmetry predicts a rich spectroscopy of new particles whose existence 
is an important test of the theory. Such a test, however, is only meaningful if the masses of the 
new particles are also predicted, at least to within an order of magnitude. Let me remind you the 
situation when the charmed particles were predicted. The motivation was the need to suppress 
unwanted processes like strangeness changing neutral current transitions. Such a suppression 
was effective only if the charmed particles were not too heavy. No precise value could be given 
but the prediction was powerful enough to be testable. We have a similar situation with 
supersymmetry and grand unified theories. One of the reasons to study supersymmetry was the need to control the bad behaviour of 
elementary scalar fields. It is the gauge hierarchy problem which plagues all
known G.U.T.s. The problem has two aspects, one physical and one
technical. The physical aspect is to understand  the profound reason why
Nature creates the two largely separated mass scales. The technical aspect is
related to renormalisation. In the notation of section \ref{secGUT}, in order
for the model to be able to sustain a gauge hierarchy, we must impose a very
precise relation among the parameters of the potential. It is this relation
which is destroyed by renormalisation effects and has to be enforced
artificially order by order in perturbation theory. As we shall see later, supersymmetry may
provide the mechanism to answer  the physical problem, but it can
certainly solve the technical one. The key is
the non-renormalisation theorem we mentioned earlier. If supersymmetry is
exact, the mass parameters of the potential do not get renormalised. What
happens is that the infinities coming from fermion loops cancel against those
coming from boson loops. When supersymmetry is broken, spontaneously or
explicitly but softly, the cancellation is not exact, but the corrections are
finite and calculable. They are of order $\sim \Delta m^2$, the square mass
splitting in the supermultiplet. For the gauge hierarchy to remain, $\Delta m$
should not be much larger than the small mass scale, namely $m_W$. A badly
broken supersymmetry is not effective in protecting the small mass scale.  Hence an upper limit on the masses of
supersymmetric particles of the order of 1 TeV.

After these remarks on the gauge hierarchy problem, one can proceed in supersymmetrizing one's 
favourite G.U.T. model. The construction parallels that of the low-energy standard model with similar 
conclusions. Again, no known particle can be the superpartner of another known particle. Furthermore, 
assuming a spontaneous symmetry breaking, we can repeat the analysis which led us to conclude that $U(1) \otimes SU(2) \otimes SU(3)$ was too small. The corresponding conclusion here will be that $SU(5)$ is too small, since 
$SU(5)$ does not contain anything larger than the group of the standard model.

Finally, we can repeat the renormalisation group estimation of the grand unification scale and the 
proton life-time. We had found in section 2 that at low energies the effective coupling constants evolve 
following, approximately, the renormalisation group equations of $U(1)$, $SU(2)$ or $SU(3)$. The same remains 
true in a supersymmetric theory, but now the values of the $\beta$-functions are different. The number of Yang-Mills gauge bosons is the same as before. They are the ones which give rise to negative $\beta$-functions. On 
the other hand supersymmetric theories have a larger number of ``matter'' fields, spinors and scalars, which give positive contributions. The net result is a smaller, in absolute value, $\beta$-function and, therefore, a 
slower variation of the asymptotically free coupling constants. The agreement
now is remarkable (see Figure \ref{figunif}). The three curves appear to come together. Expressed in terms of a prediction for the value of the weak mixing angle $sin\theta_W$, this agreement is:

\begin{displaymath}
sin^2\theta_W \mathrm{(no\ SUSY)}\sim 0.214 \hskip 0.3cm ; \hskip 0.3cm sin^2\theta_W \mathrm{(SUSY)}\sim 0.232 
\end{displaymath}

\begin{equation}
\label{eqsintheta}
sin^2\theta_W \mathrm{(exp)}=0.23149\pm0.00017
\end{equation}

The resulting value for $M$ is $M \sim 10^{16}-10^{17}$ GeV.  If nothing else contributes to proton decay, it is beyond the reach of experiment. 
Fortunately, there are other contributions, which although of higher order, turn out to be 
dominant. They are due to the exchange of the fermionic partners of
the heavy bosons and their contribution is model-dependent. Not
surprisingly, in many models the result turns out to be of the order
of $10^{33}$ years.

\subsubsection{Dualities in supersymmetric gauge theories}
\label{secsupdua}

In a previous section we had introduced the idea of duality in gauge
theories. In its simplest form it interchanges electric and magnetic
quantities as well as  weak and strong coupling regimes. We presented the Montonen-Olive conjecture which postulates the actual identity of the two descriptions, at least for the simple Georgi-Glashow model. Strange as it may sound, we saw that this conjecture passed some simple tests. We can now address the questions: How far can we trust this conjecture? In the absence of any rigorous proof can we, at least, use it as a means to define the theory in the strong coupling region? Does it apply to all gauge theories and, if not, are there models for which it comes closer to the truth? Last but not least, how can we use it in order to extract physically interesting results? In this section we shall attempt to give a partial answer to some of these questions. 

Let me first notice that the identification $g \rightarrow 1/g$ cannot be exact everywhere for a generic gauge theory. The reason is that the effective value of the coupling constant depends on the scale and if such an identification can be enforced in one scale, it won't be true in another. However, we have seen in Section \ref{secsupgaugmod} that there is a class of gauge theories for which the running of the effective coupling constant is particularly simple: They are the gauge theories with extended $N=4$ or $N=2$ supersymmetries. For these theories the duality conjecture has given novel and interesting results. 

The $N=4$ supersymmetric Yang-Mills theory is the theory with the maximum allowed number of supersymmetries in four dimensions not including gravitation. It contains a single $N=4$ vector multiplet which belongs to the adjoint representation of whichever gauge group $G$ we have chosen. The particle content of such a multiplet consists of a vector, four fermions and six scalars, giving again an equal number of fermionic and bosonic degrees of freedom. Since all fields belong to the adjoint representation of $G$, we write them as traceless matrices. Notice that they are all massless, like the gauge bosons. We can write the Lagrangian density as:

\begin{eqnarray}
\label{eqN41}
{\mathcal L}_{N=4}= & -\frac{1}{4g^2}  Tr \left(   F_{\mu \nu}F^{\mu \nu}+\sum_{i=1}^{4}\bar{\chi}^i \gamma_{\mu}D^{\mu}\chi^i + \sum_{a=1}^{6}D^{\mu}\phi_a D_{\mu}\phi_a +\right. \nonumber \\  
 & \nonumber \\
  & \left. \sum_{a,b=1}^{6}[\phi_a, \phi_b][\phi_a, \phi_b]+...\right) +\frac{\theta}{32\pi^2}TrF_{\mu \nu}\tilde{F}^{\mu \nu}
\end{eqnarray}

\noindent where the dots stand for Yukawa terms among the fermions and the
scalars. This Lagrangian has a global $SU(4)$ symmetry which play the role of
the $R$-symmetry we introduced in Section \ref{secsupbr}. The vector field is
invariant under $SU(4)$, while the fermions and the scalars belong to the 4-
and 6-dimensional representations, respectively. The potential for the scalars
is given by the trace of the commutator square and vanishes only if $\phi$ is
represented by a diagonal matrix. This means that we shall find as many
independent ground states of the system as there are diagonal matrices in the
adjoint representation of $G$. This number is called the rank of $G$. For
$SU(2)$ it is equal to one, (only one Pauli matrix can be diagonalised), for
$SU(3)$ to two, etc. Let us choose $SU(2)$, for simplicity. A non-zero vacuum
expectation value of $\phi$ breaks $SU(2)$ spontaneously to $U(1)$. Two of the
vector bosons acquire a mass and are electrically charged with respect to
$U(1)$. The $N=4$ supersymmetry is not broken, so, together with their
supersymmetric partners, they form a massive multiplet. We can check that it
is a short BPS multiplet which has the same number of states as the massless
multiplet. 

This theory has magnetic monopoles which form also $N=4$ supermultiplets. The BPS mass formula can be written as $M^2=g^2v^2|n+\tau m|^2$ where $v$ is the vacuum expectation value of the Higgs and $n$ and $m$ are integers specifying the electric and magnetic charges. $\tau$ is defined in (\ref{eqcomplt1}). The duality transformations can be generated by $\tau \rightarrow -1/\tau$, which corresponds to interchange between weak and strong coupling regime, as well as $\tau \rightarrow \tau +1$, which is the periodicity property in $\theta$. Together they form the infinite group of discreet transformations $SL(2, \IZ)$.

The most important point is that here the Montonen-Olive conjecture has good chances of being true. At least, all obvious objections one could raise against it do not apply. Vector bosons and monopoles belong to truly identical multiplets and, most importantly, the coupling constant does not run. Therefore, duality can be used to define the theory non-perturbatively for any strength of the coupling. We can perform many checks of this property which go far beyond the simple ones we presented for the non-supersymmetric case. They include a highly non-trivial calculation of the spectrum of the allowed monopole and dyon states in the theory as well as the computation of other supersymmetric invariant quantities. Needless to say that all these checks have been successful.

Let us now turn to a gauge theory with $N=2$ supersymmetry. While the $N=4$ case can be considered as a field theorist's dream, the $N=2$ theory shares many essential features with the real world. The first has vanishing $\beta$-function, no divergences whatsoever and it is scale invariant. The second is asymptotically free, it has a, presumably, complicated dynamics, but the non-renormalisation theorems allow us to bring it under some kind of control. It can be written with any gauge group, but let us again study the simplest $SU(2)$ case. The Lagrangian looks similar to the one written in (\ref{eqN41}) except that the vector multiplet contains now one vector field, two Majorana fermions instead of four and one complex scalar, all triplets of the gauge group $SU(2)$. The $R$-symmetry is now a global $U(2)$ symmetry which rotates the two spinorial generators. A non-zero vacuum expectation value of the scalar field breaks spontaneously the gauge $SU(2)$ symmetry to $U(1)$. Two of the vector bosons become massive and the third one can be identified with the photon. Supersymmetry is not broken and so both, the massive and the massless bosons form full $N=2$ BPS multiplets. So far, apart from the number of fields, there is no difference between the two cases. 

The difference comes from the running coupling. Since the theory is asymptotically free, at high scales we can use perturbation. At low energies, however, we enter the strong coupling regime. Following a suggestion of K. Wilson, let us integrate all massive degrees of freedom and obtain an effective Lagrangian describing the low energy strong interaction of the massless modes. Notice that it is precisely the kind of exercise we would have liked to solve for Q.C.D. in order to obtain an effective theory of hadrons. Such a Wilsonian effective Lagrangian is not necessarily renormalisable, in the same sense that the Fermi theory was not. For Q.C.D. we do not know how to solve this problem. Here, however, supersymmetry comes to the rescue. Integrating the heavy degrees of freedom does not break supersymmetry, so we expect the effective Lagrangian to be $N=2$ supersymmetric. We can show that the most general, non-renormalisable Lagrangian of our massless multiplet depends only on a single holomorphic function of the scalar fields, often called ``the prepotential''. If we could determine this function we would have a complete description of all the low energy dynamics. The remarkable achievement of N. Seiberg and E. Witten was to solve this problem. They showed that the duality properties of the theory, combined with the holomorphicity of the prepotential and the knowledge of the spectrum in the weak coupling regime, completely determine the Wilsonian effective action. Furthermore, the non-renormalisation theorems ensure that this determination is exact. The proof is quite involved and it will not be presented here. Let me only emphasise that it is the first time that we obtain the complete solution of a dynamical problem, not for a toy model, but for a fully interacting four dimensional field theory. 

The natural question is how to extend this solution to the real world, which may look like Q.C.D. with broken $N=1$ supersymmetry and chiral quarks. Although some progress on a number of technical aspects of this programme has been made, it is fair to say that the solution is not yet in sight.

\subsection{Supergravity}

Supergravity is the theory of local supersymmetry, i.e. supersymmetry transformations whose infinitesimal 
parameters -which are anticommuting spinors- are also functions of the space-time point $x$. There are 
several reasons to go from global to local supersymmetry:

(i) We have learned in the last years that all fundamental symmetries in nature are local (or gauge) 
symmetries.

(ii) The supersymmetry algebra contains the translations. So local supersymmetry transformations imply local translations  and we know that invariance under local 
translations leads to general relativity which, at least at the classical level, gives a perfect description of the 
gravitational interactions.

(iii) As we already noticed, local supersymmetry provided the most attractive explanation for the absence 
of a physical Goldstino.

(iv) In the last section we saw that in a supersymmetric grand unified theory the unification scale 
approaches the Planck mass ($10^{19}$ GeV) at which gravitational interactions can
no more be neglected.

 The gauge fields of local supersymmetry can be easily deduced. Let us introduce an anticommuting 
spinor $\epsilon$ for every spinorial charge $Q$ and write the basic relation ( (\ref{eqsupsym4}) as a commutator:
	
\begin{equation}
\label{eqsugra1}
[\epsilon^m Q^m ,\bar{Q}^n \bar{\epsilon}^n]=2\delta^{mn}\epsilon^m \sigma_{\mu}\bar{\epsilon}^nP^{\mu} \hskip 1cm n,m=1,...,N
\end{equation}

\noindent where no summation over $m$ and $n$ is implied. In a local supersymmetry transformation $\epsilon$ becomes a function $\epsilon (x)$.
 Eq. (\ref{eqsugra1}) implies that the product of two supersymmetry transformations with parameters $\epsilon_1 (x)$ and $\epsilon_2 (x)$ is a local translation with parameter
		
\begin{equation}
\label{eqsugra2}
\alpha_{\nu}(x)=\epsilon_1 (x)\sigma_{\nu}\bar{\epsilon}_2 (x)
\end{equation}

On the other hand we know that going from a global symmetry with parameter $\theta$ to the corresponding local 
one with parameter $\theta (x)$, results in the introduction of a set of gauge fields which have the quantum 
numbers of $\partial_{\mu}\theta (x)$. If $\theta (x)$ is a scalar function, which is the case for internal symmetries, $\partial_{\mu}\theta (x)$ is a vector and so are the corresponding gauge fields (ex. gluons, $W^{\pm}$, $Z^0$, $\gamma$). If the parameter is itself a vector, like $\alpha_{\nu}(x)$ 
of translations, $\partial_{\mu}\alpha_{\nu}(x)$ is a two-index tensor and the associated gauge field has spin two. In 
supersymmetry the parameters $\epsilon^m (x)$ have spin one-half so the gauge fields will have spin three-half. We 
conclude that the gauge fields of local supersymmetry, otherwise called supergravity, are one spin-two field and N spin-three-half ones. To those we have to add the ordinary 
vector gauge fields of whichever internal symmetry we are considering.

\subsubsection{$N$=1 supergravity}
\label{secsugr1}

This is the simplest supergravity theory. As I shall explain in the next section, I do not consider it as the 
fundamental theory of particle physics, but I believe that it provides for a good basis for a 
phenomenological analysis. The gauge fields are the metric tensor $g_{\mu \nu}(x)$ which represents the graviton and 
a spin-three-half Majorana ``gravitino'' $\psi_{\mu}(x)$. We can start by writing the Lagrangian of ``pure'' 
supergravity, i.e. without any matter fields. The Lagrangian of general relativity can be written as:

\begin{equation}
\label{eqsugra3}
{\cal L_G}=-\frac{1}{2\kappa^2}\sqrt{-g}R=\frac{1}{2\kappa^2}eR
\end{equation}

\noindent where $g_{\mu \nu}(x)$ is the metric tensor and $g = \det g_{\mu \nu}(x)$. $R$ is the curvature constructed out of $g_{\mu \nu}(x)$ and its derivatives. 
We have also introduced the vierbein field $e_m$ in terms of which  $g_{\mu
  \nu}(x)$ is given as $g_{\mu \nu}(x)=e_{\mu}^m (x)e_{\nu}^n (x) \eta_{\mu \nu}$ with $\eta_{\mu \nu}$ the 
Minkowski space metric. It is well known that if one wants to study spinor fields in general relativity the 
vierbein, or tetrad, formalism is more convenient. $e$ equals $-\sqrt{-g}$ and $\kappa^2$ is the gravitational coupling constant. 
Eq. (\ref{eqsugra3}) is the Lagrangian of the gravitational field in empty space. We add to it the Rarita-Schwinger 
Lagrangian of a spin-three-half massless field in interaction with gravitation:

\begin{equation}
\label{eqsugra4}
{\cal L_{RS}}=-\frac{1}{2}\sqrt{-g}\epsilon^{\mu \nu \rho \sigma}\bar{\psi}_{\mu}\gamma_5 \gamma_{\nu}{\cal D}_{\rho}\psi_{\sigma}
\end{equation}

\noindent where ${\cal D}_{\rho}$ is the covariant derivative

\begin{equation}
\label{eqsugra5}
{\cal D}_{\rho}=\partial_{\rho}+\frac{1}{2}\omega_{\rho}^{mn}\gamma_{mn}\hskip 0.3cm ; \hskip 0.3cm \gamma_{mn}=\frac{1}{4}[\gamma_m , \gamma_n]
\end{equation}

\noindent and $\omega_{\rho}^{mn} (x)$ is the spin connection. Although $\omega_{\rho}^{mn}(x)$ can be treated as an independent field, its equation of motion 
expresses it in terms of the vierbein and its derivatives.

The remarkable result is that the sum of (\ref{eqsugra3}) and (\ref{eqsugra4})

\begin{equation}
\label{eqsugra6}
{\cal L}={\cal L_G}+{\cal L_{RS}}
\end{equation}

\noindent gives a theory invariant under local supersymmetry transformations with parameter $\epsilon (x)$:

\begin{eqnarray}
\label{eqsugra7}
 & & \delta  e^m_{\mu}=\frac{\kappa}{2}\bar{\epsilon}(x)\gamma^m \psi_{\mu} \nonumber \\
 & & \delta  \omega_{\mu}^{mn}=0 \\
 & & \delta \psi_{\mu}=\frac{1}{\kappa}{\cal D}_{\mu}\epsilon (x)=\frac{1}{\kappa}(\partial_{\mu}+\frac{1}{2}\omega_{\mu}^{mn}\gamma_{mn})\epsilon (x) \nonumber 
\end{eqnarray}

Two remarks are in order here: First the invariance of (\ref{eqsugra6}) reminds us of the similar result obtained in 
global supersymmetry, where we found that the sum of a Yang-Mills Lagrangian and that of a set of 
Majorana spinors belonging to the adjoint representation, was automatically
supersymmetric. This means that {\it all} gauge theories, both of space-time
or internal symmetries, admit a natural supersymmetric extension. This is one
of the reasons for which many theorists consider that supersymmetry should be
part of our world. The second remark is technical.  
The transformations (\ref{eqsugra7}) close an algebra only if one uses the equations of motion 
derived from (\ref{eqsugra6}). We can avoid this inconvenience by introducing a set of auxiliary fields. In fact, we 
have partly done so, because the spin connection is already an auxiliary field.

The next step is to couple the $N = 1$ supergravity fields with matter in the form of chiral or vector 
multiplets. The resulting Lagrangian is quite complicated and will not be given explicitly here. Let me only 
mention that, in the most general case, it involves two arbitrary functions. If I call $z$ the set of complex 
scalar fields, the two functions are: $G(z,z^*)$, a real function, invariant under whichever gauge group we have used and $f_{ij}(z)$, 
 an analytic function which transforms as a symmetric product of two adjoint representations of the 
gauge group.

One may wonder why we have obtained arbitrary functions of the fields, but we must remember that, 
in the absence of gravity, we impose to our theories the requirement of renormalisability which restricts the 
possible terms in a Lagrangian to monomials of low degree. In the presence of gravity, however, 
renormalisability is anyway lost, so no such restriction exists. In view of this, it is quite remarkable that 
only the two aforementioned functions occur.

As in ordinary gauge theories, the spontaneous breaking of local supersymmetry results in a super-Higgs mechanism. The gravitino, which is the massless gauge field of local supersymmetry, absorbs the 
massless Goldstino and becomes a massive spin three-half field. At ordinary energies we can take the limit 
of the Planck mass going to infinity. In this case gravitational interactions decouple and the spontaneously 
broken supergravity behaves like an explicitly but softly broken global supersymmetry. The details of the 
final theory, like particle spectra, depend on the initial choice of the functions $G$ and $f$, but the main features remain the same. We shall discuss them shortly.

Before closing this section let me mention a famous unsolved problem, for which supergravity offers a 
new line of approach. The Einstein Lagrangian (\ref{eqsugra3}) is not the most general one. We could add a constant $\Lambda$
 with dimension [mass$]^4$ and write:

\begin{equation}
\label{eqcosmcon}
{\cal L_G}=-\frac{1}{2\kappa^2}\sqrt{-g}(R+\Lambda)
\end{equation}

$\Lambda$ is called ``the cosmological constant'' and represents the energy density of empty space, but in the 
presence of the gravitational field this is no more an unphysical quantity which one can set equal to zero. In 
fact, any matter field gives an infinite contribution to
$\Lambda$. Experimentally, $\Lambda$ is very small, although the most recent
observations favour a non-vanishing value. Even with this small value, it gives
the major part of the energy content of the Universe.
If we have exact supersymmetry $\Lambda$ vanishes identically because the infinite vacuum energy of the bosons 
cancels that of the fermions. However, in a spontaneously broken global supersymmetry vacuum energy is
always positive, as we explained before and this yields a positive cosmological  constant. 
In a spontaneously broken supergravity this is no more true and one can arrange
to have $E_{vac} = 0$ and hence $\Lambda = 0$. In a realistic theory this must be the consequence of a certain symmetry and, 
indeed, such models have been constructed. I believe that ultimately this problem will be connected to the 
way one obtains $N = 1$ supergravity as an intermediate step between low-energy phenomenology and the 
fundamental theory, whichever this one may be.

\subsubsection{$N = 8$ supergravity}

Let me remind you that one of the arguments to introduce supersymmetry was the desire to obtain a 
connection among the three independent worlds of gauge theories, the worlds of radiation, matter and Higgs 
fields. None of the models presented so far achieved this goal. They all enlarged each world separately into 
a whole supermultiplet, but they did not put them together, with the exception of an association of some of the Higgs scalars with the massive gauge vector bosons. N = 8 supergravity is the only one which 
attempts a complete unification. It is the largest supersymmetry we can consider if we do not want to introduce 
states with spin higher than two. Following the method of \ref{sec1prep} we construct the irreducible 
representation of one-particle states which contains:

\begin{eqnarray}
\label{eqN81}
 & & 1\ \ spin-2\ \ graviton \nonumber \\
 & & 8\ \ spin-3/2\ \ Majorana\ \ gravitini	\nonumber \\
 & & 28\ \ spin-1\ \ vector\ \ bosons	\\
 & & 56\ \ spin-1/2\ \ Majorana\ \ fermions \nonumber \\
 & & 70\ \ spin-0\ \ scalars \nonumber 
\end{eqnarray}

The Lagrangian which involves all these fields and is invariant under eight 
local supersymmetry transformations was constructed by E. Cremmer and
B. Julia, who also uncovered its remarkable symmetry properties. Contrary to the $N$ = 1 
case, there is no known system of auxiliary fields. Since we have 28 vector bosons we expect the natural 
gauge symmetry to be $SO(8)$. This is bad news because $SO(8)$ does not contain $U(1) \otimes SU(2) \otimes SU(3)$ as 
subgroup. The remarkable property of the theory, which raised $N$ = 8 to the status of a candidate for a truly 
fundamental theory, is the fact that the final Lagrangian has unexpected symmetries: (i) A global non-compact $E_7$ symmetry and (ii) a gauge $SU(8)$ symmetry whose gauge bosons are not elementary fields. 
They are composites made out of the 70 scalars. $SU(8)$ is large enough to contain the symmetries of the 
standard model, but this implies that all known gauge fields (gluons, $W^{\pm}$, $Z^0$, $\gamma $) are in fact composite states. 
The elementary fields are only the members of the fundamental multiplet (\ref{eqN81}). None of the particles we 
know is among them, they should all be obtained as bound states.

$N$ = 8 supergravity promised to give us a truly unified theory of all interactions, including gravitation 
and a description of the world in terms of a single fundamental multiplet. The main question was whether it 
defined a consistent field theory. The hope was that the large number of supersymmetries would ensure a 
sufficient cancellation of the divergencies of perturbation theory so that to
make the theory finite. We have no clear answer to this
question. However, the very powerful techniques which have been developed for
performing difficult perturbation theory calculations, techniques which are
often inspired by string theory and are actually used in the QCD calculations
of the LHC experiments, give us hope that the answer will be known soon.

In some sense $N=8$ supergravity can be viewed as the end of a road. As we emphasised again and again in the 
course of there lectures the response of the physicists whenever faced with a new problem was to seek the 
solution in an increase of the symmetry. This quest for larger and larger symmetry led us to the standard 
model, to grand unified theories and then to supersymmetry, to supergravity and, finally, to the largest possible supergravity, that with $N$ = 8. 
In the traditional framework we are working, namely that of local quantum field theory, there exists no 
known larger symmetry scheme. The next step had to be a very radical one. The very concept of point particle, which 
had successfully passed all previous tests, will have to be abandoned. In the next section will shall study a theory of extended objects.

\subsection{The Minimal Supersymmetric Standard Model}
\label{secMSSM}

In Tablle 1 we had given the new particles that we expect to find in a supersymmetric extension of the Standard Model. Notice, in particular, a richer Higgs system. 
  Since the symmetry is broken, an important element is the breaking mechanism which determines the mass spectrum. Unfortunately, it is the least understood sector of the theory. We believe that it is a spontaneous breaking at a scale where the effective theory is $N$=1 supergravity. In this case the Goldstone fermion  is absorbed by the spin-3/2 gravitino. Every particular model will correspond to different choices of the functions $G$ and $f$ we encountered in Section \ref{secsugr1}. The ``correct'' choice will be eventually dictated by a more fundamental theory, like string theory. We shall come back to this point later. In any case,  at lower energies the theory looks like a model with explicitly, but softly, broken global supersymmetry.  It is this framework that has been used in most phenomenological studies so far. The important point is that supersymmetry brings many new particles but no new couplings. At energies lower than the scale of grand unification we still have the three gauge couplings of $U(1)$, $SU(2)$ and $SU(3)$. With no further assumptions we must introduce a set of new arbitrary parameters describing the masses and mixing angles of all new particles. Even with massless neutrinos, this is a very large number. Notice that already in the Standard Model the masses and mixing angles of quarks and leptons are arbitrary parameters to be determined by experiment. But we have seen in Section \ref{secdynGUT} that Grand Unification may reduce this number by providing relations among masses, like, for example, the equation (\ref{eqRG2}), which was the result of the $SU(5)$ relations (\ref{eqmasrel}). Something similar was applied to supersymmetry by S. Dimopoulos and H. Georgi and, independently, by N. Sakai. In the literature one finds many variations of this idea and the most economic one is called the Minimal Supersymmetric Standard Model (MSSM). The basic assumption is that at the grand unification scale the supersymmetry breaking parameters which determine the mass splittings in the supermultiplets are the simplest possible.  From this point one uses the renormalisation group equations to derive the spectrum at present energies and compare with experiment. Remember again that these relations involve only the known gauge coupling constants.

In the MSSM the spectrum of the supersymmetric partners of ordinary particles (quarks, leptons and gauge bosons) at the GUT scale is assumed to be determined by a minimum number of parameters: A common mass parameter $m_{1/2}$ for all gauginos, a corresponding one $m_0$ for all squarks and sleptons and a common tri-linear coupling among the various scalars, denoted by $A$. This choices are dictated mainly by simplicity. The absence of flavour-changing neutral interactions sets limits on the possible mass differences among squarks and sleptons of different families, but does not force them to be zero. 
The most interesting sector is the Higgs system. We need two doublets, as we explained in Section \ref{secsupsm}. At the phenomenological level this introduces some new parameters: First, there will be two vacuum expectation values to break $U(1) \otimes SU(2)$ which we shall call $v_1$ and $v_2$. They are taken by the neutral components of the two doublets, but one has weak isospin $I_z=+1/2$ and the other  $I_z=-1/2$. It follows that no $CP$ breaking is introduced because we can rotate the two phases independently and bring both $v$'s to real values. An important parameter for phenomenology is the ratio 

\begin{equation}
\label{eqtanb}
tan \beta = v_1/v_2
\end{equation}

A second parameter is a mixing term between the two Higgs fields. At the grand unification scale of $SU(5)$ the two doublets are promoted to two chiral supermultiplets belonging to 5 and $\bar{5}$ and the mixing term is written as $\mu H_1H_2$ with $\mu$ a new arbitrary constant. Various versions of the MSSM make different assumptions regarding $\mu$. In some superstring inspired models it is set equal to zero at the grand unification scale. In others it is left arbitrary and it is determined by the requirement that the Higgs system produces the correct electroweak symmetry breaking when extrapolated to lower energies using the renormalisation group equations. In all cases this last requirement severely restricts the parameters of the Higgs system. This is because in the limit of exact supersymmetry, one cannot break a gauge symmetry by choosing the mass square of some scalar field negative, since such a choice would imply imaginary mass for the companion fermion. It is only through the breaking of supersymmetry that such a possibility arises. On the other hand, one does not want the electroweak breaking to occur at the grand unification scale, so all corresponding square masses must be positive or zero at that scale. In extrapolating from $M_{GUT}$ down to present energies,  one requires the correct breaking at $m_W$, no breaking in between, as well as the maintaining of the perturbative nature of the theory everywhere. This means that all couplings should remain smaller than one and the effective potential bounded from below in the entire region. It turns out that all these requirements leave a relatively narrow window for the possible values of the Higgs parameters which we shall compare with experiment in the next section. Let me just mention here that they provide a crucial test of this scheme. 
The attractive point of this scenario is the introduction in a ``natural'' way
of the large separation between $M_{GUT}$ and  $m_W$. It is simply due to the
logarithmic running of the parameters. In practice the running of the
effective Higgs mass is dominated by the $t$-quark loop because of the
corresponding very large Yukawa coupling. The typical renormalisation group
equations give a relation of the form:

\begin{equation}
\label{eqRGMm}
m_W \sim M_{GUT}~e^{-\frac{1}{\alpha_t}} \hskip 0.3cm ; \hskip 0.3cm \alpha_t = \frac{\lambda_t^2}{4\pi}
\end{equation}

\subsection{Supersymmetry and Experiment}
\label{secsupsymexp}

As we said earlier, there  is still no concrete evidence for
supersymmetry in particle physics. This in spite of the fact that
supersymmetry makes well defined predictions which can be put to
experimental test. Some of these predictions are very general and test
the entire scheme, others depend on the particular model one
considers. There exists only one, admittedly indirect, evidence: As we
said in \ref{secdynGUT}, the renormalisation group equations for the
evolution of the three coupling constants of the Standard Model do not
satisfy the requirements of grand unification (Figure \ref{figunif}). We saw in
Section \ref{secsupGUT} that the situation changes if we include supersymmetry.  The agreement between MSSM and experiment is impressive. However, it is fair to say that this agreement establishes a connexion between the idea of supersymmetry and that of grand unification, but it does not prove either of them. 

It is natural for theorists to attempt to interpret any, real or
hypothetical, departure from the Standard Model predictions as
evidence for supersymmetry. The famous Brookhaven result for the muon
anomalous magnetic moment caused for a while such an excitement.  I am afraid
it is too early to say whether a disagreement exists and even earlier to
speculate on its possible significance and I can only regret the interruption
of the experimental programme.

Strictly speaking, the only general prediction  of supersymmetry is the existence of the particles of Table 1.  Furthermore, if supersymmetry is meant to protect the electroweak scale from all higher scales, the masses of all these particles cannot exceed 1 TeV. This already puts supersymmetry in the range of LHC.
 A simple, hand-waving argument shows that ordinary particles are expected to be lighter than their supersymmetric partners. The reason is that the former take their masses solely through the Higgs mechanism while the latter through both the supersymmetry breaking and the Higgs mechanisms. Similar theoretical 
arguments almost always predict squarks and gluinos heavier than sleptons and other gauginos. The reason is that in most 
models the masses are set equal at the grand unification scale and the differences are due to the 
strong interactions of squarks and gluinos. For the same reason the masses of sneutrinos are 
predicted to be of the same order as those of the corresponding charged
 sleptons. 

Some simple relations follow from these general assumptions of the
 Minimal Model. The gauginos must obey the renormalisation group equations of
 the gauge couplings. 

\begin{equation}
\label{gauginomasses}
m_i(\mu)=m \frac{\alpha_i(\mu)}{\alpha_i(M)}
\end{equation}
where $m$ is the common mass at the GUT scale $M$ and $\mu$ is the low
scale. This gives a ``prediction'' for the three gauginos of $U(1)$, $SU(2)$
and $SU(3)$ at present energies:

\begin{equation}
\label{gauginomasses1}
m_1~~:~~m_2~~:~~m_3~~=~~1~~:~~2~~:~~7
\end{equation}

This picture may be slightly complicated because of mixings with higgsinos. Notice also that in the MSSM $R$-parity is conserved, therefore all new particles are produced in pairs and the lightest among them is stable. It is usually denoted by $LSP$ (Lightest Supersymmetric Particle). In almost all models it is identified with a linear combination of the neutral gauginos and Higgsinos. In this case its interactions are comparable to those of the neutrinos and it leaves no trace in the detector. Since all new particles will eventually end up giving $LSP$'s, a precise determination of missing transverse momentum is an essential  handle in the search of 
supersymmetric particles. Furthermore, the $LSP$ offers an excellent candidate for cold dark matter, a necessary ingredient in cosmological models. For all these reasons, 
 $m_{LSP}$ is a very important phenomenological parameter, although no precise predictions for its value can be given. The cosmological arguments mentioned above give a rather loose bound $m_{LSP} < O(200)$ GeV. I remind you that the Goldstino, 
which, if it exists, is massless, is absent from theories derived from supergravity. 

Let us now briefly discuss some results on masses and decay properties. In the absence of 
any concrete experimental evidence I can only quote limits and expected signatures. The mass 
spectrum is very model dependent but some general features can be extracted.
In the MSSM the analysis is made as a function of the parameters we introduced in the previous Section and the result should be given as a multi-dimensional plot. I shall try here to summarise the most important points:

(i) The Higgs system. It is probably the most sensitive test of the MSSM in
the sense that the predictions are very close to present experimental limits
with little room left. Five scalars are predicted, a pair of charged ones and
three neutrals. The requirements of the correct symmetry breaking we explained
in the last Section allow for rather narrow windows in the mass of the
lightest neutral, the analogue of the Standard Model Higgs. At the tree level
these predictions give a limit for $m_{\phi}$ of the order of the $Z$-mass,
already excluded by experiment. Radiative corrections, especially the
$t$-quark loops, raise this limit considerably. In most models we find
$m_{\phi}\leq $ 130 GeV. The present LEP limit is 114 GeV with a tantalising
possible signal at 115 GeV. Never before the limits of an accelerator were
more painfully felt. The overall
agreement of the Standard Model with experiment  implies the limits on the
Standard Model Higgs mass shown in Figure \ref{limhiggs}. In the
plot  the $\chi^2$ of the fit raises sharply when $m_{H}$ exceeds
200 GeV. With or without supersymmetry, the Tevatron and/or the LHC will solve
the puzzle of the Higgs sector. 

(ii) Scalar partners of quarks and leptons (squarks, sleptons). There exists one such partner for every left- or right-handed quark and lepton. The breaking of $U(1) \otimes SU(2)$ causes mixings among the partners of opposite chirality fermions, so the final mass spectrum is the result of several diagonalisations. For this reason squarks do not necessarily follow the mass hierarchy of their quark partners. In particular, the scalar partners of the $t$-quark may turn out to be lighter than the others.  Squarks are produced in hadron collisions either in pairs or in association with gluinos (R must be 
conserved). Their decay modes are of the form $\tilde{q}\rightarrow q+LSP$
(quark + $LSP$) or, if phase space permits, $\tilde{q}\rightarrow q+
\tilde{g}$ (quark + gluino). The signature is missing $P_T$ plus
jets. Sleptons behave similarly and give $\tilde{l}\rightarrow l+LSP$. The
signal is again missing energy and momentum. 

The most direct limits on the masses of the 
charged ones come from LEP. With small variations they are of the order of 100 GeV depending slightly on the values of the other parameters of the MSSM, provided the mass difference between the sparticle and $LSP$ is not too small. The  Tevatron
 results suggest that squarks may be at least as heavy as 250-300 GeV, but the limits depend on the other parameters of the MSSM, such as the gluino masses. Notice also that if one of the squarks is much lighter than the others the Tevatron gives no limits for its mass. 

(iii) Gluinos. They may decay into a gluon and an $LSP$ or into a quark-antiquark pair and an $LSP$. The Tevatron gives limits for the gluino masses similar to those for squarks.

(iv) Gauginos and Higgsinos. They mix among themselves and must be analysed together. The charged ones are the supersymmetric partners of $W^{\pm}$ and $H^{\pm}$ and are described by a $2\times 2$ mass matrix. LEP gives a limit of 90-100 GeV for the mass of the lighter of the two. Among the neutral ones, the partners of $W^3$, $B$ and the two $CP$-even neutral Higgses mix in a $4 \times 4$ matrix. The lightest of them is assumed to be the $LSP$. 

The picture that emerges is that supersymmetric particles may be spread all over from 50 GeV to 1 
TeV. The mass ratios we presented depend all on the minimal hypothesis which
was made only for convenience and has no solid theoretical base. However, if
sparticles are discovered and their masses and mixing angles measured, we can
easily go back and compute the symmetry breaking pattern at GUT energies. This
in turn will give us a hint on the breaking mechanism which, as we explained,
is probably related to the fundamental way gravity is unified with the other
interactions. Looking for supersymmetric particles will be an important part of experimental search in the years to come. I hope that it will be both exciting and rewarding and, in any case, we shall know soon whether supersymmetry is a fundamental symmetry of particle forces.

\section{Beyond local quantum field theory}
\subsection{Introduction}

The concept of point particle has been challenged several times in the past and people have often tried 
to write theories of extended objects. However, it was only very recently that the motivation for such a 
radical step appeared to be compelling. As we explained in the last section this was due to the apparent 
failure of all attempts to write a consistent quantum field theory of gravity. 
Strings are the simplest 
extended objects. Although theories of higher dimensional objects have been studied (membranes, etc) and are incorporated in most models to-day, stings remain always a fundamental ingredient. 

For somebody who is used to the technology of local quantum field theory, the obvious 
generalisation to a theory of quantised strings would be to consider a theory in which the fields, instead 
of being operator-valued functions (in fact, distributions) of the space-time point $x$, they would be 
operator-valued functionals of the string function. We could call such a theory ``quantum-field theory of 
strings''. Nobody has succeeded in writing such a theory and it is not clear whether it can be written 
using the available mathematical tools. Today we start having some feeling of what such a theory may be. Let me note in passing that one of the most attractive features of 
string theory to the eyes of many theoretical physicists, is precisely the fact that it meets with the most 
advanced research in modern mathematics.

The approach which has been followed in string theory corresponds to a first quantised theory and 
it is the generalisation of particle mechanics. The classical mechanics of a freely moving relativistic 
string can be obtained by extremising the invariant area of its trajectory which is a two-dimensional 
world-surface whose points are parametrised by $X_{\mu}(\sigma ,\tau)$. The index $\mu$ runs from 0 to $d-1$, where $d$ is 
the dimensionality of the embedding space in which the string is moving and $\sigma$ and $\tau$ are the coordinates on the surface. On the other hand  $X_{\mu}$ can be 
viewed as an ordinary field in a two-dimensional space-time. It is this equivalence between string 
theory and quantum field theory in 1 + 1 dimensions that allowed to make progress in the theory of 
quantised strings. In this picture the particles are the excitation
modes of the vibrating string. String theory has been a central theme of theoretical high energy physics in the last twenty years and I am sure all of you have heard about it probably more than you ever wanted to know, so I will not go into any details. There exist several introductory texts in the literature. Let me just list some important results. I shall try to put the emphasis on the most recent ones, but the choice  reflects my incomplete understanding of the subject. 

Like every theory of extended objects, string theory contains a fundamental length $l_s$, the length of the string. Its size is certainly smaller than anything we have measured so far, which, including evidence from radiative corrections, is of the order of $10^{-17}$cm, or, equivalently, one inverse TeV. Since string theory includes gravity, the natural value of $l_s$ is the Planck length $l_P$, which is $10^{-33}$cm, sixteen orders of magnitude smaller. For years it was assumed that $l_s \sim l_P$ and this doomed string theory to be outside the reach of any conceivable accelerator. It was only recently that we understood that there is no compelling logical connexion between the two lengths and $l_s$ can, in fact, be anywhere. 

\begin{figure}[!th]
\epsffile{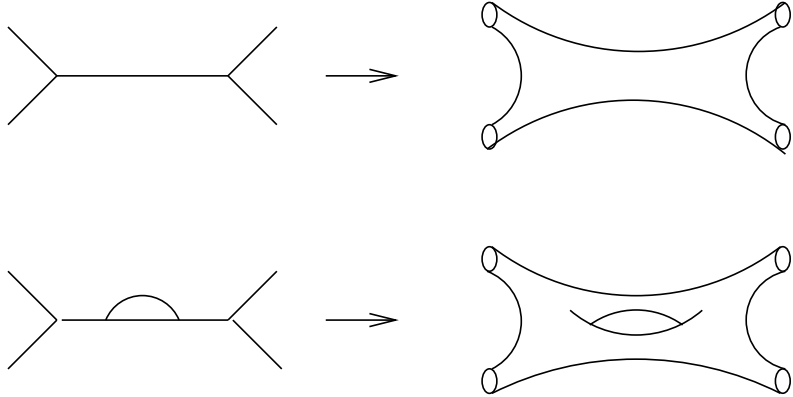}
\caption {Diagrams showing a scattering process for particles and strings. In the figure we have used closed strings, but it is clear that, even if we start with open strings, we shall generate closed ones in the loops.} \label{figstr}
\end{figure}

According to string theory, at distances of the order of $l_s$ the
geometry of space-time changes. For example, Feynman diagrams become
extended, as in Figure \ref{figstr}. Interactions  no more occur at a point but
extend to a finite region of space-time. Through this mechanism the
ultraviolet singularities, which used to plague field theories, get
smoothed out. The theory is finite at all distances. A second feature
we can see in the figure is that the perturbation expansion, which
counted the number of loops in a Feynman diagram, counts now the
number of holes in a closed surface, $i.e.$ we have traded
combinatorics for topology. 

Here are some of the main results obtained from string theory:

(i) Quantum string theory contains quantum gravity. At the classical
limit it gives the General Theory of Relativity. 

This result is not easy to explain. We saw that string theory can be
formulated as a non-linear quantum field theory in a two-dimensional
space-time. The connexion with four-dimensional gravity is not
obvious. At the technical level, this connexion comes from the
conformal invariance of the two-dimensional theory which, as an
algebra, has an infinite number of generators. We can show that, when
gauged fixed, this symmetry contains the four-dimensional
reparametrisation invariance of general relativity. A simpler way to
arrive at the same conclusion is to notice that quantum string theory
contains in the spectrum of excitations a massless spin two
particle and we know that general relativity is the only consistent way
to describe its interactions.

(ii) Quantum string theory is finite at all distances.

(iii) A very interesting result is that quantum string theory requires
supersymmetry for its mathematical consistency. In fact, only
supersymmetric string theories, also called super-string theories,
satisfy the finiteness condition stated in (ii).

(iv) The string moves in an ambient space. Classically this space can
have any number of dimensions greater than two (1+1). One of the most
remarkable results, known already from the early days of string
theory, is that, at the quantum level, this number can take only one
value. For the superstring it is 10=9+1. The technical reason is that
the symmetries which are necessary to ensure the mathematical
consistency of the theory break down and become anomalous at the
quantum level. The coefficient of the anomaly is a function of $d$,
the dimensionality of the ambient space and it vanishes only for
$d=10$. It follows that our world must contain six hidden space-like
dimensions. Two questions arise immediately: First, is this picture
consistent with the underlying dynamics of the string and, second,
which are the phenomenological consequences. A great deal of this
lecture will try to address these questions. So, let me postpone this
discussion for the moment.

(v) There exist five distinct superstring theories. This again is a
surprising result. We are used in field theory to have an infinity of
possible theories. For example, we can write a Yang-Mills theory for
any Lie group $G$. The novel feature of string theory has to do again with various symmetries which become anomalous at the quantum level. The great discovery of M. Green and J. Schwarz in the early eighties, which triggered the explosion of superstring theory, was the proof that, for very particular theories, the anomalies cancel and mathematical consistency is recovered. The complete list of the consistent supersting theories is: Type I, Type IIA, Type IIB, $E_8\times E_8$ heterotic, and $SO(32)$ heterotic. They differ by containing, or not, open strings, by their symmetry group, by their supersymmetry content, as well as by the particular way they use to combine together fermionic and bosonic degrees of freedom.

(vi) A few years ago I would have ended the discussion here. But in
the late nineties took place what people call ``the second superstring
revolution'' which changed radically our perception of string
theory. It has several interrelated aspects but the most important
ones are the application of duality ideas in string theory and the
realization that strings generate naturally objects of higher
dimensionalities, like membranes. 

\begin{figure}
\epsffile{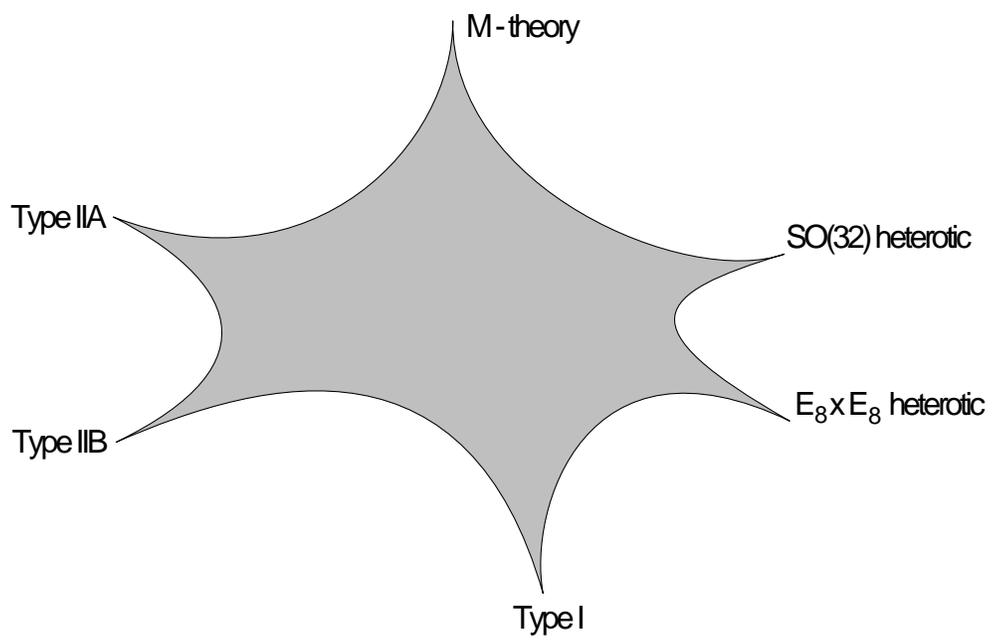}
\caption {Polchinski's diagram for the space of string vacua. In each
cusp the corresponding string model admits a weak coupling
expansion. Nothing is known for the M theory cusp.} \label{figstrdual}
\end{figure}

The consequences of duality are best illustrated by the Figure \ref{figstrdual}. It is an artist view of the idea that all five superstring theories are different manifestations of a single fundamental theory, which, in the absence of any better name, we call ``M theory''. This big theory depends on a few parameters, like coupling constants. By varying the values of the parameters we can move continuously from one string theory to the other. The sharp peaks illustrate the fact that, each time, at most one theory is in a weakly coupled regime. For most values of the parameters, the centre of the figure, all coupling constants are large and we have no weak coupling expansion at all. We know preciously little on this mysterious M theory. We do not know its fundamental equations, not even the dynamical  variables in terms of which they should be written. It may provide the ``field theory of strings'' which people have sought for many years. We believe that it lives in eleven space-time dimensions and its low energy effective theory is the $N=8$ supergravity theory of Cremmer and Julia. When one spatial dimension is compactified on a circle of radius $R$, M theory yields the Type IIA string theory. The effective coupling constant is proportional to $R$, so, at weak coupling, the theory is effectively ten dimensional. Strong $\leftrightarrow$ weak coupling duality is here equivalent to a duality $R \leftrightarrow 1/R$, or M theory $\leftrightarrow$ Type IIA string theory.  Alternatively, we can compactify the eleventh dimension on an interval with appropriate boundary conditions at the end points. This compactification gives the heterotic string. Again, the weak coupling regime is obtained when the interval is small. This is an interesting case, first analysed by P. Horava and E. Witten. The geometrical picture consists of two ten dimensional spaces separated by the interval in the eleventh dimension. They are called ``end-of-the-world branes''. It turns out that mathematical consistency requires the introduction of new degrees of freedom living on these boundary worlds. They can move in the ten dimensions, so from the string theory point of view, they are ordinary degrees of freedom, but they cannot extend in the eleventh dimension, so they are special from the point of the M theory. This picture will be used again later. 

Type I string theory is special because it contains both open and closed strings. Looking at the diagrams of Figure \ref{figstr}, we see that we can have a consistent theory with only closed strings but, if we start with open strings, unitarity will force us to introduce closed strings at higher orders. The presence of open strings brings a new element, as was first realised by J. Polchinski. Open strings have end points which, normally, can move everywhere. However, they can also get stuck on membranes, or other higher dimensional objects. Strings have a tension, so they cannot get stressed at arbitrary lengths. This means that their degrees of freedom are confined to move on the membrane, or very close to it. This brings a new physical picture: The degrees of freedom of closed strings propagate everywhere in the ambient space. The spin two graviton belongs to them. On the other hand the degrees of freedom associated to open strings may be confined in a subspace. Standard Model particles may behave this way. We shall come back to this point shortly. 

This remark brought in the picture objects of any dimensionality. They are
called ``Dp-branes''. ``D'' stands for Dirichlet and denotes the kind of
boundary conditions we impose on them. ``p'' is the dimensionality of the
object. Thus D0 are particles, D1 strings, D2 membranes, etc. We can look at
D-branes in two complementary ways: As manifolds on which open strings end,
or, as solitonic solutions of the string theory equations. This second approach
reminds us of the situation in field theory where also we have solutions
describing extended objects such as magnetic monopoles, flux tubes, domain
walls, etc. This also brings the idea of duality in which the ``elementary''
fields and these extended objects may exchange roles. 

A ten, or eleven, dimensional space may contain all sorts of D-branes. Some may lay on top of each other, some may stay a certain distance apart, some may intersect. The study of the stability properties of such configurations is a very difficult dynamical question, only very partially understood. This difficulty not withstanding, such configurations may describe interesting physical situations. For example, let us consider $n$ identical branes, each one having three spatial dimensions, stuck together. The open strings that start and end on them will have zero tension and they will give rise to massless particles, confined on these branes. Since each string can start and end on any of these branes, we can represent them with $n \times n$ matrices. It is easy to show that the spin one excitations will generate a $U(n)$ gauge group on the branes. If now we move $n_1 < n$ branes a distance $l$ away, leaving the other $n-n_1$ behind, we see that the strings which are stretched at length $l$ will have a tension and the excitations coming from them will be massive. In other words, we have a description of the spontaneous breaking $SU(n) \rightarrow SU(n-n_1) \otimes SU(n_1) \otimes U(1)$. There has been a lot of activity recently in such a brane engineering of the Standard Model. 

Before closing this introductory section let me mention a remarkable and, probably not yet fully understood, result. It is again an equivalence, in a sense I shall try to explain,  between a string theory formulated in a ten dimensional curved space and a gauge field theory in four dimensional flat space. We have reasons to believe that such an equivalence may be quite general, but the example which is analysed in some detail is due to a conjecture of J. M. Maldacena. The string theory is of the Type IIB and the ten dimensional space is $AdS_5 \times S^5$, which stands for the product of five dimensional Anti de Sitter space and a five sphere. This space is a solution of the equations of motion of the low energy effective theory of Type IIB. A few words about Anti de Sitter space: It is a maximally symmetric solution of the Einstein equations  with negative cosmological constant. It is called Anti de Sitter because the original de Sitter solution corresponds to positive cosmological constant. In order to avoid any misunderstanding, let me remind you that the recent supernovae data indicate that the cosmological constant in our space may have a small, non-zero, positive value. Coming back to $AdS_5$, we can choose a coordinate system to bring the five dimensional metric to the form:

\begin{equation} 
\label{eqAdS}
ds^2=dr^2+e^{2r}(\eta_{\mu \nu}dx^{\mu}dx^{\nu})
\end{equation}

\noindent where $\mu$ and $\nu$ run from 0 to 3, $\eta_{\mu \nu}$ is the four
dimensional Mincowski metric and the fifth coordinate $r$ is radial. The
important point which we can see from (\ref{eqAdS}), is that, for $r
\rightarrow \infty$,  Anti de Sitter space has a boundary which is Minkowski
space. It is precisely in this space that the gauge theory lives. This theory
is our old $N=4$ super Yang-Mills theory with an $SU(m)$ gauge group. For one
thing, the counting of symmetries is correct: The five dimensional Anti de
Sitter group has fifteen generators (same number as $O(6)$) and we have seen
that $N=4$ super Yang-Mills theory has vanishing $\beta$-function and it is
known to be conformally invariant. Furthermore this ``$AdS/CFT$'' equivalence
has been checked by several calculations in the level of the spectrum of
states in the two theories, that of correlation functions, etc. 

This equivalence exemplifies two  important and far-reaching theoretical ideas. The first, formulated long ago by G. 't Hooft, is known as holography. 't Hooft's motivation was the information paradox connected with the presence of an horizon around a black hole. He conjectured that the underlying theory must be such that the degrees of freedom sitting at the boundary surface around the black hole must exactly match those in the interior. The second is an old idea, according to which a field theory may become simple at the limit when the number of fields goes to infinity. For a gauge theory this limit was first studied also by 't Hooft. We know to-day that an $SU(m)$ Yang-Mills theory in $d$ space-time dimensions in a suitable $m \rightarrow \infty$ limit, becomes a classical theory in $d+2$ dimensions in which the commutator in the Yang-Mills interaction term is replaced by a classical Poisson bracket. The additional two dimensions describe a surface. Taking this limit we can again check Maldacena's conjecture. 
       
\subsection{String theory and Physics}

So far we have introduced in a qualitative way some of the concepts used in string theory. Let me now move into whatever physical results we can obtain, or hope to obtain. I see two fields in which string theory should play a central role: astrophysics and cosmology on the one hand and particle physics on the other. In this School I shall only talk about the second, but let me mention that probably the result which comes closer to real physics that string theory can claim, is the first and only exact computation of the Bekenstein-Hawking entropy of a black hole. The first such calculation is due to A. Strominger and C. Vafa in a simple model and was extended to more realistic cases by K. Sfetsos and K. Skenderis. Although their black hole is still not the one that may exist in the centre of our galaxy, it is, nevertheless a remarkable result. 

Coming to particle physics, the first question we must ask is the way to go to a four dimensional space-time.    
In order to introduce the
ideas, let me present the simple example of Th. Kaluza and O. Klein. 

Back in the early years of general relativity, the only known
fundamental interactions were electromagnetism and gravitation. They
were both gauge theories, the first described by the $U(1)$ Maxwell
theory and the second by the reparametrisation invariant Einstein's
General Theory of Relativity. They are both geometric theories, but
while gravity's geometry is the one of our space-time, that of
electromagnetism refers to an internal $U(1)$ space. Kaluza's idea of
unification was a very simple, but ingenious one. He assumed that only
space-time geometry has a physical meaning and, therefore, he tried to
promote the internal space into a real one. This naturally leads to
the study of  gravity in a space-time with five dimensions. The
dynamical variables are given by the five dimensional metric tensor
$g_{MN}$, where the indices $M,N$ go from 0 to 4 and $g$ satisfies the
five dimensional Einstein's equations. For empty space the action is simply:

\begin{equation}
\label{eqKK11}
S=-\frac{1}{4{\kappa}^2}\int d^5x \sqrt{-g}R
\end{equation}

\noindent where $g$ is the determinant of $g_{MN}$ and $R$ the five dimensional scalar curvature. It is clear that the equations derived from (\ref{eqKK11}) admit the five dimensional Minkowski  space $M_5$ as a solution, but also a space $M_4 \times S^1$, $i.e.$ a four dimensional
Mincowski space and a circle, since both correspond to zero curvature. In this case the action can be written as:

\begin{equation}
\label{eqKK12}
S=-\frac{1}{4{\kappa}^2}\int d^4x \int_{0}^{2\pi R} dx^4 \sqrt{-g}R
\end{equation}

This choice implies  that all fields must
be periodic functions in the fifth coordinate $x^4$: $x^4 \rightarrow
x^4+2 \pi R$. In the general case we look for solutions of the form:

\begin{equation}
\label{eqKK1}
ds^2=g_{MN}dx^M dx^N=g_{\mu
\nu}dx^{\mu}dx^{\nu}+g_{44}(dx^4+A_{\mu}dx^{\mu})^2
\end{equation}

\noindent where $\mu$ and $\nu$ run from 0 to 3. So far (\ref{eqKK1})
is very general and we can always parametrise the metric tensor this
way. Let us first assume that no function, such as the metric
components, or whatever field variables we may have, depends on the
coordinate $x^4$. We still allow for reparametrisations of the form
$x^{\mu} \rightarrow {x'} ^{\mu}(x^{\nu})$ and ${x'}^4 =
x^4+\lambda (x^{\mu})$. It is easy to check that the part of the metric
that we have called $A_{\mu}$ transforms as:

\begin{equation}
\label{eqKK2}
{A'}_{\mu} = A_{\mu}-\partial _{\mu}\lambda
\end{equation}

\noindent $i.e.$ $A_{\mu}$ transforms as a $U(1)$ gauge field. The
reparametrisations in the fifth dimension look like gauge
transformations when restricted in the four non compact coordinates. 

Let us now relax our assumption of $x^4$ independence and impose
instead on all fields periodicity with period $2\pi R$. The previous
case corresponds to the limit $R \rightarrow 0$. Let us take the
example of a massless scalar field $\phi (x^M)$ in five dimensions and choose, for
simplicity, $g_{44}=1$. The momentum in the fifth dimension take only
discreet values $p_4=n/R$. We can expand $\phi$ in modes:

\begin{equation}
\label{eqKK3}
\phi (x^M)=\sum _{n=-\infty}^{\infty}\phi _n (x^M)e^{\frac{inx^4}{R}}
\end{equation}

\noindent in other words, we obtain an infinity of fields in four dimensions.
The five dimensional Klein-Gordon equation yields, for the
four-dimensional fields $\phi _n$:

\begin{equation}
\label{eqKK4}
\partial _M \partial ^M \phi (x^M)=0 \hskip 0.3cm \Rightarrow \hskip
0.3cm \partial _{\mu}
\partial ^{\mu} \phi_n (x^{\mu})=\frac{n^2}{R^2}\phi_n (x^{\mu})
\end{equation}

The mass spectrum in four dimensions consists of a massless mode
$\phi_0$ and an infinite tower of massive states, the Kaluza-Klein
states,  with equal spacing
$m_n=n/R$. At energies much lower than $1/R$ only the massless mode
can be excited and all the higher ones decouple.
The charge of the symmetry (\ref{eqKK2}) is given by the
fifth component of the momentum. In this example, the massless mode is
neutral, but this is a property of the particular example we
considered. It is also clear that the $U(1)$ gauge symmetry we obtained is a consequence of our choice of toroidal compactification.  $U(1)$ is the group of the circle. Other groups can be obtained if we have larger compact spaces. The moral of the story is twofold: (i) Space-time symmetries of the higher dimensions may appear as internal gauge symmetries to a four dimensional observer. (ii) For every low mass particle we obtain an infinite tower of Kaluza-Klein states with masses which are integer multiplets of the inverse compactification size. 

This example was presented in the framework of quantum field theory. Strings may add one new element. A string can get wrapped around the compact space. Therefore, string configurations can be classified according to a topological property, namely the number of times the string gets around the space. We call this number ``the winding number''. We again obtain a tower of states, the winding states. Their masses are proportional to the energy of the corresponding string configuration, therefore they are integer multiplets of the compactification size. Kaluza-Klein states carry conserved charges. Winding states carry topological charges. It won't come as a surprise to realise that under an $R\rightarrow 1/R$ duality the Kaluza-Klein and the winding states exchange roles. Let me only add that the topological property of winding around a compact space can be found also with more general p-branes. 

Coming now to physics, it is clear that we need to consider compact spaces of six, if we start from string theory, or seven, if from M theory, dimensions. In either case the important parameter is the compactification size. Going back to the Kaluza-Klein example, we see that nothing determines $R$. This is also the case for general superstring theories. Different compactification sizes often correspond to degenerate but inequivalent states of the system. They are labeled by the values of massless scalar fields called ``moduli''. Unless we develop a much more profound understanding of the non-perturbative dynamics of string theory, including the mechanism of supersymmetry breaking, we shall not be able to compute the values of all the moduli and the compactification size will remain a phenomenological parameter. Although it can take any value, for presentation purposes I shall distinguish two cases: Small size and large size. 

\subsubsection{Small compact dimensions}

Here by ``small'' I mean smaller than anything we can measure in any
foreseeable future. In practice I shall assume that $R$ is of the order
of the Planck length $10^{-33}$ cm. This is the old fashion way, the
one used before the advent of dualities and branes but I still believe
that it yields more realistic models. In particular, it is only in
this class of models that we obtain a prediction concerning the
experimentally found gauge coupling unification of Figure \ref{figunif}.  We start from a ten dimensional superstring theory. Since we do not know how to solve the theory, we cannot rigorously extract its phenomenological consequences. For this reason we have invented a new term, we talk about ``superstring inspired models''. If the programme
 is correct, one should find that in the ground state of the string theory   the topology of space-time is given by $M_4 \otimes \tilde{M}_6$, where $M_4$ is the usual 4-dimensional Minkowski space-time and $\tilde{M}_6$ is a compact six-dimensional 
manifold. Needless to say that nobody has ever been able to find the ground state of any realistic theory 
and string theories are no exception. Assuming compactification does occur, it is obvious that the 
symmetries of $\tilde{M}_6$ will appear to the effective four-dimensional theory as internal symmetries. If, 
furthermore we restrict ourselves to the lowest excitation of the string, i.e. the massless sector, we shall obtain a good honest quantum field theory in four dimensions. It is clear from these 
considerations that, even if the initial ten-dimensional superstring theory is unique, we are still  
unable to deduce a unique effective low-energy theory. The result  depends on the choice of $\tilde{M}_6$ 
and/or the choice of strategy. Here we shall restrict ourselves to two models, each one exemplifying a 
different approach. In the first we shall assume that the compactification from ten to four dimensions 
takes place at the string level. We thus obtain a four-dimensional string theory whose zero-mass limit 
will give an effective low-energy field theory. Here ``low-energy'' is meant with respect to the Planck-mass and could well encompass the scale of grand unification. In the second the order of the two 
operations is reversed: We first consider the zero-mass limit of the ten-dimensional string theory thus 
obtaining an effective ten-dimensional field theory. Then, as a second step, we consider a Kaluza-Klein 
compactification to four dimensions. The two approaches are not equivalent because in each case the 
compactification takes place in a different space. Since this determines the internal symmetries of the 
resulting effective field theory, it follows that we shall end up with different grand unified models. 
None will turn out to be identical to the ones we examined so far. One of the reasons is that these 
methods do not seem to produce models with Higgs fields belonging to high representations of the 
internal symmetry group.

{\it 1) The $SU(5) \otimes U(1)$ model:} There are at least two ways to study a grand unified theory based on the group $SU(5) \otimes U(1)$. The first is 
purely traditional and string theory never enters. In fact, that is how historically this model was 
first proposed. Starting from the $SO(10)$ grand unified theory, one can notice that there exist two 
inequivalent ways to break $SO(10)$ to $SU(5)$. One is the usual one studied in Section \ref{secso10}. There exist, 
however, a second one corresponding to the two inequivalent ways to embed $SU(5)$ into $SO(10)$. This 
breaking is $SO(10) \rightarrow SU(5) \otimes U(1)$ and the 16-dimensional representation of $SO(10)$ decomposes into:

\begin{equation} 
\label{eqph1}
16 \rightarrow (10, \frac{1}{2})\oplus (\bar{5}, -\frac{3}{2})	\oplus (1,\frac{5}{2})
\end{equation}

\noindent where the first number denotes the $SU(5)$ representation and the second the $U(1)$ charge. The electric 
charge operator, which is always a generator of $SO(10)$, is no more one of
$SU(5)$. It is rather a combination of the $U(1)$ generator and the neutral component of $SU(5)$. It follows 
that the sum of the charges of the members of an $SU(5)$ representation no longer vanishes. The fermion 
assignment in the $\bar{5}$ and 10 representations is now flipped with respect to ordinary $SU(5)$ with $u \leftrightarrow d$ 
and $\nu \leftrightarrow e$. For example, the  $\bar{5}$ representation is: 

\begin{equation}
\label{eqph2} 
\bar5 = {\left( \begin{array}{c}
{u^c}_1 \\ {u^c}_2 \\ {u^c}_3 \\ {\nu}_e  \\ e^-
\end{array}
\right)}_L
\end{equation}

\noindent and similarly for the 10.

This model appears in a sort of natural way when one starts from a four dimensional string theory 
and considers the zero-mass limit. The connection is not rigorous but it is suggested by the following 
considerations: (i) Through the string compactification one encounters orthogonal symmetry groups. So 
one expects to obtain a supersymmetric grand unified theory with a gauge group which is the result of 
breaking of some orthogonal group. (ii) As we mentioned before, in the zero-mass limit of a string 
theory we do not find matter multiplets belonging to adjoint, or higher, representations. Almost all 
traditional grand unified theories require Higgs fields in such representations. $SU(5) \otimes U(1)$ is an 
exception:
it can be broken to the standard model using only complex 5's and l0's. The vacuum expectation value 
of the 10 breaks $SU(5) \otimes U(1)$ to $U(1) \otimes SU(2) \otimes SU(3)$ and that of the 5 breaks $U(1) \otimes SU(2)$ to 
$U(1)_{em}$.

Let me now mention the main features of this model:

\noindent (i) A natural triplet-doublet splitting of the Higgs fields. Let me first remind you that in ordinary $SU(5)$ there was no way to obtain 
such a splitting without fine-tuning the parameters of the model. This was part of the problem of gauge 
hierarchy. Here this splitting is natural. The 5 will get its mass through the coupling with 10 which has 
a large vacuum expectation value. It is easy to verify that the structure of the invariant couplings is such 
that only the triplet becomes heavy. Similarly, we give large Dirac masses to the colour-triplet 
fermionic partners of the 5-Higgs fields.

\noindent (ii) With respect to neutrino masses the situation is similar to that of $SO(10)$ since we have an $SU(5)$ 
singlet field (see (\ref{eqph1})). Again, we can have large Majorana masses which yield naturally very light 
physical neutrinos.

\noindent (iii) Finally, let me mention that the prediction for proton decay is different from that of ordinary $SU(5)$. 
Like in any supersymmetric theory, the expected life-time is longer. 
Furthermore, because of the flipped assignment, the main decay mode is $p \rightarrow \pi^+ \bar{\nu}$ which is expected to 
be twice as large as the ordinary  $p \rightarrow \pi^0 e^+$ mode.

{\it 2) The $[SU(3)]^3$ model:} We now come to the alternative strategy, namely we first consider a limiting 10-dimensional effective field theory and then compactify. The most interesting superstring model to start with is the 
heterotic string which has a gauge symmetry based on the group $E_8 \otimes E_8$. The important step, which 
will determine the properties of the 4-dimensional effective theory, is the choice of the six-dimensional 
manifold we shall use for compactification. We are guided by two phenomenological requirements: (i) The 4-dimensional theory must admit chiral fermions and (ii) an N = 1 supersymmetry must be 
preserved. It turns out that, at least to low orders in string perturbation theory, these requirements 
determine the structure of the manifold: it must be Ricci flat and have $SU(3)$ holonomy. Such manifolds have been studied mathematically and are known as ``Calabi-Yau'' manifolds. (Notice that if we start from M theory we must compactify in a seven dimensional manifold for which the holonomy group is $G_2$. Such models have not been analysed in detail.) Identifying the $SU(3)$ 
group with a subgroup of the original $E_8 \otimes E_8$ symmetry, reduces the observable gauge group to $E_6 \otimes E_8$. We shall assume that all matter is singlet under $E_8$ which will constitute a ``hidden'' sector in the 
theory. It follows that the grand unification gauge group will be a subgroup
of $E_6$.

From now on everything depends on the particular Calabi-Yau manifold we choose. Obviously, it is 
not easy to visualise such manifolds but they can often be defined as the set of zeros of systems of 
algebraic equations. A particular example which has been studied in the literature and seems to 
reproduce many features of low energy phenomenology, is the so-called ``Tian-Yau'' manifold. In order 
to construct it explicitly we proceed in two steps: We start with the simply connected Calabi-Yau 
manifold $K_0$ defined as the complete intersection of the following three equations in $CP^3 \otimes CP^3$:

\[\sum_{i=0}^{3}X_i^3+a_1X_0X_1X_2+a_2X_0X_1X_3=0\]

\begin{equation} 
\label{eqCY}
\sum_{i=0}^{3}Y_i^3+b_1Y_0Y_1Y_2+b_2Y_0Y_1Y_3=0
\end{equation}

\[\sum_{i=0}^{3}c_iX_iY_i +c_4X_2Y_3+c_5X_3Y_2=0\]

\noindent where $X_i$ and $Y_i$ ($i$ = 0, 1, 2, 3) are the homogeneous coordinates of the two $CP^3$'s. The complex 
parameters $a_i$, $b_i$, ($i$ = 1, 2) and $c_i$ ($i$ = 0, . . . ,5) are restricted by some transversality conditions. The second 
step consists in identifying a group $Z_3$ of discreet transformations which act freely on $K_0$ and define the 
multiply connected Tian-Yau manifold $K$ by dividing $K_0$ by $Z_3$. Since $K$ is multiply connected we can consider non-trivial Wilson loops on it. It is assumed that, in 
a non-perturbative way, they break $E_6$ to a certain subgroup $H$. The important point is that the geometry of $K$ determines also the number of matter supermultiplets that 
are allowed. They are given by $\chi (K)/2$, where $\chi (K)$ is the Euler characteristic of $K$. For the 
Tian-Yau manifold $\chi (K)$ = 6, so the model allows for just three families of quarks and leptons. 
The most interesting class of models have $H = SU(3) \otimes SU(3) \otimes SU(3)$ and a surviving group of 
discrete symmetries of $K$. It turns out that there exist two inequivalent ways of breaking $E_6$ to 
$[SU(3)]^3$ through Wilson loops. In each case a complete classification of all possible groups of 
discrete symmetries has been carried out. The result is that we can construct forty five different 
models using one breaking and twenty one models using the other, thus obtaining a total of sixty 
six possible models. They all have three complete 27's, a number of incomplete 27's and $\bar{27}$'s and 
the discrete symmetries determine the Yukawa couplings and the Higgs potential. Not all of 
them have been analysed but we can find among them some phenomenologically viable models. 
This is extremely encouraging since it is the first time that the number of families is determined 
by geometrical considerations.

\subsubsection{Large compact dimensions}

Here by large we mean of order 1-100 inverse TeV, distances you may hope to explore, directly or indirectly, during your life-time as physicists. We assume that the string scale is small, much smaller than the Planck scale. Such models have been considered seriously only during the last years. The reason is that large compactification radius corresponds, for the heterotic string, to large coupling, where no perturbation expansion is reliable. It is only through strong $\leftrightarrow $ weak coupling duality that we can approach this region. 

Claiming that the string scale may be as low as 1 TeV sounds at first absurd, since it means that gravitational interactions will become strong at that scale. Of course, the answer is in the extra compact dimensions. Gravity is weak because it spreads over many extra dimensions. Let us assume that there are $n$ of them and let us take, for simplicity, the case in which all have size $R$. In this  $(4+n)$-dimensional space-time  the gravitational potential of a mass, $m,$
is given by 

\begin{equation}
V(r)=\frac{Km}{M_{\ast }^{2+n}}\text{ }\frac{1}{r^{1+n}},\text{ }r<R
\label{eqpot1}
\end{equation}

\noindent where $M_{\ast }$ is the fundamental mass scale of the $(4+n)$ dimensional
theory and $K$ a numerical constant. At distances larger than the compactification size of the new dimension,
space-time effectively becomes four dimensional and the potential has the
usual $1/r$ behaviour

\begin{equation}
\label{eqpot2}
V(r)=\frac{Km}{M_{\ast }^{2+n}R^{n}}\text{ }\frac{1}{r}\equiv \frac{Gm}{r},%
\text{ }r>R
\end{equation}

\noindent which means that Newton's law is modified only at distances smaller or equal to $R$. 
Identifying the last two terms in (\ref{eqpot2}) shows that the usual four dimensional form of
gravity applies with the Planck mass given by

\begin{equation}
\label{eqPl1}
M_{P}^{2}=\frac{2}{K}M_{\ast }^{2+n}R^{n}
\end{equation}

The extreme case is to choose $M_{\ast }$=1TeV $i.e.$ close to the electroweak
breaking scale. Then $n$=1 would give $R$ of the order of one million kms,
clearly excluded by all sorts of terrestrial or solar system measurements, but
for $n$=2 we obtain $R \sim 0.1$ mm. This is dangerously close to present
limits, but it is clear that the scenario is perfectly viable, provided we
choose $n>2$ and/or slightly higher values for $M_{\ast }$. 

This simple picture gives the main experimental consequences of this idea: Modification of Newton's law at distances $r\leq R$ according to (\ref{eqpot1}) and gravitational interactions which become comparable in strength to other interactions at energies $E \geq M_{\ast }$. The first test will be hard to do. As we saw, $R$ decreases very rapidly with increasing $n$ and/or $M_{\ast }$ and we must be extremely lucky to fall in the experimentally accessible range. Direct tests of Newton's law at subsubmillimiter distances will be hard because gravitation is in competition with the badly known van der Waals forces. At very short distances one should consider even the Casimir attraction due to vacuum fluctuations. We are left with high energy experiments, mainly at the LHC. If $M_{\ast }$ is really of order 1 TeV, the signatures will be spectacular. Gravitons and even black holes, will be produced. The latter will decay giving ordinary particles, such as photons, or gravitons. Missing energy and momentum will be an important signature. On the other hand we shall also see the first excitations of the known particles in their respective Kaluza-Klein or winding towers and they also will provide signals you cannot miss. But again, we must be lucky. Let me remind you that no solid theoretical argument determines the value of $M_{\ast }$ which may be everywhere, from here to the Planck mass. We have chosen the value of 1 TeV essentially because it is the present experimental lower limit. The only argument is a prejudice that supersymmetry breaking may be triggered by compactification. Nevertheless, it makes experimental search even more exciting. For a moment it had even caused worries connected with the possibility of producing black holes. Are they going to be dangerous? The answer is no for a very simple reason: only extremely massive ones are stable. We can see this using a very simple argument. 

A four-dimensional black hole of mass $M$ has a Schwarzschild radius $R_S=2GM$ and a temperature $T_{BH}=M_P^2 /M$. Consequently, its thermal decay rate, which is proportional to its area, is given by $\Gamma_D \sim T_{BH}^4 R_S^2$. Unless the accretion rate is greater than this decay rate the black hole
will decay harmlessly. An upper bound to the accretion rate is given by the
energy density in the volume swept out by the black hole in one second.
Assuming the limiting case where black hole is moving with relativistic
velocity the accretion rate is $\Gamma _{A}\sim \pi R_{S}^{2}\rho$, where $\rho $ is the mean density of the matter through which the
black hole passes. Thus the condition for growth
of the black hole, $\Gamma _{A}>\Gamma _{D},$ implies $M>M_P^2 /\rho^{1/4}$ in CeV which is of order $10^{42}$ GeV!

We can easily repeat this estimation for a (4+$n$)-dimensional black hole. Replacing the four dimensional Newton's potential with that of (\ref{eqpot1}), we find for the Schwarzchild radius

\begin{equation} 
\label{eqBH1}
R_{S}=\frac{K^{\prime }}{M_{\ast }}\left( \frac{M}{M_{\ast }}\right) ^{%
\frac{1}{1+n}} 
\end{equation}

\noindent with $K^{\prime }$ another numerical coefficient. The temperature is given by 

\begin{equation}
T_{BH}=\left( \frac{M_{\ast }}{M}\right) ^{\frac{1}{1+n}}M_{\ast }
\label{bhtempd}
\end{equation}

The decay rate in this case is 
\begin{eqnarray}
\Gamma _{D} &\approx &T_{BH}^{4+n}R_{S}^{2+n}  \notag \\
&=&M_{\ast }^{2}\left( \frac{M_{\ast }}{M}\right) ^{\frac{2}{2+n}}
\label{decay}
\end{eqnarray}

Since normal matter lives in four dimensions, the accretion rate has the same form as before, so we
immediately obtain the bound for a stable black hole given by

\begin{equation}
\label{BHbound}
M>M_{\ast }\left( \frac{10^{4}M_{\ast }}{GeV}\right) ^{1+n}
\end{equation}

\noindent which, even for the extreme case $n=2$ and $M_{\ast }$=1 TeV, gives $M>10^{24}$ GeV. We can show that this rough estimation remains valid even if we take into account other effects, such as the possibility of normal matter propagating in the extra dimensions, black holes carrying conserved charges, etc. It is conceivable, although not very probable, that black holes offer the greatest discovery an accelerator can make, but they do not present any imaginable threat.

Let me now present very briefly some of the models that exemplify these ideas. There is a great variety of them and, to my taste, none imposes itself by predictive power, simplicity and/or aesthetic beauty. In particular, contrary to some claims in the literature, none incorporates the prediction of coupling constant unification we had in ordinary supersymmetric grand unified theories. This is not surprising, since they all introduce completely new physics above 1 TeV. They can still fit the data but they have no prediction. They all use the brane world hypothesis in which we are living on some collection of extended objects. D-branes are the favourite ones, although, in some models other extended objects, like orientifolds, are introduced in addition. From this point everyone can use his/her imagination and, indeed, practically any conceivable configuration has been used. We can distinguish roughly two classes according to whether gravity is spread everywhere in the bulk, or it is also localised on a brane. The simplest version starts from Type I string theory with different collections of D-branes. The branes we are living on must have at least three spatial dimensions and the unbroken gauge symmetries of the Standard Model are generated by putting together $l$ identical branes for $SU(l)$. The Higgs mechanism is generated by pulling some of them a certain distance apart. 

Let us consider a space with nine spatial dimensions in which we have introduced a collection of Dp-branes. If p is larger than three, the extra dimensions must be also compact. We call them ``parallel'' dimensions p=3+$n_{\parallel}$ and they are compactified with size $R_{\parallel}$. The remaining  dimensions are called ``transverse'' $n_{\perp}$=9-p and have a compactification size $R_{\perp}$. In this space there are several kinds of strings: 

(i) Closed strings. They contain the graviton as well as the other string modes. They have Kaluza-Klein excitations in each compact dimension with mass-spacings $1/R_{\parallel}$ and $1/ R_{\perp}$, as well as winding excitations with spacings $R_{\parallel}/l_s^2$ and $R_{\perp}/l_s^2$ respectively, where $l_s$ is the length of the string. 

(ii) Open strings which start and end on the same D-brane. They give massless
states, such as gauge bosons of an unbroken group, living on the brane and
have Kaluza-Klein excitations on $n_{\parallel}$ and winding on $n_{\perp}$.

(iii) Open strings which stretch between branes that have been pulled apart. They give massive string modes representing gauge bosons of spontaneously broken gauge groups and have all the corresponding Kaluza-Klein and winding excitations.

(iv) If branes intersect, there may be open strings starting and ending on an intersection, which may correspond to our three-dimensional space.  They give massless string states, have winding states on $n_{\perp}$ and no Kaluza-Klein states. 

With these ingredients we can construct a huge variety of phenomenological models. If we choose $l_s$ and the compactification radii large enough, all this plethora of new states may become observable. If Nature is extremely kind to us, experiments at the LHC will be a tremendous fun with a new discovery every few minutes. The reverse side of the story is that we have enough parameters to accommodate practically any result, positive or negative. In the meantime it helps to be optimistic.

\section{Epilogue}

In these lectures I tried to give my personal view of the large class of theories that come under the general name ``Beyond the Standard Model''. They represent thirty years of theoretical high energy physics, thirty years of efforts to understand our world. The trouble is that during all this period theorists have worked with very little experimental input. The enormous complexity of modern high energy experiments has stretched the time between the conception, the design and the completion of an experiment to decades. We are approaching the limit of the professional life of a physicist. This is probably  the greatest danger of our field. We have been extremely frustrated during all these years and we cannot hide our excitement now that the long awaited experiments are at last in sight. Never in the past a new experimental facility was loaded with so many expectations. We are confident that great and exciting discoveries lay ahead.

\section{Bibliography}

These lectures have touched at so many topics, that a complete list of references would fill a volume. I just give a short list of suggested further reading. Only text books and review articles are included and the selection criterion is my familiarity with them and their easy availability.
\vskip 0.5cm

\noindent 1) The subject has been taught in practically every CERN School of
the last years. The corresponding Yellow Reports offer a very complete picture.

\noindent 2) For an introduction to Quantum Field Theory and Gauge Theories see:

-Cl. Itzykson and J.B. Zuber, Quqntum field theory, McGraw-Hill Inc. 1980.

-S. Weinberg, The quantum theory of fields, Vol. 1,2, Cambridge Univ. Press,
 1995.

-M. E. Peskin and D. V. Schroeder, An introduction to Quantum Field Theory,
 Addison-Wesley (1995) 

\noindent 3) For an introduction to the Standard Model see:

-R. Kleiss, Lectures at this School

-J. Stirling,  Lectures at this School 

\noindent 4) For neutrino physics see:

-H. Murayama, Lectures at this School

\noindent 5) For magnetic monopoles see:

-S. Coleman, in Gauge theories in high energy physics, M.K. Gaillard
and R. Stora, eds. Les Houches summer school 1981.

-P. Goddard and D. Olive, Rep. Prog. Phys. {\bf 41} (1978) 1357.

\noindent 6) For Supersymmetry and Supergravity see:

-J. Wess and J. Bagger, Supersymmetry and Supergravity, Princeton Univ. Press, 1992.

-S. Weinberg, The quantum theory of fields, Vol. 3, Cambridge Univ. Press,
 1995.

-M. E. Peskin, Supersymmetry in Elementary Particle Physics, hep-ph 0801.1928

\noindent 7) For Superstring theory see:

-M. Green, J. Schwarz and E. Witten, Superstring Theory, Vol. 1,2, Cambridge Univ. Press, 1988.

-J. Polchinski, String Theory, vol. 1,2, Cambridge Univ. Press, 1998.

\noindent 8) For dualities in field theory and string theory see:

-K. Intriligator and N. Seiberg, Nucl. Phys. Proc, Suppl. {\bf 45BC},
1996, hep-th/9509066

-E. Kiritsis, Supersymmetry and duality in field theory and string
theory, 1999 Cargese summer school, hep-ph/9911525

-A. Bilal, Duality in $N=2$ SUSY $SU(2)$ Yang-Mills theory. A
pedagogical introduction to the work of Seiberg and Witten. hep-th/9601007.

-M. Peskin, Duality in supersymmetric Yang-Mills theory. hep-th/9702094.

\noindent 9) For the experimental consequences of low mass compactification see:

-I. Antoniadis, Physics with large extra dimensions, CERN-TH/2001-318.; Topics
 on string phenomenology, Les Houches 2007, hep-th 0710.4267

\noindent 10) For the latest compilation of $\alpha_s$ computed from
$\tau$-decays see:

-M. Davier, S. Descotes-Genon, A. Hocker, B. Malaescu and Z. Zhang, The
 Determination of alpha(s) from Tau Decays Revisited, hep-ph 0803.0979

\end{document}